\documentclass[preprint,3p,12pt]{elsarticle}
\usepackage{amsmath}
\usepackage{stmaryrd}
\usepackage{bbding}
\usepackage{dcolumn}
\usepackage{graphicx}
\usepackage{amsfonts}
\usepackage{amssymb}
\usepackage{psfrag}
\usepackage{wrapfig}
\usepackage{subfigure}
\usepackage{makeidx}
\usepackage{bm}
\usepackage{epsf}
\usepackage{epsfig}
\usepackage{setspace}
\usepackage{color}
\usepackage{epstopdf}
\begin{document}
\title{High-order gas-kinetic scheme in general curvilinear coordinate for iLES of compressible wall-bounded turbulent flows}

\author[SUSTech1]{Guiyu Cao}
\ead{caogy@sustech.edu.cn}
\author[BNU]{Liang Pan\corref{cor}}
\ead{panliang@bnu.edu.cn}
\author[SUSTech2]{Minping Wan}
\ead{wanmp@sustech.edu.cn}
\author[HKUST1,HKUST2]{Kun Xu}
\ead{makxu@ust.hk}
\author[SUSTech1,SUSTech2]{Shiyi Chen\corref{cor}}
\ead{chensy@sustc.edu.cn}

\address[SUSTech1]{Academy for Advanced Interdisciplinary Studies, Southern University of Science and Technology, Shenzhen, China}
\address[BNU]{Laboratory of Mathematics and Complex Systems, School of Mathematical Sciences, Beijing Normal University, Beijing, China}
\address[SUSTech2]{Department of Mechanics and Aerospace Engineering, Southern University of Science and Technology, Shenzhen, China}
\address[HKUST1]{Department of Mathematics, Hong Kong University of Science and Technology, Clear Water Bay, Kowloon, Hong Kong}
\address[HKUST2]{Shenzhen Research Institute, Hong Kong University of Science and Technology, Shenzhen, China}
\cortext[cor]{Corresponding author}

\begin{abstract}
In this paper, a high-order gas-kinetic scheme in general curvilinear coordinate (HGKS-cur) is developed for the numerical simulation of compressible turbulence. 
Based on the coordinate transformation, the Bhatnagar-Gross-Krook (BGK) equation is
transformed from physical space to computational space. 
To deal with the general mesh given by discretized points, the geometrical metrics need to be constructed by the dimension-by-dimension Lagrangian interpolation.  
The multidimensional weighted essentially non-oscillatory (WENO) reconstruction is adopted in the computational domain for spatial accuracy, where the reconstructed variables are the cell averaged Jacobian and the Jacobian-weighted conservative variables.  
The two-stage fourth-order method, which was developed for spatial-temporal coupled flow solvers, is used for temporal discretization. 
The numerical examples for inviscid and laminar flows validate the accuracy and geometrical conservation law of HGKS-cur. 
As a direct application, HGKS-cur is implemented for the implicit large eddy simulation (iLES) in compressible wall-bounded turbulent flows, including the compressible turbulent channel flow and compressible turbulent flow over periodic hills.
The iLES results with HGKS-cur are in good agreement with the refereed spectral methods and high-order finite volume methods. 
The performance of HGKS-cur demonstrates its capability as a powerful tool for the numerical simulation of compressible wall-bounded turbulent flows and massively separated flows.
\end{abstract}

\begin{keyword}
high-order gas-kinetic scheme, general curvilinear coordinate,
implicit large eddy simulation, wall-bounded turbulent flows, compressible turbulence.
\end{keyword}

\maketitle

\section{Introduction}
The understanding and prediction of multi-scale turbulent flows is one of the most difficult problems in both mathematics and physical sciences. 
With the development of numerical methods and super computers, great successes have been achieved by the numerical simulation of unsteady turbulent flows. 
Conceptually, the direct numerical simulation (DNS) \cite{kim1987turbulence, moin1998direct} is supposed to resolve turbulent structures above the Kolmogorov dissipation scale by using resolved grid size and time step, which solves the
Navier-Stokes equations directly and eliminates modeling entirely.
But the extremely expensive computational costs prohibit its application in high-Reynolds number turbulent flows. 
In order to study turbulent flows on the unresolved grids, the large eddy simulation (LES) \cite{smagorinsky1965,sagaut2006large} have been developed. 
LES solves the filtered Navier-Stokes equations with resolvable turbulent structures above the inertial scale. 
For unsteady separated turbulent flows, LES has gradually become an indispensable tool to obtain high-resolution turbulent flow fields. 
The high-order numerical schemes play a key role in the numerical simulation of turbulence.
In the past few decades, the spectral method \cite{kim1987turbulence} and the pseudo-spectral method \cite{wang1996examination} have been well established as a powerful DNS tool for the incompressible turbulent flows \cite{lee2015direct}.  
For the simulation of near incompressible turbulence, the lattice Boltzmann method \cite{chen1998lattice,yu2005lattice} is widely used. 
Unfortunately, for the simulation of compressible turbulence with discontinuity, the shocklets may appear in the flow fields and all of them suffer from numerical instability. With the properties of high-order accuracy in smooth region and no oscillation near
shocks, the high-order finite difference method \cite{lele1992compact, fu1997high, GCL-1} have been widely developed and utilized for compressible turbulence simulation with
discontinuities \cite{samtaney2001direct, wang2010hybrid}.

Due to the significance of engineering applications and the study on fundamental physical mechanism of compressible boundary layer, the compressible wall-bounded turbulent flows have been extensively simulated using the high-order schemes. 
The representative research are briefly presented as follows: the compressible turbulent channel flow from the supersonic to hypersonic regime
\cite{coleman1995numerical,morinishi2004direct,yu2019genuine}, the supersonic and hypersonic flat plate turbulence
\cite{pirozzoli2004direct,lixinliangma8,franko2013breakdown}, the compressible separated turbulent flow over periodic hills
\cite{peller2004dns,ziefle2008large,breuer2009flow,balakumar2015dns}, and the compression ramp \cite{adams2000direct,wu2007direct,chen2017constrained} with shock-boundary interactions. 
The high-order finite difference methods are dominated in the simulation of  compressible wall-bounded turbulence, except the temporal supersonic turbulent channel flow can be simulated by spectral method \cite{coleman1995numerical}. 
Even for the hypersonic flat plate turbulence with free-stream Mach number $Ma = 8.0$, the maximum turbulent Mach number $M_t$ is less than $0.5$ \cite{lixinliangma8}, which means no strong shock-lets in such cases.  
This is the key reason why the high-order finite difference methods are the main workhorse in compressible wall-bounded turbulence simulations. 
However, due to the numerical instability when encountering strong shocklets, the highest
turbulent Mach number for high-order finite difference scheme is still limited, and the critical threshold of simulating supersonic flow remains, i.e., turbulent Mach number $M_t \leq 1.2$ for DNS of supersonic isotropic turbulence \cite{wang2010hybrid}. 
Besides, to simulate the hypersonic flows robustly, the complicated artificial viscosity and artificial heat conductivity are usually constructed in high-order
finite difference method \cite{franko2013breakdown}. 
Because of the excellent conservative properties and favorable ability in capturing
strong discontinuities, high-order finite volume scheme may pave a new way for high-resolution simulation of turbulent flows in all  flow regimes from subsonic to supersonic ones \cite{GKS-high-cao-1,GKS-high-cao-2,GKS-high-cao-3}.

In the past decades, the finite-volume gas-kinetic scheme (GKS) based on the Bhatnagar-Gross-Krook (BGK) model \cite{BGK-1,BGK-2} have been developed systematically for computations from low speed flows to supersonic ones \cite{GKS-Xu1,GKS-Xu2}.  
The GKS presents a gas evolution process from kinetic scale to hydrodynamic scale, where
both inviscid and viscous fluxes are recovered from a time-dependent and multi-dimensional gas distribution function at a cell interface.
Based on the time-dependent flux function, a reliable two-stage framework was provided for developing the GKS into fourth-order and even higher-order accuracy \cite{GRP-high-1,GKS-high-pan-1,GKS-high-ji-1,GKS-high-zhao-1}.  
More importantly, the high-order GKS (HGKS) is as robust as the second-order scheme and works perfectly from the subsonic to the hypersonic viscous flows. 
With the advantage of the finite-volume GKS, it has been developed as a powerful tool to simulate turbulent flows. 
For high-Reynolds number engineering turbulence, the GKS coupled with traditional eddy viscosity turbulence model has been developed and implemented in turbulence simulations
\cite{GKS-tub-1,GKS-tub-2,GKS-tub-3}. 
For low-Reynolds number turbulent flows, the HGKS has been directly used as a DNS tool
\cite{GKS-high-cao-1, GKS-high-cao-2}. 
Recently, a parallel code of HGKS is developed for the lare-scale DNS, where the domain decomposition and message passing interface (MPI) is used for parallel implementation \cite{GKS-high-cao-2}. 
The computational cost is comparable with the high-order finite difference method. 
For the nearly incompressible turbulent flows, the performance of HGKS is also comparable with the finite difference method.
More importantly, HGKS shows special advantages for the supersonic turbulence due to the accuracy and robustness, i.e., the supersonic isotropic turbulence with turbulent Mach number $M_t = 2.0$ has been simulated successfully \cite{GKS-high-cao-3}. 
It can be concluded that the HGKS provides a valid tool for the numerical simulation of turbulence, which is much less reported in the framework of finite volume scheme.

In previous study \cite{GKS-high-pan-2}, the high-order gas-kinetic scheme has been developed in the curvilinear coordinate for laminar flows, in which the coordinate transformations are given analytically. 
However, for more turbulent cases, the grid points are given by the discretized points and there is no analytical transformation.
In this paper, the HGKS in general curvilinear coordinate (HGKS-cur) will  be presented within the two-stage fourth-order framework. 
The curvilinear meshes can be given analytically or in the form of discretized grid points without analytical transformation.  
With the discretized grid points, the geometric metrics can be constructed by the dimension-by-dimension Lagrangian interpolation, and the
geometrical conservation law can be preserved.  
The weighted essentially non-oscillatory (WENO) reconstruction \cite{WENO-Liu, WENO-JS} is adopted in the computational domain for spatial accuracy, where the reconstructed variables are the cell averaged Jacobian and the Jacobian-weighted conservative variables.  
The two-stage fourth-order method \cite{GRP-high-1}, which was developed for spatial-temporal coupled flow solvers, is used for temporal discretization. 
Due to the lower computational costs and reasonable performance, the HGKS-cur is implemented for the implicit large  eddy simulation (iLES) \cite{iles-1,grinstein2007implicit,hu2011scale,fernandez2017hybridized,iles-2}. 
The built-in numerical dissipation acts as the subgrid-scale (SGS) dissipation, thus no explicit SGS model is utilized in iLES \cite{iles-1,grinstein2007implicit}.
The compressible wall-bounded turbulent flows, including the compressible turbulent
channel flow and compressible turbulent flow over periodic hills, are simulated. 
The performance of HGKS-cur shows its great potential for the numerical simulation of compressible wall-bounded turbulent flows. More challenging compressible
wall-bounded turbulence problems, such as the supersonic and hypersonic flat plate
turbulent boundary layer, will be investigated in the future.

This paper is organized as follows. The high-order gas-kinetic scheme in general curvilinear coordinate will be provided in Section 2. Numerical examples and discussions are included in Section 3. The last section is the conclusion.

\section{High-order gas-kinetic scheme in general curvilinear coordinate}
\subsection{BGK equation and coordinate transformation}
The three-dimensional BGK equation \cite{BGK-1} can be written as
\begin{equation}\label{bgk}
f_t+uf_x+vf_y+wf_z=\frac{g-f}{\tau},
\end{equation}
where $\boldsymbol{u}=(u,v,w)^T$ is the particle velocity, $f$ is the
three-dimensional gas distribution function, $g$ is the
three-dimensional Maxwellian distribution and $\tau$ is the
collision time. The collision term satisfies the compatibility
condition
\begin{equation}\label{compatibi}
\int \frac{g-f}{\tau} \boldsymbol{\psi} \text{d}\Xi=0,
\end{equation}
where
$\displaystyle \boldsymbol{\psi}=(1,u,v,w,\frac{1}{2}(u^2+v^2+w^2+\varsigma^2))^T$,
the internal variables
$\varsigma^2=\varsigma_1^2+\cdots+\varsigma_N^2$,
$\text{d}\Xi=\text{d}u\text{d}v\text{d}w\text{d}\varsigma^1 \cdots \text{d}\varsigma^{N}$,
$\gamma$ is the specific heat ratio and  $N=(5-3\gamma)/(\gamma-1)$
is the internal degrees of freedom for three-dimensional flows.
According to the Chapman-Enskog expansion for BGK equation, the
Euler and Navier-Stokes equations can be derived \cite{BGK-2,GKS-Xu1}.

To construct the numerical scheme in general curvilinear coordinate, a
coordinate transformation from the physical domain $(x,y,z)$ to the
computational domain $(\xi,\eta,\zeta)$ is considered as
\begin{align}\label{jacobian}
\Big(\frac{\partial(x,y,z)}{\partial(\xi,\eta,\zeta)}\Big)=\begin{pmatrix}
x_\xi & x_\eta &x_\zeta\\
y_\xi & y_\eta &y_\zeta\\
z_\xi & z_\eta &z_\zeta\\
\end{pmatrix}.
\end{align}
With above transformation, the BGK equation Eq.\eqref{bgk} can
be transformed as
\begin{align}\label{bgk2}
\frac{\partial}{\partial t}(\mathcal{J} f)
+\frac{\partial}{\partial\xi}([u\widehat{\xi}_x+v\widehat{\xi}_y+w\widehat{\xi}_z]f)
&+\frac{\partial}{\partial\eta}([u\widehat{\eta}_x+v\widehat{\eta}_y+w\widehat{\eta}_z]f)\nonumber\\
&+\frac{\partial}{\partial\zeta}([u\widehat{\zeta}_x+v\widehat{\zeta}_y+w\widehat{\zeta}_z]f)
=\frac{g-f}{\tau}\mathcal{J},
\end{align}
where $\mathcal{J}$ is the Jacobian of transformation and the
metrics above are given as follows
\begin{align}\label{jacobian_inv}
\begin{pmatrix}
\widehat{\xi}_x    & \widehat{\xi}_y   & \widehat{\xi}_z\\
\widehat{\eta}_x   & \widehat{\eta}_y  & \widehat{\eta}_z\\
\widehat{\zeta}_x   & \widehat{\zeta}_y  & \widehat{\zeta}_z\\
\end{pmatrix}=
\begin{pmatrix}
 y_\eta z_\zeta-z_\eta y_\zeta &  z_\eta x_\zeta-x_\eta z_\zeta &  x_\eta y_\zeta-y_\eta x_\zeta \\
 y_\zeta z_\xi-z_\zeta y_\xi &  z_\zeta x_\xi-x_\zeta z_\xi &  x_\zeta y_\xi-y_\zeta x_\xi \\
 y_\xi z_\eta-z_\xi y_\eta &  z_\xi x_\eta-x_\xi z_\eta &  x_\xi y_\eta-y_\xi x_\eta \\
\end{pmatrix}.
\end{align}
Taking moments and integrating Eq.\eqref{bgk2} over the control volume $V_{ijk}$, the
semi-discretized finite volume scheme reads
\begin{align}\label{gksolver}
\frac{\text{d}\widehat{Q}_{ijk}}{\text{d}t}=\mathcal{L}(\widehat{Q}_{ijk})=-\frac{1}{|V_{ijk}|}\Big[
&\int_{\eta_j-\Delta \eta/2}^{\eta_j+\Delta \eta/2}\int_{\zeta_k-\Delta \zeta/2}^{\zeta_k+\Delta \zeta/2}(\widehat{\mathbb F}_{i+1/2,j,k}-\widehat{\mathbb F}_{i-1/2,j,k})\text{d}\eta\text{d}\zeta\nonumber\\
+&\int_{\xi_i-\Delta \xi/2}^{\xi_i+\Delta \xi/2}\int_{\zeta_k-\Delta \zeta/2}^{\zeta_k+\Delta \zeta/2}(\widehat{\mathbb G}_{i,j+1/2,k}-\widehat{\mathbb G}_{i,j-1/2,k})\text{d}\xi\text{d}\zeta\nonumber\\
+&\int_{\xi_i-\Delta \xi/2}^{\xi_i+\Delta \xi/2}\int_{\eta_j-\Delta
\eta/2}^{\eta_j+\Delta \eta/2}(\widehat{\mathbb H}_{i,j,k+1/2}-\widehat{\mathbb H}_{i,j,k-1/2})\text{d}\xi\text{d}\eta\Big],
\end{align}
where the mesh is uniformly distributed in the computational domain
for simplicity, $|V_{ijk}|=\Delta \xi\Delta \eta\Delta \zeta$ and
the Jacobian weighted conservative variable in Eq.\eqref{gksolver} is defined as
\begin{align*}
\widehat{Q}_{ijk}=\frac{1}{|V_{ijk}|}\int_{V_{ijk}}\int \boldsymbol{\psi} \mathcal{J} f \text{d}\Xi\text{d}\xi\text{d}\eta\text{d}\zeta.
\end{align*}

\subsection{Gas-kinetic solver}
For the finite volume method, the key procedure is updating the
conservative flow variables inside each control volume through the
numerical fluxes. The flux in $\xi$-direction is given as an example
and Gaussian quadrature is used as
\begin{align}\label{semi-finite}
\widehat{\boldsymbol{F}}_{i+1/2,j,k}=\int_{\eta_j-\Delta
\eta/2}^{\eta_j+\Delta \eta/2}\int_{\zeta_k-\Delta
\zeta/2}^{\zeta_k+\Delta
\zeta/2}\widehat{\mathbb F}_{i+1/2,j,k}\text{d}\eta\text{d}\zeta=\Delta\eta\Delta\zeta\sum_{m,n=1}^2\omega_{mn}
S_{mn}F(\boldsymbol{\xi}_{i+1/2,j_m,k_n},t).
\end{align}
For each Gaussian quadrature point of cell interface, the geometrical metric
$S=\sqrt{\widehat{\xi}_x^2+\widehat{\xi}_y^2+\widehat{\xi}_z^2}$,
and the local particle velocity $\widetilde{\boldsymbol{u}}$ is given by
\begin{align*}
\widetilde{\boldsymbol{u}} = (\widetilde{u},\widetilde{v},\widetilde{w})=(u,v,w)\cdot(\boldsymbol{n}_x,\boldsymbol{n}_y,
\boldsymbol{n}_z),
\end{align*}
where
$\boldsymbol{n}_x=(\widehat{\xi}_x,\widehat{\xi}_y,\widehat{\xi}_z)/\sqrt{\widehat{\xi}_x^2+\widehat{\xi}_y^2+\widehat{\xi}_z^2}$
is the normal direction and $\boldsymbol{n}_y, \boldsymbol{n}_z$ are
two orthogonal tangential directions at each Gaussian quadrature
point. For gas-kinetic solver, the time dependent numerical flux can be given by
\begin{align*}
F(\boldsymbol{\xi}_{i+1/2,j_m,k_n},t)=\int\widetilde{u} \boldsymbol{\psi}
f(\boldsymbol{x}_{i+1/2,j_m,k_n},t,\widetilde{\boldsymbol{u}},\varsigma)\text{d}\widetilde{\Xi},
\end{align*}
where $\text{d}\widetilde{\Xi} =
\text{d}\widetilde{u}\text{d}\widetilde{v}\text{d}\widetilde{w}\text{d}\widetilde{\varsigma}^1 \cdots \text{d}\widetilde{\varsigma}^{N}$
and the gas distribution function
$f(\boldsymbol{x}_{i+1/2,j_m,k_n},t,\boldsymbol{u},\varsigma)$  
can be given by the integral solution of BGK equation as Eq.\eqref{bgk}
\begin{equation}\label{intergal}
f(\boldsymbol{x}_{i+1/2,j_m,k_n},t,\boldsymbol{u},\varsigma)=\frac{1}{\tau}\int_0^t
g(\boldsymbol{x}',t',\boldsymbol{u},\varsigma)e^{-(t-t')/\tau}\text{d}t'+e^{-t/\tau}f_0(-\boldsymbol{u}t,\varsigma),
\end{equation}
where
$\boldsymbol{x}'=\boldsymbol{x}_{i+1/2,j_m,k_n}-\boldsymbol{u}(t-t')$
is the trajectory of particles, $\widetilde{\boldsymbol{u}}$ is
denoted as $\boldsymbol{u}$ for simplicity, $f_0$ is the initial gas
distribution function and $g$ is the corresponding equilibrium
state. For a multi-dimensional second-order gas-kinetic solver
\cite{GKS-Xu2}, $g$ and $f_0$ can be constructed as
\begin{equation*}\label{gks_g}
    \begin{aligned}
        g = g_0(1 + \overline{a} x + \overline{b} y + \overline{c} z + \overline{A} t),
    \end{aligned}
\end{equation*}
and
\begin{equation*}\label{gks_f0}
    \begin{aligned}
        f_0 =
        \begin{cases}
            g_l [1 +  (a_l x + b_l y + c_l z) - \tau (a_l u + b_l v + c_l w + A_l)], &x \leq 0, \\
            g_r [1 +  (a_r x + b_r y + c_r z) - \tau (a_r u + b_r v + c_r w + A_r)], &x > 0,
        \end{cases}
    \end{aligned}
\end{equation*}
where $g_l$ and $g_r$ are the initial equilibrium gas distribution
functions on both sides of a cell interface, and $g_0$ is the
initial equilibrium state located at cell interface, which can be
determined through the compatibility condition as Eq.\eqref{compatibi}. Substituting $g$ and
$f_0$ into Eq.\eqref{intergal}, the second-order gas distribution
function at cell interface can be constructed as
\begin{align}\label{formalsolution_neq}
    f(\boldsymbol{x}_{i+1/2,j_m,k_n},t,\boldsymbol{u}, \varsigma) = &(1-e^{-t/\tau})g_0+
    ((t+\tau)e^{-t/\tau}-\tau)(\overline{a} u + \overline{b} v + \overline{c} w)g_0\nonumber\\
    +&(t-\tau+\tau e^{-t/\tau}){\bar{A}} g_0 \nonumber\\
    +&e^{-t/\tau}g_l[1-(\tau+t)(a_l u + b_l v + c_l w)-\tau
    A^l)] H(u)\nonumber\\
    +&e^{-t/\tau}g_r[1-(\tau+t)(a_r u + b_r v + c_r w)-\tau A^r)](1-H(u)).
\end{align}
Eq.\eqref{formalsolution_neq} presents a gas evolution process from kinetic scale to hydrodynamic scale, where both inviscid and viscous 
fluxes are recovered from a time-dependent and multi-dimensional gas distribution function at a cell interface.
More details of the second-order gas-kinetic solver can be found in refereed paper
\cite{GKS-Xu1,GKS-Xu2}.  To achieve high-order accuracy in space and
time, the high-order spatial reconstruction and the multi-stage time
discretization will be provided in the following subsections.

\subsection{Spatial reconstruction}
High-order gas-kinetic scheme has been developed in the curvilinear coordinate \cite{GKS-high-pan-2}, where the coordinate transformations are given analytically. 
For these cases, the terms in Eq.\eqref{jacobian} at quadrature points
can be calculated by taking derivatives of the transformation
directly, and the geometrical conservation law can be preserved
automatically. In general curvilinear coordinate, the grid points are given by
the discretized points and there is no analytical transformation.
In addition, the reconstruction of geometrical metrics is also needed to achieve
the spatial accuracy and geometrical conservation law.

As preparation, the derivative terms can be given by the Lagrangian
interpolation at each grid point
\begin{equation}\label{deri}
\begin{aligned}
(\boldsymbol{x}_\xi)_{ijk}=&\frac{1}{12\Delta\xi} \Big(8(\boldsymbol{x}_{i+1,j,k}-\boldsymbol{x}_{i-1,j,k})-(\boldsymbol{x}_{i+2,j,k}-\boldsymbol{x}_{i-2,j,k})\Big) ,\\
(\boldsymbol{x}_\eta)_{ijk}=&\frac{1}{12\Delta\eta} \Big(8(\boldsymbol{x}_{i,j+1,k}-\boldsymbol{x}_{i,j-1,k})-(\boldsymbol{x}_{i,j+2,k}-\boldsymbol{x}_{i,j-2,k})\Big) ,\\
(\boldsymbol{x}_\zeta)_{ijk}=&\frac{1}{12\Delta\zeta} \Big(8(\boldsymbol{x}_{i,j,k+1}-\boldsymbol{x}_{i,j,k-1})-(\boldsymbol{x}_{i,j,k+2}-\boldsymbol{x}_{i,j,k-2})\Big),
\end{aligned}
\end{equation}
where $\boldsymbol{x}_{ijk}=(x,y,z)_{ijk}$ is the coordinate of each
grid point. To preserve the geometric conservation law (GCL) \cite{GCL-1,
GCL-2}, each term in Eq.\eqref{jacobian_inv} should be evaluated by the
symmetric conservative forms, and $ (\widehat{\xi}_x,
\widehat{\xi}_y,\widehat{\xi}_z)$ is given as an example
\begin{equation}\label{deri2}
\begin{aligned}
\widehat{\xi}_x=&\frac12\big((zy_\eta)_\zeta-(yz_\eta)_\zeta+(zy_\zeta)_\eta-(yz_\zeta)_\eta\big),\\
\widehat{\xi}_y=&\frac12\big((xz_\eta)_\zeta-(zx_\eta)_\zeta+(zx_\zeta)_\eta-(zx_\zeta)_\eta\big),\\
\widehat{\xi}_z=&\frac12\big((yx_\eta)_\zeta-(xy_\eta)_\zeta+(yx_\zeta)_\eta-(xy_\zeta)_\eta\big),
\end{aligned}
\end{equation}
where the terms $zy_\eta, yz_\eta, zy_\zeta, yz_\zeta,\cdots$ at the
grid point can be prepared by $\boldsymbol{x}_{ijk}$ and Eq.\eqref{deri} for
$(\boldsymbol{x}_\xi)_{ijk}$, $(\boldsymbol{x}_\eta)_{ijk}$,
$(\boldsymbol{x}_\eta)_{ijk}$. The next step is  the
dimension-by-dimension Lagrangian interpolation from the grid points
to the quadrature points, and two-point Gaussian quadrature is used
for spatial accuracy. The interpolated variables and their spatial
derivatives can be given by
\begin{equation*}
\begin{aligned}
\boldsymbol{\alpha}_1=&\frac{1}{216} \big((-9-\sqrt{3}) \boldsymbol{\alpha}_{i-1}+(117+39\sqrt{3})\boldsymbol{\alpha}_{i}
+(117-39 \sqrt{3})\boldsymbol{\alpha}_{i+1}+ (\sqrt{3}-9)\boldsymbol{\alpha}_{i+2}\big),\\
\boldsymbol{\alpha}_2=&\frac{1}{216} \big((\sqrt{3}-9)\boldsymbol{\alpha}_{i-1}+(117-39 \sqrt{3})\boldsymbol{\alpha}_{i}
+(117+39 \sqrt{3})\boldsymbol{\alpha}_{i+1}+(-9-\sqrt{3})\boldsymbol{\alpha}_{i+2}\big),
\end{aligned}
\end{equation*}
and
\begin{equation*}
\begin{aligned}
(\boldsymbol{\alpha}_\eta)_1=&\frac{1}{12\Delta \eta} \big(-\sqrt{3}\boldsymbol{\alpha}_{i-1}-(12-\sqrt{3})\boldsymbol{\alpha}_{i}
+(12+ \sqrt{3})\boldsymbol{\alpha}_{i+1}-\sqrt{3}\boldsymbol{\alpha}_{i+2}\big),\\
(\boldsymbol{\alpha}_\eta)_2=&\frac{1}{12\Delta \eta} \big(\sqrt{3}\boldsymbol{\alpha}_{i-1}-(12+\sqrt{3})\boldsymbol{\alpha}_{i}
+(12- \sqrt{3})\boldsymbol{\alpha}_{i+1}+\sqrt{3}\boldsymbol{\alpha}_{i+2}\big),
\end{aligned}
\end{equation*}
where $\boldsymbol{\alpha}$ represents the variables for
interpolation, i.e. $zy_\eta, yz_\eta, zy_\zeta, yz_\zeta, \cdots$. Thus, the variables in Eq.\eqref{deri2} can be given at Gaussian quadrature point.

In the computation, the cell averaged Jacobian and Jacobian weighted
conservative variables are needed for spatial reconstruction, and
both of them are given according to the following quadrature rule
\begin{align*}
\widehat{\mathcal{J}}_{ijk}&=\int_{V_{ijk}}\mathcal{J} \text{d}\xi\text{d}\eta\text{d}\zeta=\sum_{l,m,n}\mathcal{J}_{l,m,n}\Delta \xi\Delta \eta\Delta \zeta,\\
\widehat{Q}_{ijk}=&\int_{V_{ijk}}\mathcal{J} Q\text{d}\xi\text{d}\eta\text{d}\zeta=\sum_{l,m,n}(\mathcal{J}Q)_{l,m,n}\Delta \xi\Delta \eta\Delta \zeta,
\end{align*}
where the subscripts $(l,m,n)$ represent the index of three-dimensional Gaussian
quadrature points for cell $V_{ijk}$. For the high-order spatial
accuracy, the fifth-order WENO method \cite{WENO-Liu, WENO-JS} is adopted, and the dimension-by-dimension reconstruction is applied for the three-dimensional computation. 
With the WENO reconstruction of $\widehat{\mathcal{J}}$ and $\widehat{Q}$, the point value of $(\mathcal{J} Q), \mathcal{J}$ can be reconstructed at each
Gaussian quadrature points of cell interface, and  the point value $Q$ can be
calculated by
\begin{align*}
Q=\frac{(\mathcal{J} Q)}{\mathcal{J}}.
\end{align*}

For the numerical scheme with Riemann solvers, the numerical fluxes
can be fully given by reconstructed conservative variables at both
side of cell interface. However, for the gas-kinetic solver, the
spatial derivatives of the conservative variables at Gaussian
quadrature points are also needed for the time dependent evolution. 
The spatial reconstruction is performed in the computational
space, and $Q_\xi, Q_\eta, Q_\zeta$ can be obtained by the chain
rule
\begin{align*}
\frac{(\mathcal{J}Q)_{\xi}-Q\mathcal{J}_\xi}{\mathcal{J}}=Q_\xi&=Q_x x_\xi+Q_y y_\xi+Q_z z_\xi,\\
\frac{(\mathcal{J}Q)_{\eta}-Q\mathcal{J}_\eta}{\mathcal{J}}=Q_\eta&=Q_x x_\eta+Q_y y_\eta+Q_z z_\eta,\\
\frac{(\mathcal{J}Q)_{\zeta}-Q\mathcal{J}_\zeta}{\mathcal{J}}=Q_\zeta&=Q_x x_\zeta+Q_y y_\zeta+Q_z z_\zeta.
\end{align*}
The directional derivatives can be normalized as follows
\begin{align*}
Q_{\xi'}=Q_\xi/|\boldsymbol{x}_\xi|,~~&\boldsymbol{\tau}_1=(x_\xi, y_\xi,z_\xi)/|\boldsymbol{x}_\xi|,\\
Q_{\eta'}=Q_\eta/|\boldsymbol{x}_\eta|,~~&\boldsymbol{\tau}_2=(x_\eta, y_\eta,z_\eta)/|\boldsymbol{x}_\eta|,\\
Q_{\zeta'}=Q_\zeta/|\boldsymbol{x}_\zeta|,~~&\boldsymbol{\tau}_3=(x_\zeta, y_\zeta, z_\zeta)/|\boldsymbol{x}_\zeta|,
\end{align*}
where $\boldsymbol{\tau}_1, \boldsymbol{\tau}_2,
\boldsymbol{\tau}_3$ can be obtained from the coordinate
transformation. For the general curvilinear coordinate, they are not orthogonal and
$\boldsymbol{\tau}_i$ can be presented as
\begin{align*}
\boldsymbol{\tau}_i=(\boldsymbol{\tau}_i,\boldsymbol{n}_x)\boldsymbol{n}_x+(\boldsymbol{\tau}_i,\boldsymbol{n}_y)\boldsymbol{n}_y
+(\boldsymbol{\tau}_i,\boldsymbol{n}_z)\boldsymbol{n}_z.
\end{align*}
The spatial derivatives in the local orthogonal coordinate are fully
determined by the following relation
\begin{equation*}
\begin{aligned}
Q_{\xi'}&=(\boldsymbol{\tau}_1,\boldsymbol{n}_x)\frac{\partial Q}{\partial\boldsymbol{n}_x}
+(\boldsymbol{\tau}_1,\boldsymbol{n}_y)\frac{\partial Q}{\partial\boldsymbol{n}_y}
+(\boldsymbol{\tau}_1,\boldsymbol{n}_z)\frac{\partial Q}{\partial\boldsymbol{n}_z},\\
Q_{\eta'}&=(\boldsymbol{\tau}_2,\boldsymbol{n}_x)\frac{\partial Q}{\partial\boldsymbol{n}_x}
+(\boldsymbol{\tau}_2,\boldsymbol{n}_y)\frac{\partial Q}{\partial\boldsymbol{n}_y}
+(\boldsymbol{\tau}_2,\boldsymbol{n}_z)\frac{\partial Q}{\partial\boldsymbol{n}_z},\\
Q_{\zeta'}&=(\boldsymbol{\tau}_3,\boldsymbol{n}_x)\frac{\partial Q}{\partial\boldsymbol{n}_x}
+(\boldsymbol{\tau}_3,\boldsymbol{n}_y)\frac{\partial Q}{\partial\boldsymbol{n}_y}
+(\boldsymbol{\tau}_3,\boldsymbol{n}_z)\frac{\partial Q}{\partial\boldsymbol{n}_z}.
\end{aligned}
\end{equation*}
More details about spatial reconstruction can be found in previous
work \cite{GKS-high-pan-1,GKS-high-cao-1,GKS-high-pan-2}.

\subsection{Temporal discretization}
With the time dependent flux function, the two-stage fourth-order
time-accurate method \cite{GRP-high-1,GKS-high-pan-1} can be adopted
for temporal discretization. Consider the time dependent numerical
flux as Eq.\eqref{semi-finite}, the state $\widehat{Q}^{n+1}$ at
$t_{n+1}=t_n+\Delta t$ can be updated with
\begin{equation}\label{two-stage}
    \begin{split}
        &\widehat{Q}^*=Q^n+\frac{1}{2}\Delta t\mathcal {L}(\widehat{Q}^n)+\frac{1}{8}\Delta
        t^2\partial_t\mathcal{L}(\widehat{Q}^n), \\
        \widehat{Q}^{n+1} = &\widehat{Q}^n+\Delta t\mathcal {L}(\widehat{Q}^n)+\frac{1}{6}\Delta
        t^2\big(\partial_t\mathcal{L}(\widehat{Q}^n)+2\partial_t\mathcal{L}(\widehat{Q}^*)\big),
    \end{split}
\end{equation}
where the subscripts are omitted. For hyperbolic equations, it can be proved that the above temporal discretization Eq.\eqref{two-stage} provides a
fourth-order time accurate solution for $\widehat{Q}^{n+1}$. To
implement two-stage fourth-order method for Eq.\eqref{semi-finite},
a linear function is used to approximate the time dependent
numerical flux
\begin{align}\label{expansion-1}
    \widehat{\boldsymbol{F}}_{i+1/2,j,k}(t)\approx\widehat{\boldsymbol{F}}_{i+1/2,j,k}^n+ \partial_t
    \widehat{\boldsymbol{F}}_{i+1/2,j,k}^n(t-t_n).
\end{align}
Integrating Eq.\eqref{expansion-1} over $[t_n, t_n+\Delta t/2]$ and
$[t_n, t_n+\Delta t]$, the following two equations read
\begin{align*}
    \widehat{\boldsymbol{F}}_{i+1/2,j,k}^n\Delta t&+\frac{1}{2}\partial_t
    \widehat{\boldsymbol{F}}_{i+1/2,j,k}^n\Delta t^2 =\int_{t_n}^{t_n+\Delta t}\widehat{\boldsymbol{F}}_{i+1/2,j,k}(t)\text{d}t, \\
    \frac{1}{2}\widehat{\boldsymbol{F}}_{i+1/2,j,k}^n\Delta t&+\frac{1}{8}\partial_t
    \widehat{\boldsymbol{F}}_{i+1/2,j,k}^n\Delta t^2 =\int_{t_n}^{t_n+\Delta t/2}\widehat{\boldsymbol{F}}_{i+1/2,j,k}(t)\text{d}t.
\end{align*}
The coefficients at the initial stage can be determined by solving
the linear system, and the flow variables $\widehat{Q}^*$ at the
intermediate stage can be updated. Similarly,
$\mathcal{L}(\widehat{Q}^*)$ and $\partial_t\mathcal
{L}(\widehat{Q}^*)$ at the intermediate state can be constructed and
$\widehat{Q}^{n+1}$ can be updated as well. More details of the
two-stage fourth-order temporal discretization can be found in
refereed paper \cite{GRP-high-1, GKS-high-pan-1}. 
Up to this point, the so-called HGKS in general curvilinear coordinate is presented with the second-order gas-kinetic solver, as well as the fifth-order spatial
reconstruction and two-stage fourth-order time discretization.

\section{Numerical simulation and discussion}
In this section, numerical tests from the nearly incompressible flow
to the supersonic one will be presented to validate the HGKS-cur. 
For the numerical examples of this section, the grid points are
given by analytical transformations or discretized points. While, the
dimension-by-dimension Lagrangian interpolation is used for spatial
accuracy in all the meshes.
For following smooth flows without discontinuities, the collision time takes
\begin{align*}
\tau=\frac{\mu}{p},
\end{align*}
where $\mu$ is the dynamic viscous coefficient and $p$ is the
pressure at the cell interface. The ideal gas is assumed and the
ratio of specific heat $\gamma = 1.4$ is adopted.  It is well known that the BGK scheme corresponds to unit Prandtl number. To achieve the targeted Prandtl number, the Prandtl number is modified  by modifying energy flux as previous work \cite{GKS-Xu2}.

Due to the explicit computation of HGKS, a parallel strategy has been developed, where the two-dimensional domain decomposition is used \cite{GKS-high-cao-2}.
The procedure is the only data communication of the algorithm, which
is handled by the MPI libraries. The total number of cells is
$N_x\times N_y\times N_z$, and the computational domain is divided
into $n_y$ parts in $y$-direction, $n_z$ parts in $z$-direction and
no division is used in $x$-direction. The processor $P_{jk},
j=0,\cdots,n_y-1, k=0,\cdots,n_z-1$ handles a sub-domain with $N_x \times
ny_j\times nz_k$ cells.
The scalability of our MPI code is examined
by measuring the wall clock time against the number of processors,
which scales properly with the number of processors used. 
It is indicated that the data communication
crossing nodes costs a little time and the computation for flow
field is the dominant one.
Thus, the same parallel strategy is applied in current HGKS-cur for following numerical tests.

\begin{figure}[!h]
	\centering
	\includegraphics[width=0.465\textwidth]{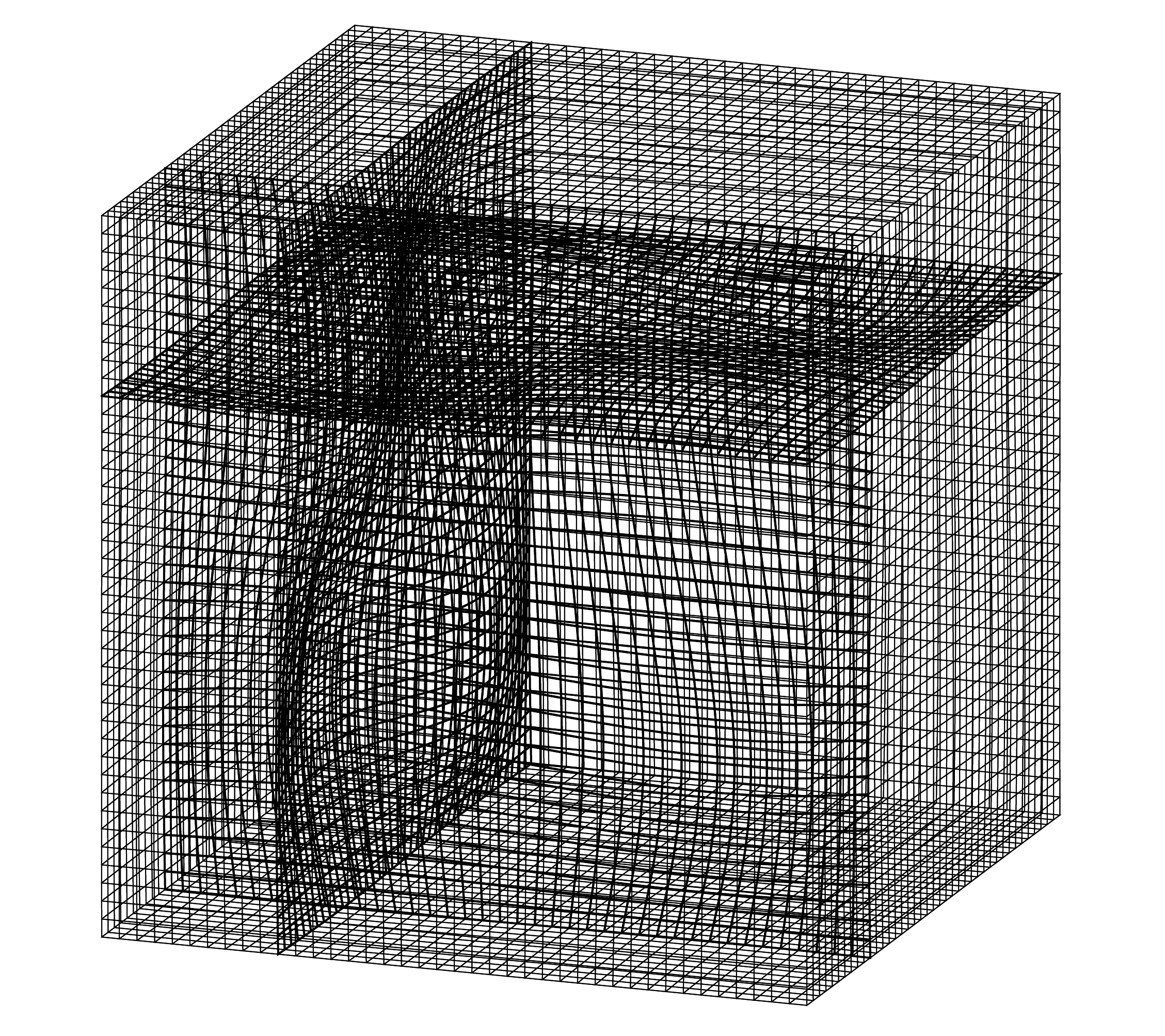}
	\includegraphics[width=0.465\textwidth]{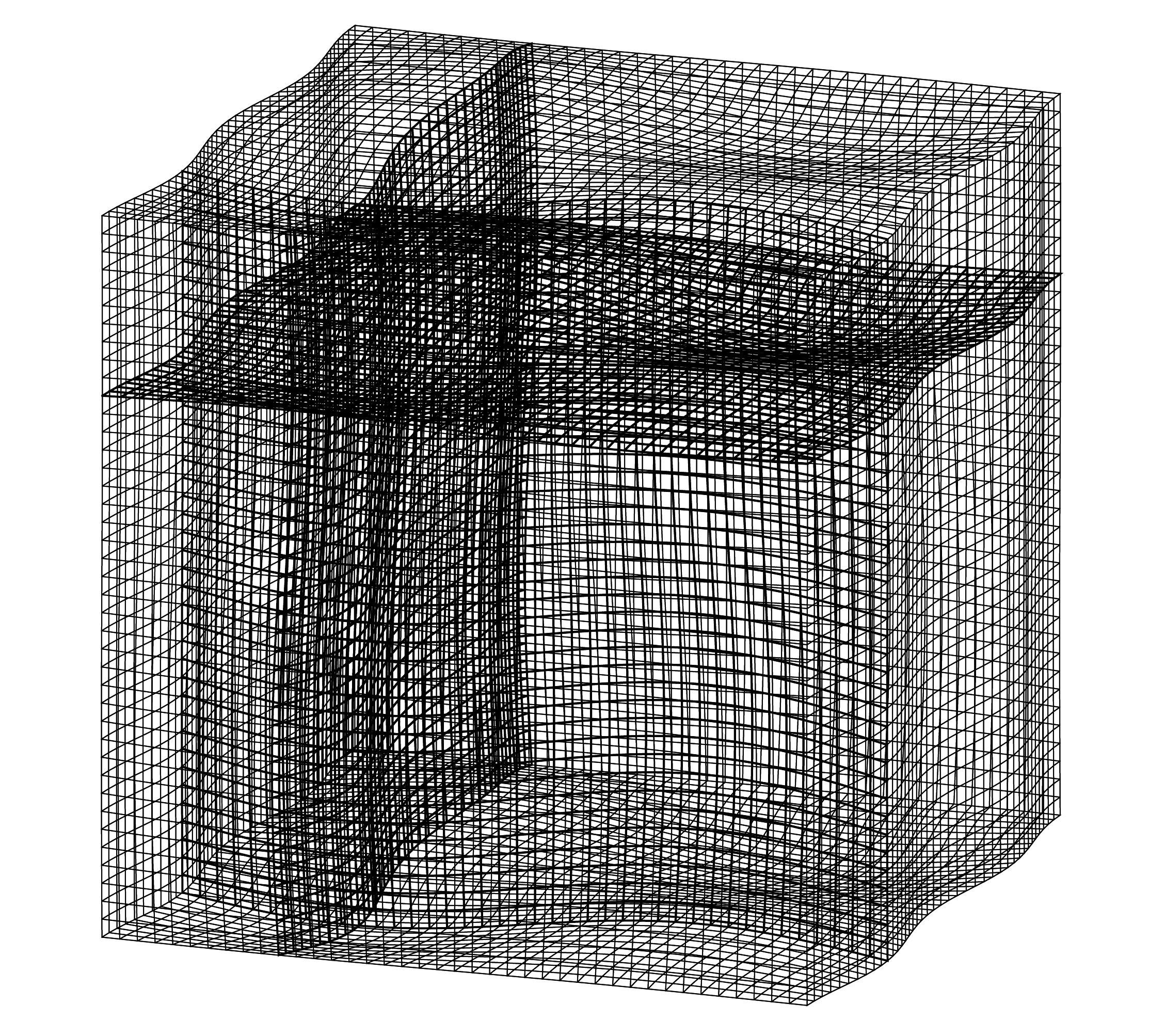}
	\caption{\label{accuracy-mesh} Accuracy test: the curvilinear physical meshes for mesh 1 (left) and mesh 2 (right) with $40^3$ cells.}
\end{figure}
\begin{table}[!h]
\centering
\begin{tabular}{c|cc|cc}
\hline 
\hline
mesh   1  & $L^1$ error  &    order   &  $L^2$ error  &   order  \\
\hline
$10^3$    &   2.8315E-02 & ~~     &  1.1198E-02  &  ~~     \\
$20^3$    &  1.3524E-03 & 4.3879 &  5.7195E-04  &  4.2912 \\
$40^3$    &  5.7464E-05 & 4.5567 &  2.3528E-05  &  4.6034 \\
$80^3$    &  2.7384E-06 & 4.3912 &  1.0882E-06 &  4.4344 \\
$160^3$   &  1.5286E-07 & 4.1630 &  6.0120E-08  &  4.1779 \\
\hline 
\hline
\end{tabular}
\caption{\label{tab-3d-1} Accuracy test: 3D  advection of density
perturbation for mesh $1$.}
\end{table}
\begin{table}[!h]
\centering
\begin{tabular}{c|cc|cc}
\hline 
\hline
mesh   2  & $L^1$ error  &    order   &  $L^2$ error  &   order  \\
\hline
$10^3$  &   2.7971E-02 &   ~~~     &  1.1473E-02  &   ~~~    \\
$20^3$  &   1.1418E-03 &   4.6145 &  4.9546E-04  &   4.5334 \\
$40^3$  &   4.5344E-05 &   4.6543 &  2.0510E-05  &   4.5943 \\
$80^3$  &   2.5800E-06 &   4.1354 &  1.0528E-06  &   4.2839 \\
$160^3$  &   1.5776E-07 &   4.0315 &  6.2671E-08  &   4.0703 \\
\hline 
\hline
\end{tabular}
\caption{\label{tab-3d-2} Accuracy test: 3D  advection of density
perturbation for mesh $2$.}
\end{table}

\subsection{Accuracy tests}
In this case, the advection of density perturbation is presented for accuracy tests and the validation of geometric conservation law \cite{GCL-1,GCL-2}. For the three-dimensional (3D) case, the
initial condition is set as
\begin{align*}
\rho_0&(x, y, z)=1+0.2\sin(\pi(x+y+z)),~p_0(x,y,z)=1,\\
&U_0(x,y,z)=1,~V_0(x,y,z)=1,~W_0(x,y,z)=1.
\end{align*}
In the computation, the physical domain is $[0,2]\times[0,2]\times[0,2]$.
The periodic boundary conditions are applied at all boundaries, and the
exact solutions are
\begin{align*}
\rho(x,y&,z,t)=1+0.2\sin(\pi(x+y+z-t)),~p(x,y,z,t)=1,\\
&U(x,y,z,t)=1,~V(x,y,z,t)=1,~W(x,y,z,t)=1.
\end{align*}
For the curvilinear mesh, two types of mesh are tested, which are given as follows
\begin{align*}
\text{mesh 1:}&
\begin{cases}
\displaystyle x=\xi+0.05\sin (\pi \xi)\sin (\pi \eta)\sin (\pi \zeta),\\
\displaystyle y=\eta+0.05\sin (\pi \xi)\sin (\pi \eta)\sin (\pi \zeta),\\
\displaystyle z=\zeta+0.05\sin (\pi \xi)\sin (\pi \eta)\sin (\pi \zeta),
\end{cases}\\
\text{mesh 2:}&
\begin{cases}
\displaystyle x=\xi+0.05\sin (\pi \eta)\sin (\pi \zeta),\\
\displaystyle y=\eta+0.05\sin (\pi \zeta)\sin (\pi \xi),\\
\displaystyle z=\zeta+0.05\sin (\pi \xi)\sin (\pi \eta),
\end{cases}
\end{align*}
where $(\xi, \eta, \zeta)\in[0,2]\times[0,2]\times[0,2]$ and the
uniform cells are used in the computational domain, and the above
meshes with $40^3$ cells are shown in Fig.\ref{accuracy-mesh} as
an example. The $L^1$ and $L^2$ errors and orders of accuracy at $t=2$
with $N^3$ cells are given in Table.\ref{tab-3d-1} and
Table.\ref{tab-3d-2}. The expected accuracy can be achieved for the
current HGKS-cur.

The GCL is also tested by the above meshes. The GCL is mainly about
the maintenance of a uniform flow passing through a non-uniform
non-orthogonal mesh. The initial condition for the three-dimensional
case is
\begin{align*}
\rho_0&(x, y,
z)=1,~p_0(x,y,z)=1,~U_0(x,y,z)=1,~V_0(x,y,z)=1,~W_0(x,y,z)=1.
\end{align*}
The periodic boundary conditions are adopted as well. The $L^1$ and
$L^2$ errors at $t=2$  are given in Table.\ref{tab-3d-3}. The
results show that the errors reduce to the machine zero, and the
geometric conservation law is well preserved by the HGKS-cur.
\begin{table}[!h]
	\centering
	\begin{tabular}{c|cc|cc}
		\hline 
		\hline
		~      &   mesh 1  &~ &   mesh 2  &  ~     \\
		\hline
		mesh        & $L^1$ error  &      $L^2$ error &  $L^1$ error  &      $L^2$ error    \\
		\hline
		$10^3$     &    6.2119E-15  & 2.7861E-15  &  5.8406E-15 &  2.6278E-15 \\
		$20^3$     &    8.2257E-15  & 3.6661E-15  &  7.4312E-15 &  3.3097E-15 \\
		$40^3$     &    1.2293E-14  & 5.4848E-15  &  1.1961E-14  & 5.3458E-15 \\
		$80^3$     &    2.1767E-14  & 9.7864E-15  &  2.1670E-14  & 9.7434E-15 \\
		$160^3$    &    4.6088E-14  & 2.0787E-14  &  4.5892E-14  & 2.0716E-14 \\
		\hline 
		\hline
	\end{tabular}
	\caption{\label{tab-3d-3} Accuracy test: geometric conservation law for 3D meshes.}
\end{table}

\subsection{Lid-driven cavity flow}
\begin{figure}[!h]
	\centering
	\includegraphics[width=0.485\textwidth]{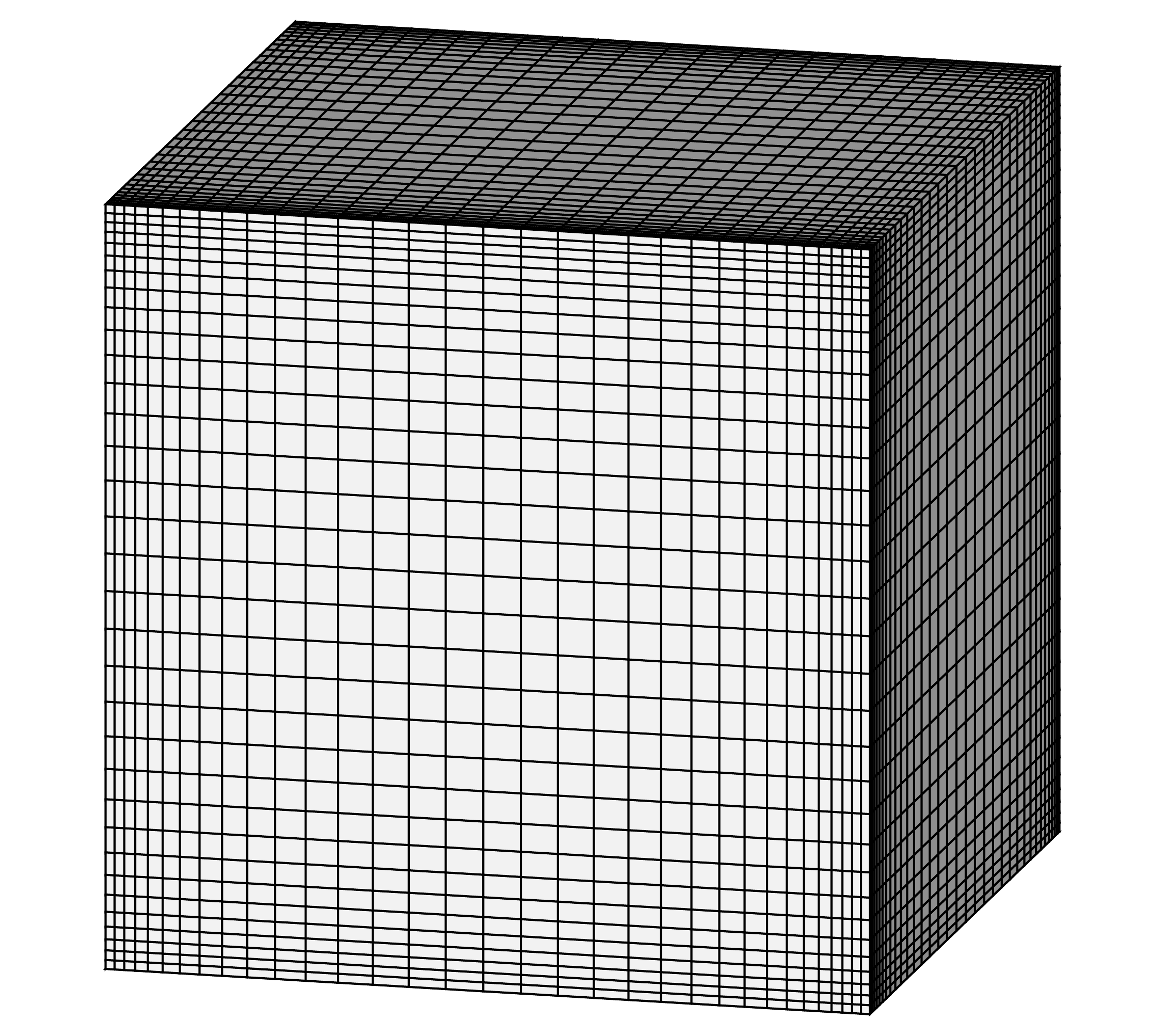}
	\caption{\label{cavity-1} Lid-driven cavity flow: the non-uniform physical mesh with $33^3$ cells.}
\end{figure}

\begin{figure}[!h]
	\centering
	\includegraphics[width=0.485\textwidth]{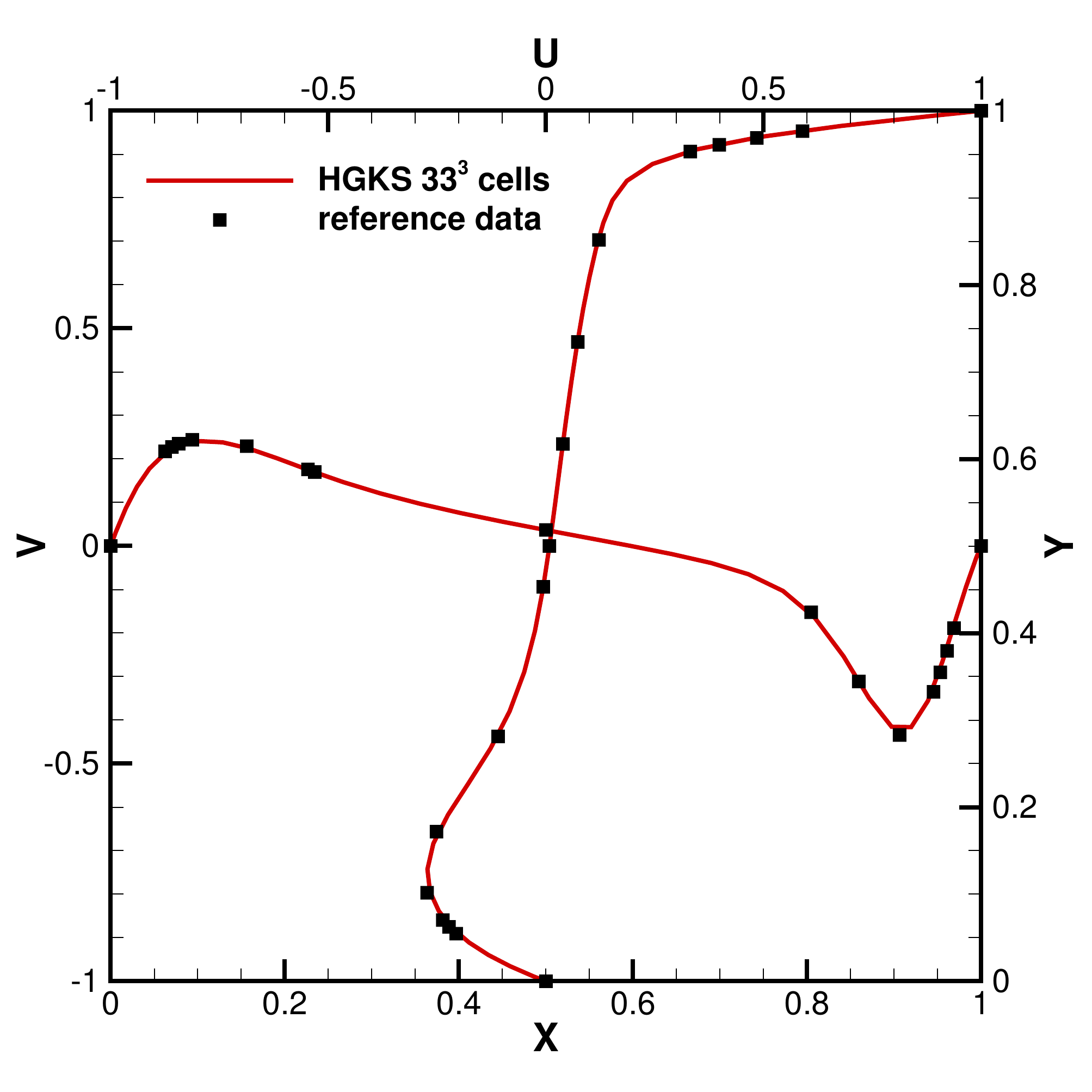}
	\includegraphics[width=0.485\textwidth]{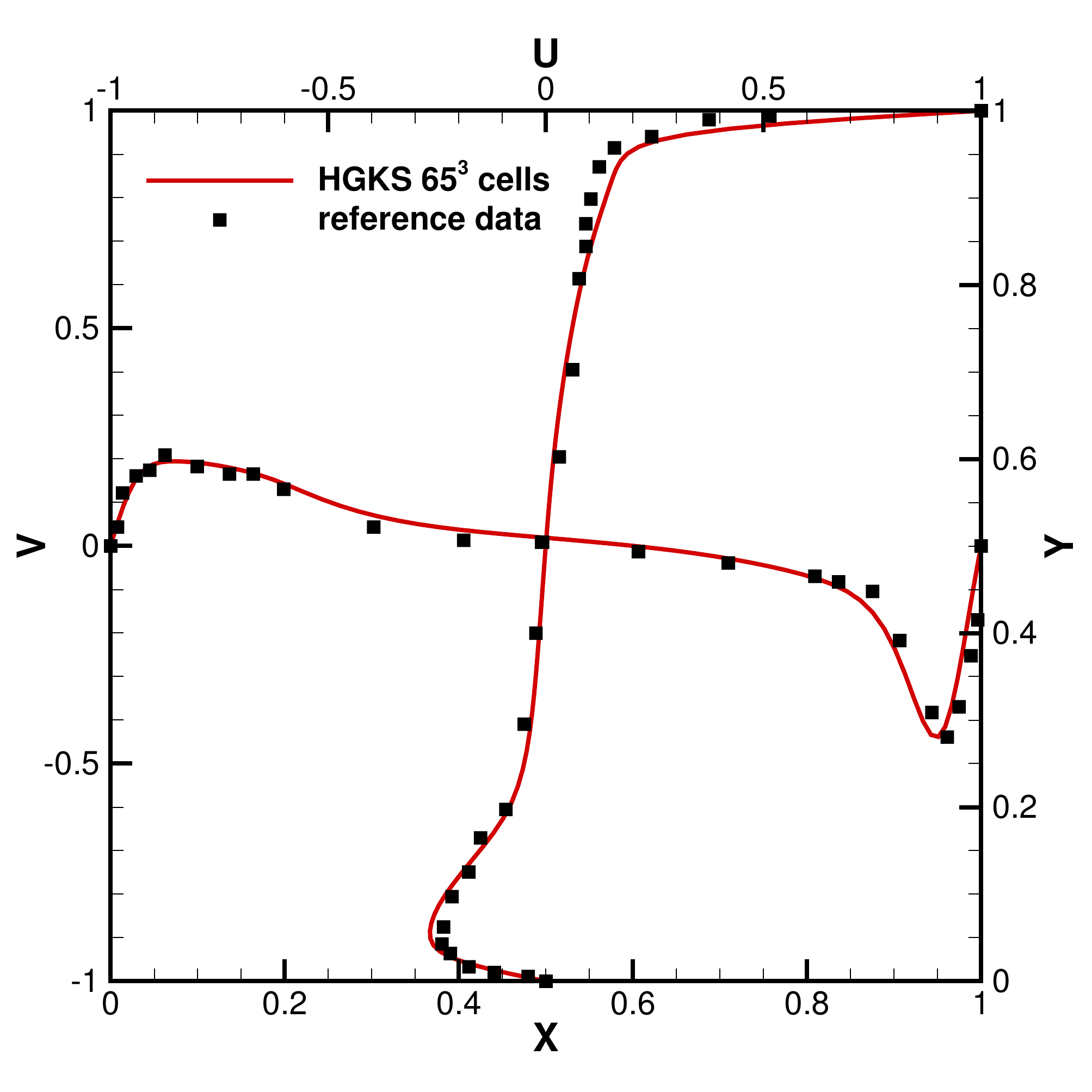}
	\caption{\label{cavity-2} Lid-driven cavity flow: $U$-velocity
		profiles along the vertical centerline  and $V$-velocity profiles
		along the horizontal centerline for $Re=1000$ (left) and $3200$ (right).}
\end{figure}

The lid-driven cavity problem is a benchmark for laminar flow simulations.
The fluid is bounded by a unit cubic $[0, 1]\times[0, 1]\times[0,
1]$ and driven by a uniform translation of the top boundary with
$Y=1$. Three-dimensional cavity-flow calculations have been
carried out early \cite{Case-Goda}. 
In this case, the flow is simulated with Mach
number $Ma=0.15$ and all the boundaries are isothermal and nonslip.
To well resolve the boundary layer, the following local refined
meshes are used
\begin{align*}
\begin{cases}
\displaystyle x=\xi-0.1\sin (2\pi \xi),\\
\displaystyle y=\eta-0.1\sin (2\pi \eta),\\
\displaystyle z=\zeta-0.1\sin (2\pi \zeta).
\end{cases}
\end{align*}
Numerical simulations are conducted with Reynolds numbers $Re=1000$
and $3200$. For the case with $Re=1000$, the convergent solution is
obtained and the uniform mesh in the computational domain with
$33^3$ cells is used. 
The non-uniform physical meshes with $33^3$ cells is shown in Figure.\ref{cavity-1}.
The flow at $Re=3200$ corresponds to unsteady solution, which have been studied
extensively \cite{Case-Albensoeder,Case-Prasad}. The uniform mesh in the
computational domain with $65^3$ cells is used and the
numerical results are averaged in 250 time period. The $U$-velocity
profiles along the vertical centerline, $V$-velocity profiles
along the horizontal centerline in the symmetry $X-Y$ plane are shown in
Figure.\ref{cavity-2}. 
For these two cases, the results from 
the Chebyshev-collocation method \cite{Case-Albensoeder} on a Gauss-Lobatto grid of size $96^3$ for $Re = 1000$  and the experimental data \cite{Case-Prasad}  for $Re = 3200$ are adopted as the benchmark data, respectively. 
The agreement between them shows that current HGKS-cur is capable of
simulating three-dimensional steady and unsteady laminar flows.

\subsection{Compressible turbulent channel flow}
\begin{figure}[!h]
\centering
\includegraphics[width=0.6\textwidth]{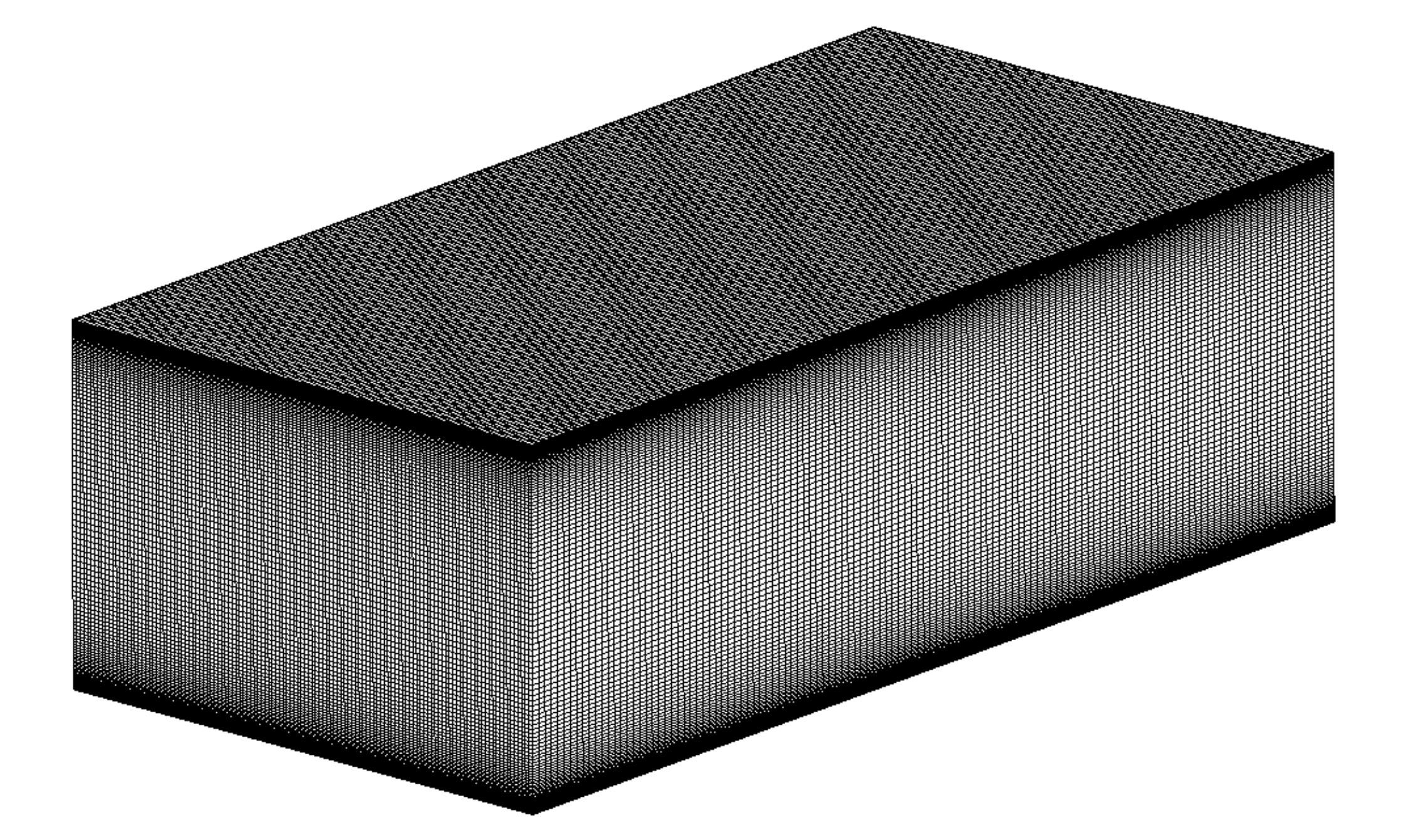}
\caption{\label{channel-ini} Compressible turbulent channel flow: the non-uniform physical mesh with $128^3$ cells.}
\end{figure}

Considering the simplicity of geometry and boundary conditions, the faithful computational studies of incompressible to
hypersonic turbulent channel flow
\cite{kim1987turbulence,lee2015direct,coleman1995numerical,morinishi2004direct,yu2019genuine}
have been carried out to study the mechanism of turbulent boundary
layer. In this section, the compressible turbulent channel flow
\cite{coleman1995numerical,morinishi2004direct} with bulk Mach
number $Ma = 1.5$ and bulk Reynolds number $Re = 3000$ is tested
with non-uniform mesh. In the computation, the physical domain is
$(x,y,z)\in[0,4\pi H]\times[-H,H]\times[0,4\pi H/3]$ and the
computational domain takes
$(\xi,\eta,\zeta)\in[0,4\pi H]\times[-1.5\pi H,1.5\pi H]\times[0,4\pi H/3]$. In the
computation, the coordinate transformation is given by
\begin{align*}
\begin{cases}
\displaystyle x=\xi,\\
\displaystyle y=\tanh(b_g(\frac{\eta}{1.5\pi}-1))/\tanh(b_g),\\
\displaystyle z=\zeta,
\end{cases}
\end{align*}
where $b_g=2$.  
The mesh with $128^3$ cells is given in
Fig.\ref{channel-ini} as an example. This case addresses the
performance of HGKS-cur in non-uniform mesh for compressible
wall-bounded turbulent flows. The periodic boundary conditions are
used in streamwise $x$-direction and spanwise $z$-directions, and
the non-slip and isothermal boundary conditions are used in wall-normal
$y$-direction.

\begin{table}[!h]
    \centering
    \begin{tabular}{c|c|c|c|c|c|c|c}
        \hline \hline
        Case     &Run   &$Pr$ &Physical domain  &$N_x \times N_y \times N_z$   &$\Delta Y^{+}_{min}$/$Y^{+}_{N10}$   &$\Delta X^{+}$   &$\Delta Z^{+}$    \\
        \hline
        Ref$_1$  &DNS          &0.70  &$4 \pi H \times 2H \times 4 \pi H/3$  &$144 \times 90 \times 60$    &0.10/8   &19   &12   \\
        \hline
        Ref$_2$   &DNS         &0.72 &$4 \pi H\times 2H \times 4 \pi H/3$  &$120 \times 180 \times 120$    &0.36/-      &23       &7.60    \\
        \hline
        Ref$_3$   &CLES         &0.70 &$4 \pi H\times 2H \times 4 \pi H/3$  &$64 \times 65 \times 64$    &0.50/-      &43       &14    \\
        \hline
        $G_1$     &iLES        &0.70  &$4 \pi H\times 2H \times 4 \pi H/3$  &$128 \times 128 \times 128$    &0.50/12.66   &21.18    &7.06   \\
        \hline
        $G_2$    &iLES         &0.70  &$2 \pi H\times 2H \times 4 \pi H/3$  &$128 \times 128 \times 128$   &0.50/12.66   &10.59    &7.06   \\
        \hline
        $G_3$    &iLES         &1.0  &$4 \pi H\times 2H \times 4 \pi H/3$  &$128 \times 128 \times 128$     &0.50/12.66   &21.18    &7.06  \\
        \hline
        $G_4$    &DNS         &0.70  &$4 \pi H\times 2H \times 4 \pi H/3$  &$160 \times 160 \times 160$   &0.40/9.50   &16.92    &5.64   \\
        \hline \hline
    \end{tabular}
    \caption{\label{channel_parameters} Compressible turbulent channel flow: Prandtl number and numerical parameters of the
    present and the reference simulations. $``-"$ means that the data can not be find in the refereed paper.}
\end{table}

In current study, the fluid is initiated with density $\rho = 1$ and
the initial streamwise velocity $U(y)$ profile is given by the perturbed Poiseuille flow profile
\begin{align*}
U(y) = 1.5(1-y^{2}) + \text{white noise},
\end{align*}
where the white noise is added with $10\%$ amplitude of local
streamwise velocity. The spanwise and wall-normal velocity is initiated with white noise. The initial non-dimensional parameters bulk Mach number $Ma$
and bulk Reynolds number $Re$ are defined as
\begin{align*}
Ma  = \frac{U_b}{c_w}, ~
Re = \frac{\rho_b U_bH}{ \mu_w},
\end{align*}
where $H = 1$ is the half height of the channel,  $c_w =
\sqrt{\gamma R T_w}$ is the wall sound speed, $\mu_w$ the wall
molecular viscosity, $T_w$ the wall temperature and $R$ the
gas constant. The viscosity $\mu$ is determined by the power law as
$\mu(T) \propto T^{0.7}$. 
The Prandtl number is defined as $Pr = \mu c_p / \kappa$, where $c_p$ is the specific heat at constant pressure and the $\kappa$ is
the heat conductivity. The bulk velocity $U_b$ and bulk-averaged density $\rho_b$ are defined as
\begin{align*}
	U_b =  \int_{-H}^{H} U(y) \text{d}y, 
	~\rho_b =  \int_{-H}^{H} \rho(y) \text{d}y.
\end{align*}
The plus unit $Y^+$ and plus velocity $U^+$ are defined as
\begin{align*}
Y^+ = \frac{\rho u_\tau y}{\mu}, ~U^+ = \frac{U}{u_{\tau}},
\end{align*}
with the friction velocity $u_\tau$ and the wall shear stress $\tau_{wall}$ as
\begin{align*}
u_\tau = \sqrt{\frac{\tau_{w}}{\rho_w}}, ~\tau_{w}=\mu_w \frac{\partial U}{\partial y}\big|_{w}.
\end{align*}
The friction Mach number $Ma_\tau$ and the friction Reynolds
number $Re_{\tau}$ are given by
\begin{align*}
Ma_\tau = \frac{u_{\tau}}{c_w}, ~ Re_{\tau} = \frac{H}{\delta_v}, ~ \delta_v = \frac{\mu_w}{\rho_w u_{\tau}}.
\end{align*}
The heat flux $q_{w}$ and the non-dimensional heat flux $B_q$ of the
wall are defined as
\begin{align*}
q_{w} = - \kappa \frac{\partial T}{\partial y}\big|_{w}, ~ B_q =\frac{q_w}{\rho_w c_p u_{\tau} T_w}.
\end{align*}
In this computation, the details of Prandtl number and numerical
parameters are given in Table.\ref{channel_parameters}. 
The numerical results of DNS in refereed paper \cite{coleman1995numerical}
and \cite{morinishi2004direct} are denoted as Ref$_1$ and Ref$_2$, constrained large-eddy simulation (CLES) approach \cite{jiang2013constrained} is denoted as Ref$_3$, and four cases $G_1-G_4$ are implemented by current HGKS-cur.
CLES is implemented on the coarsest grid, which has succeeded in predicting compressible turbulent flows \cite{hong2014constrained, chen2017constrained}.
The spectral method and B-spline collocation method is used by Ref$_1$ and Ref$_2$, respectively. Compared with the set-up of case
$G_1$, the half length of streamwise direction is used in case
$G_2$. In addition, the unit Prandtl number $Pr = 1$ is used for
case $G_3$, and the finer mesh with $160^3$ cells is applied in case
$G_4$. Specifically, $\Delta Y^{+}_{min}$ is the first grid space
off the wall in the wall-normal direction, and $Y^{+}_{N10}$ is the
plus unit for the first ten points off the wall. $\Delta X^{+}$ and
$\Delta Z^{+}$ are the equivalent plus unit for uniform streamwise
and spanwise grids, respectively.
For current HGKS-cur, cases $G_1 - G_3$ are implemented as iLES, and case $G_4$ is for DNS study.
\vspace{-4.5mm}
\begin{figure}[!h]
	\centering
	\includegraphics[width=0.55\textwidth]{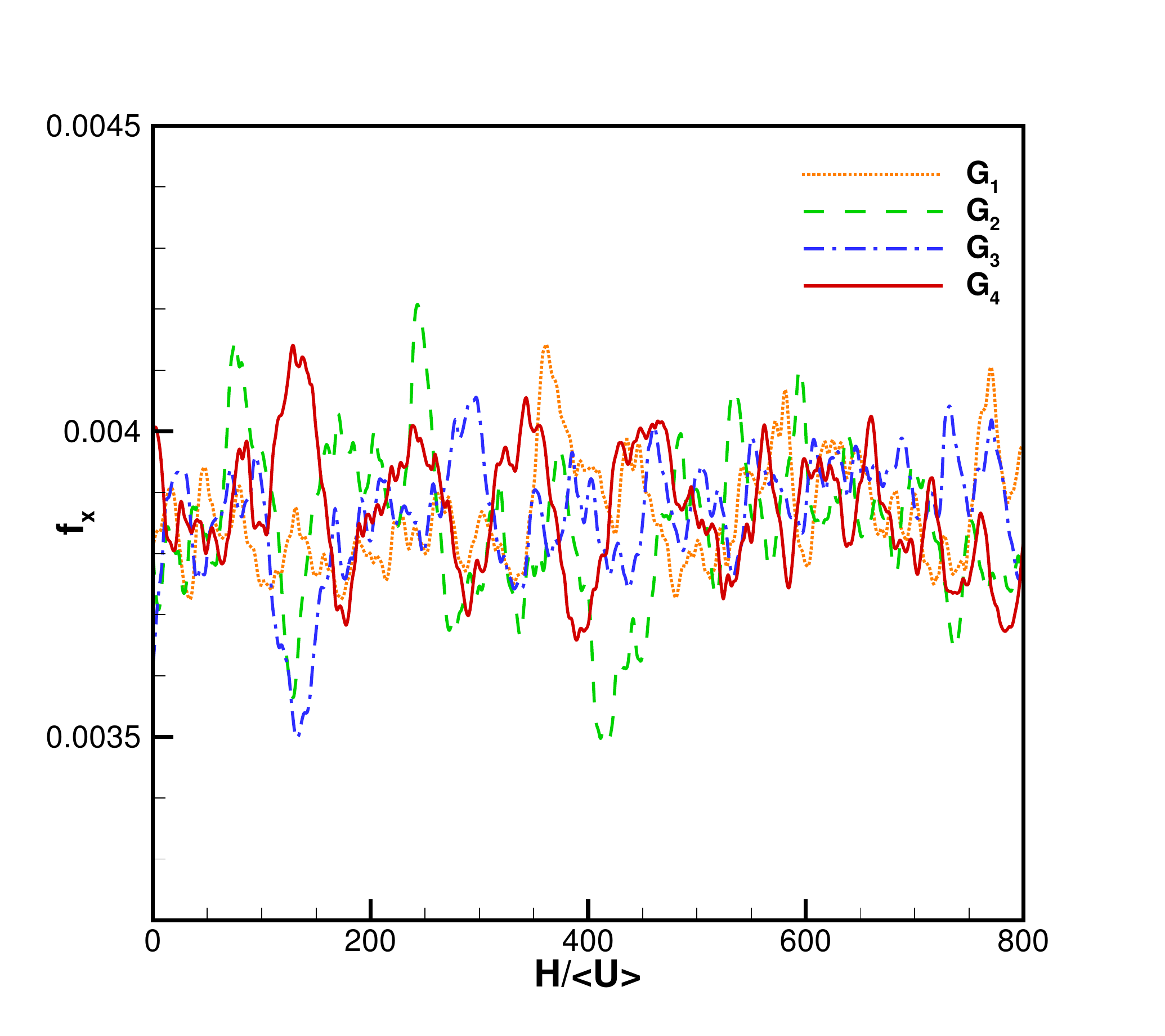}
	\vspace{-4mm}
	\caption{\label{channel_force} Compressible turbulent channel flow: the external  force $f_x$ for case $G_1 - G_4$ after transition.}	
\end{figure}

\begin{figure}[!h]
    \centering
    \includegraphics[width=0.75\textwidth]{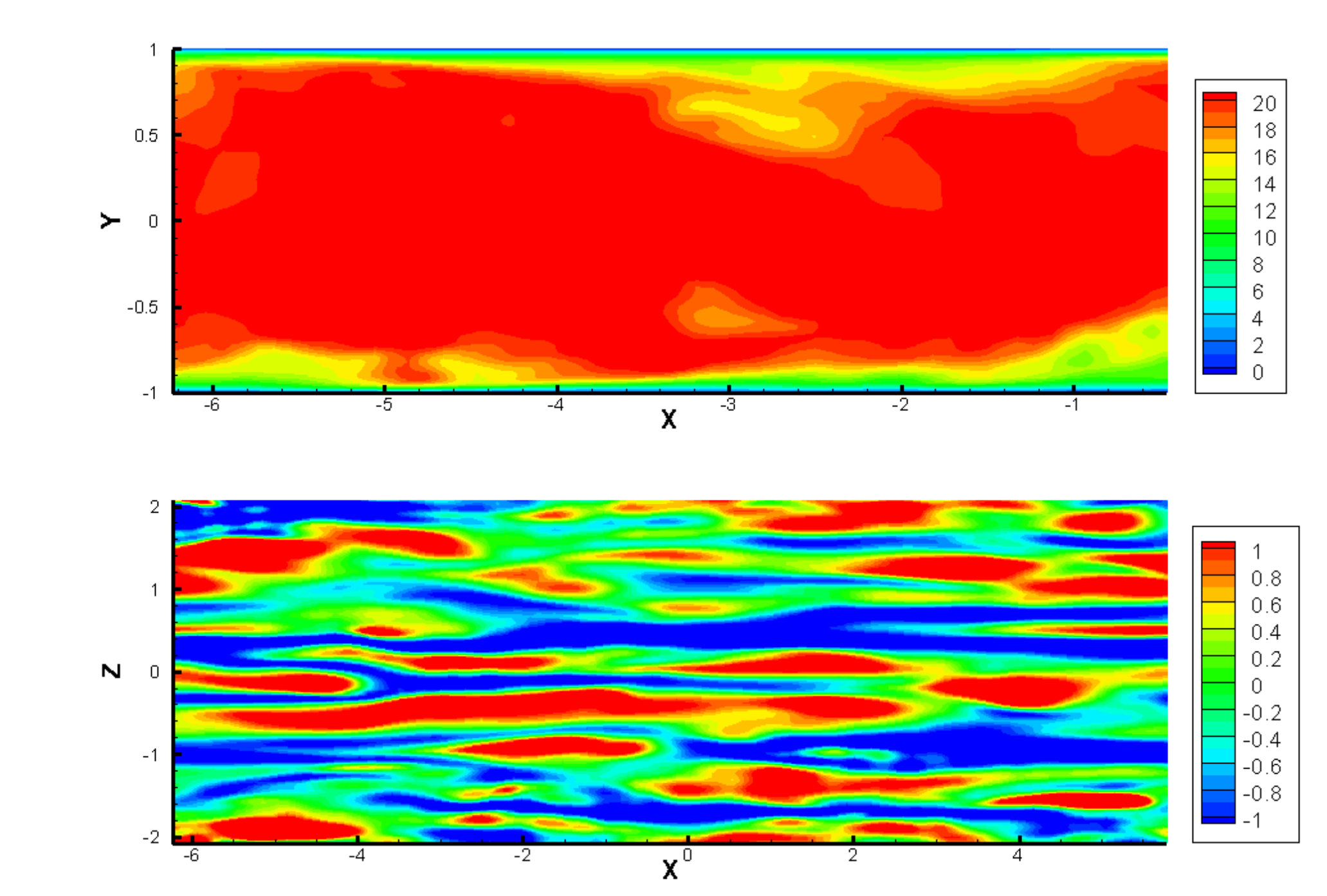}
    \vspace{-4mm}
    \caption{\label{channel_sketch}  Compressible turbulent channel flow: instantaneous contour for case $G_1$, the streamwise velocity is normalized by friction velocity $u_{\tau}$. The upper is contour of instantaneous streamwise velocity at $Z = 0$, and the lower contour represents the instantaneous streamwise velocity at $Y^+ = 3.2$ with extracting the mean velocity.}
\end{figure}
\begin{table}[!h]
	\centering
	\begin{tabular}{c|c|c|c|c|c|c|c}
		\hline \hline
		Case         &Ref$_1$   &Ref$_2$  &Ref$_3$  &$G_1$  &$G_2$     &$G_3$   &$G_4$  \\
		\hline
		${\left\langle u_{\tau} \right\rangle}$   &0.054     &-        &0.054   &0.053      &0.053     &0.051   &0.053\\
		\hline 
		${\left\langle Ma_{\tau} \right\rangle}$  &0.082     &0.080    &0.080   &0.079      &0.079     &0.076   &0.079\\
		\hline
		${\left\langle Re_{\tau} \right\rangle}$  &222       &218      &218     &211        &212       &221     &213\\
		\hline
		${\left\langle \rho_w \right\rangle}$  &1.355     &-           &1.354    &1.355     &1.363      &1.476      &1.356\\
		\hline
		${\left\langle q_w \right\rangle}$     &-0.0089    &-          &-     &-0.0084     &-0.0084    &-0.0085    &-0.0085\\
		\hline
		${\left\langle B_q \right\rangle}$       &-0.049     &-0.048   &-0.048         &-0.047   &-0.047   &-0.045   &-0.048\\
		\hline \hline
	\end{tabular}
	\caption{\label{channel_re_tau} Compressible turbulent channel flow: statistical quantities at the wall. $``-"$ means that the data can not be find in the refereed paper.}
\end{table}

To excite channel flow to turbulence, an fixed external force $f_x$ is exerted
in the streamwise direction initially. After transition, the
constant moment flux is used to determine the external force. More
details of the implementation of external force can be found in Ref
\cite{GKS-high-cao-2}. The external force after transition for cases
$G_1 - G_4$ are presented in Figure.\ref{channel_force}, which fluctuates to balance the wall shear stress. In the following analysis, $800$ characteristic periodic time is used to obtain the statistically stationary turbulence. The averaging
time is longer than that in the reference paper
\cite{coleman1995numerical}. In what follows, note that the mean
average over time and the $X$- and $Z$-directions is represented by
$\langle\cdot \rangle$. Instantaneous slides of normalized
streamwise velocity at $Z = 0$ and $Y^+ = 3.2$ for case $G_1$ are
shown in Figure.\ref{channel_sketch}, where the streamwise velocity
is normalized by friction velocity $u_{\tau}$. The mean
velocity is extracted for the slide at $Y^+ = 3.2$, and the
high-speed streaks and low-speed streaks are clearly presented. The
key statistical quantities at the wall are presented in
Table.\ref{channel_re_tau}. For current iLES with HGKS-cur, the cases $G_1$
and $G_4$ agree well with the refereed solutions, and $G_1$ converges
to $G_4$. Compared with the effect of large Prandtl number as
case $G_3$, the smaller streamwise computational size as case $G_2$ almost dose
not affect the statistical variables at the wall. Table.\ref{channel_re_tau}
shows that the large Prandtl number enlarges the mean
friction Reynolds number $Re_\tau$, the density at the wall $\rho_w$, and
the friction non-dimensional heat flux $B_q$. It is known from
dimensional analysis that the mean velocity and temperature profiles
depend on the non-dimensional heat flux $B_q$, and the friction Mach
number $M_{\tau}$ \cite{morinishi2004direct}. As the ratio of
specific heats $\gamma$ and the specific heat at constant pressure $c_p$ are constants, the mean velocity and temperature profiles depend on the Prandtl number,  and this will be validated in the following part.
\begin{figure}[!h]
	\centering
	\includegraphics[width=0.455\textwidth]{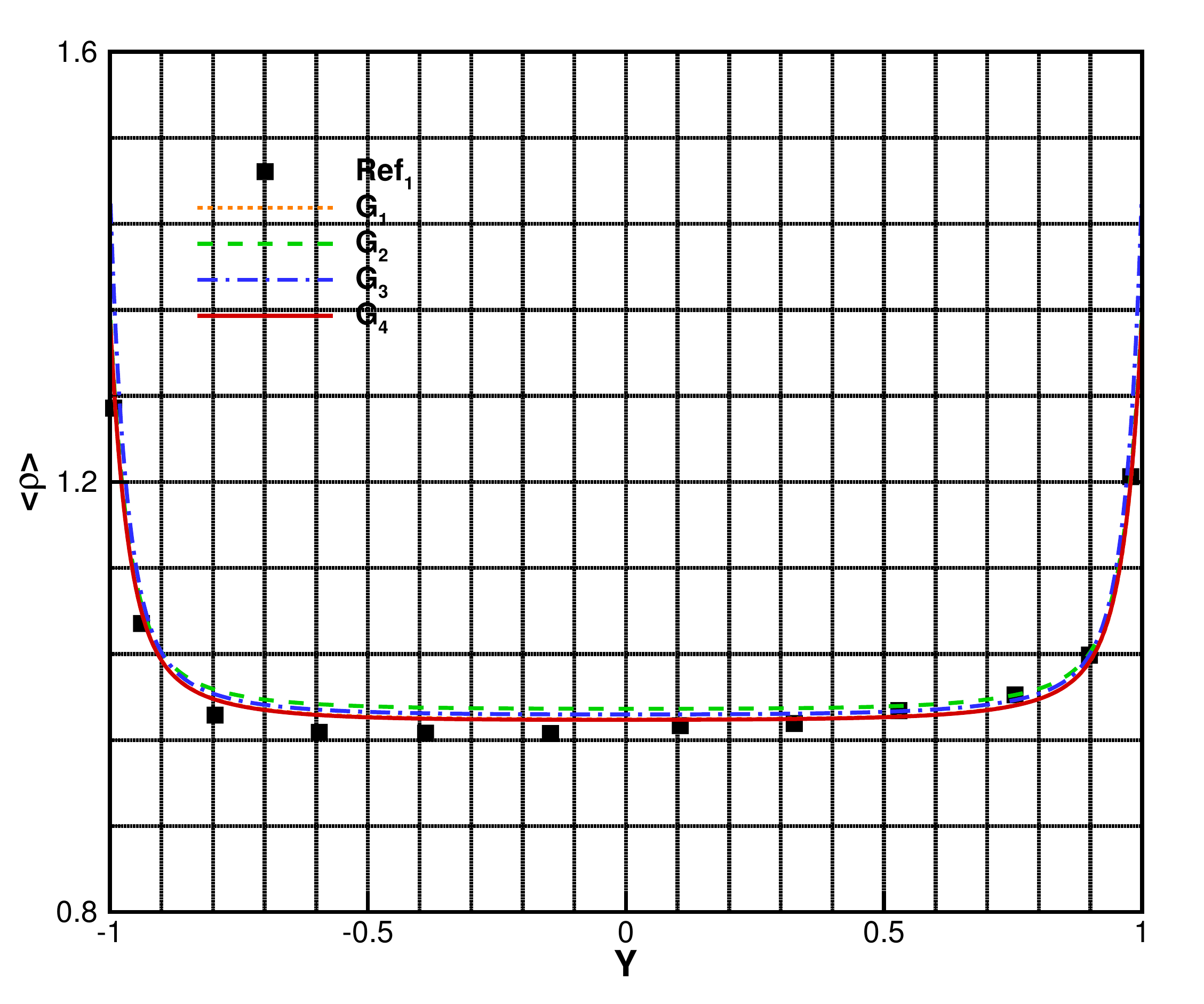}
	\includegraphics[width=0.455\textwidth]{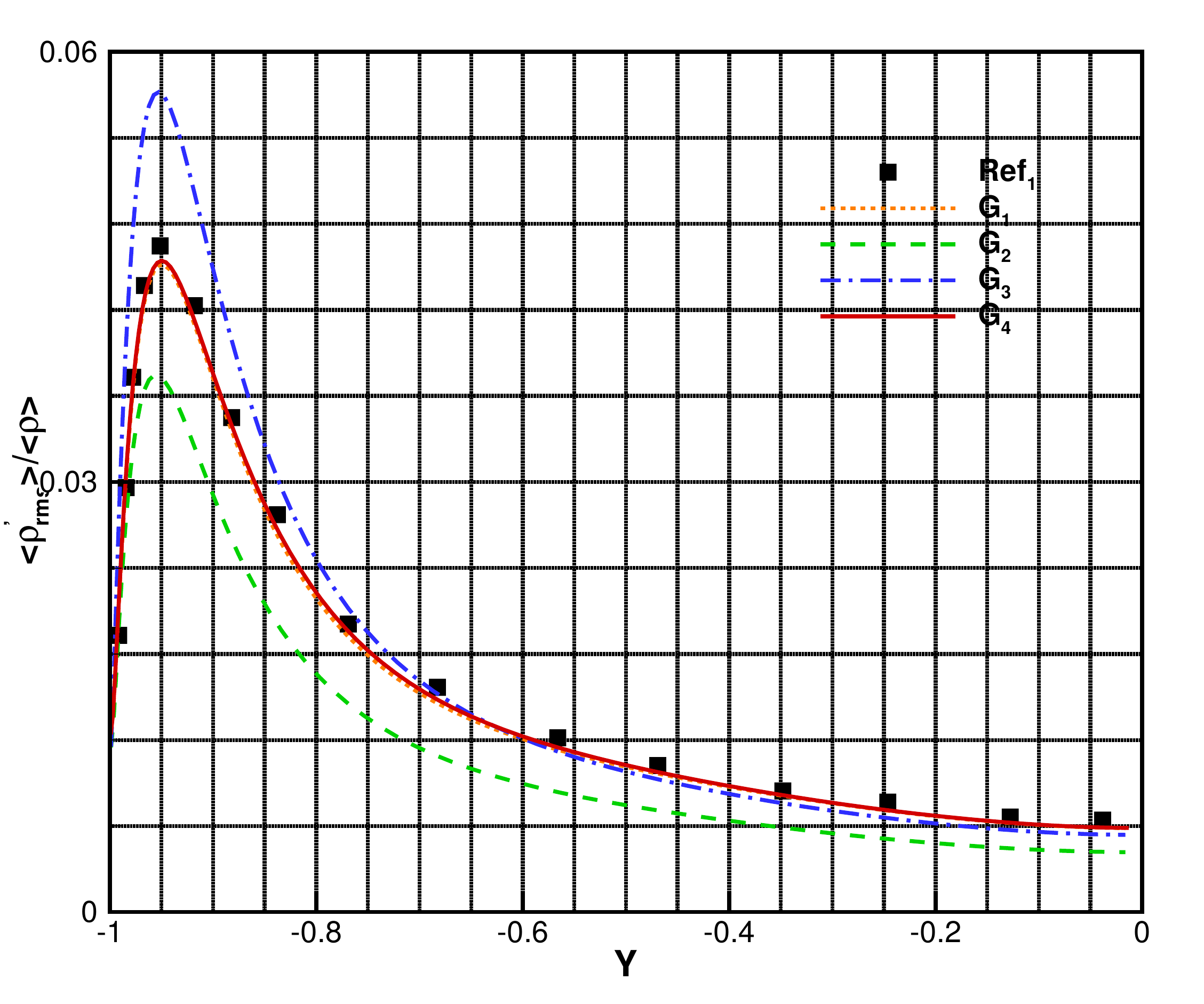}
	\includegraphics[width=0.455\textwidth]{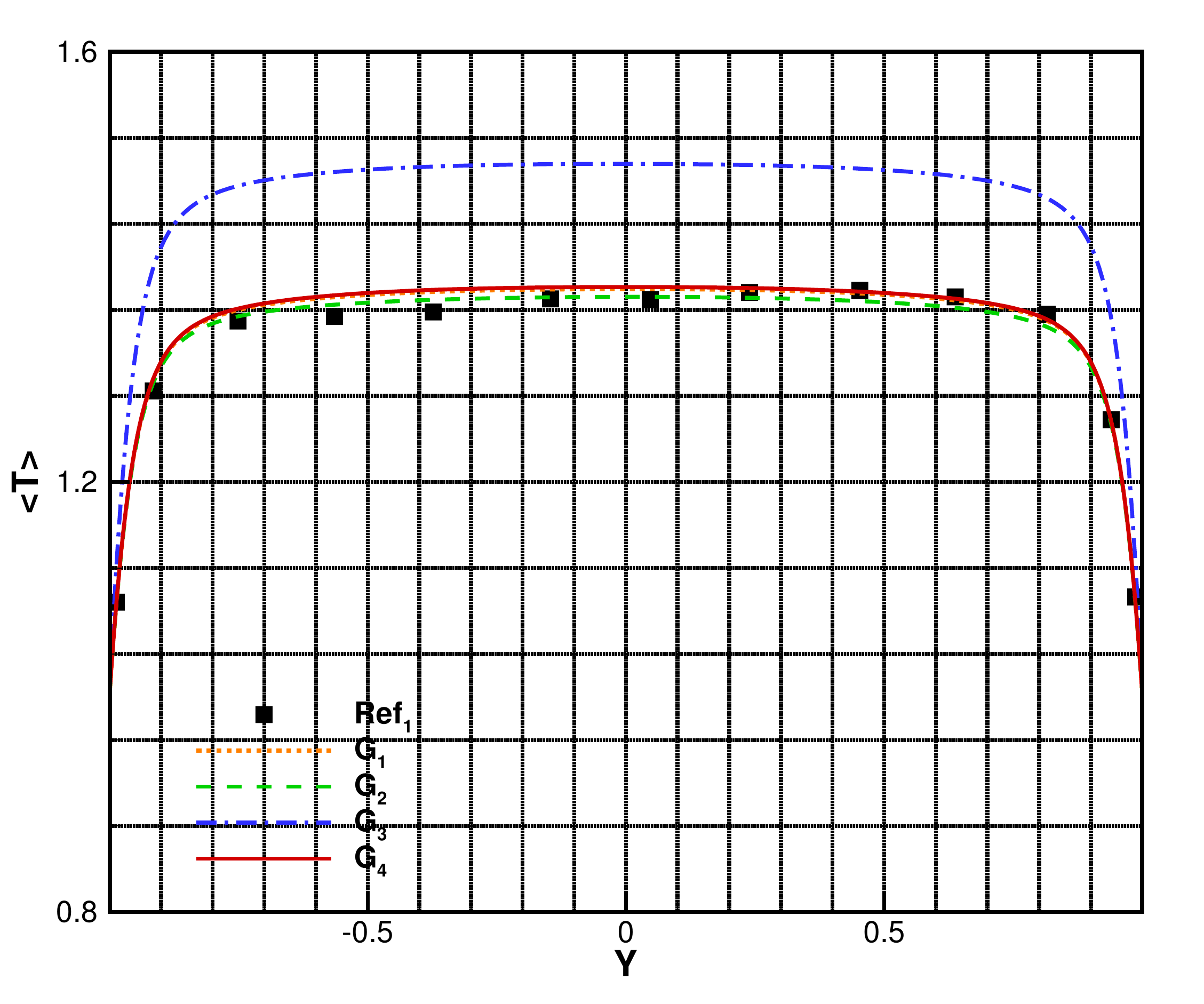}
	\includegraphics[width=0.455\textwidth]{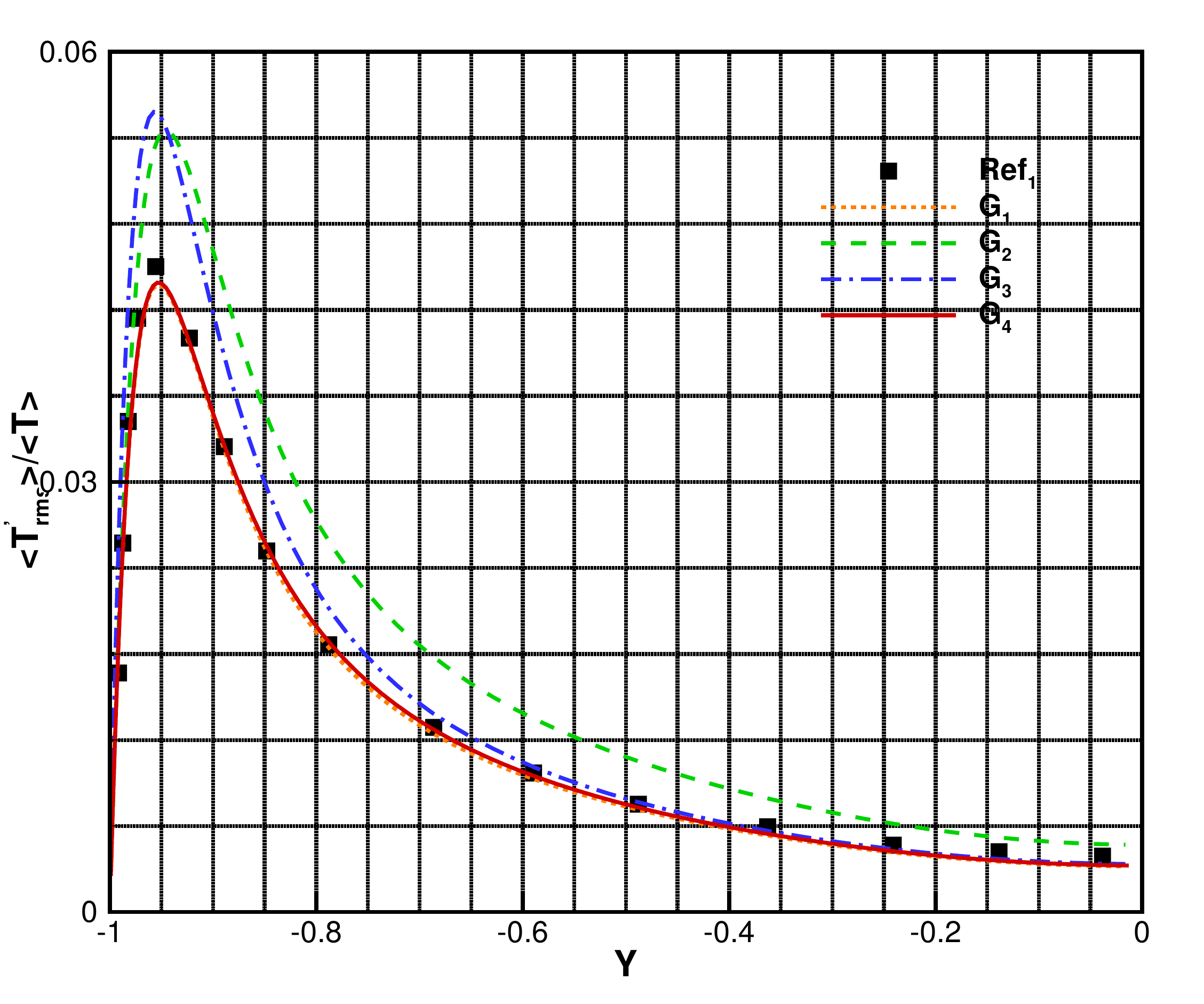}
	\includegraphics[width=0.455\textwidth]{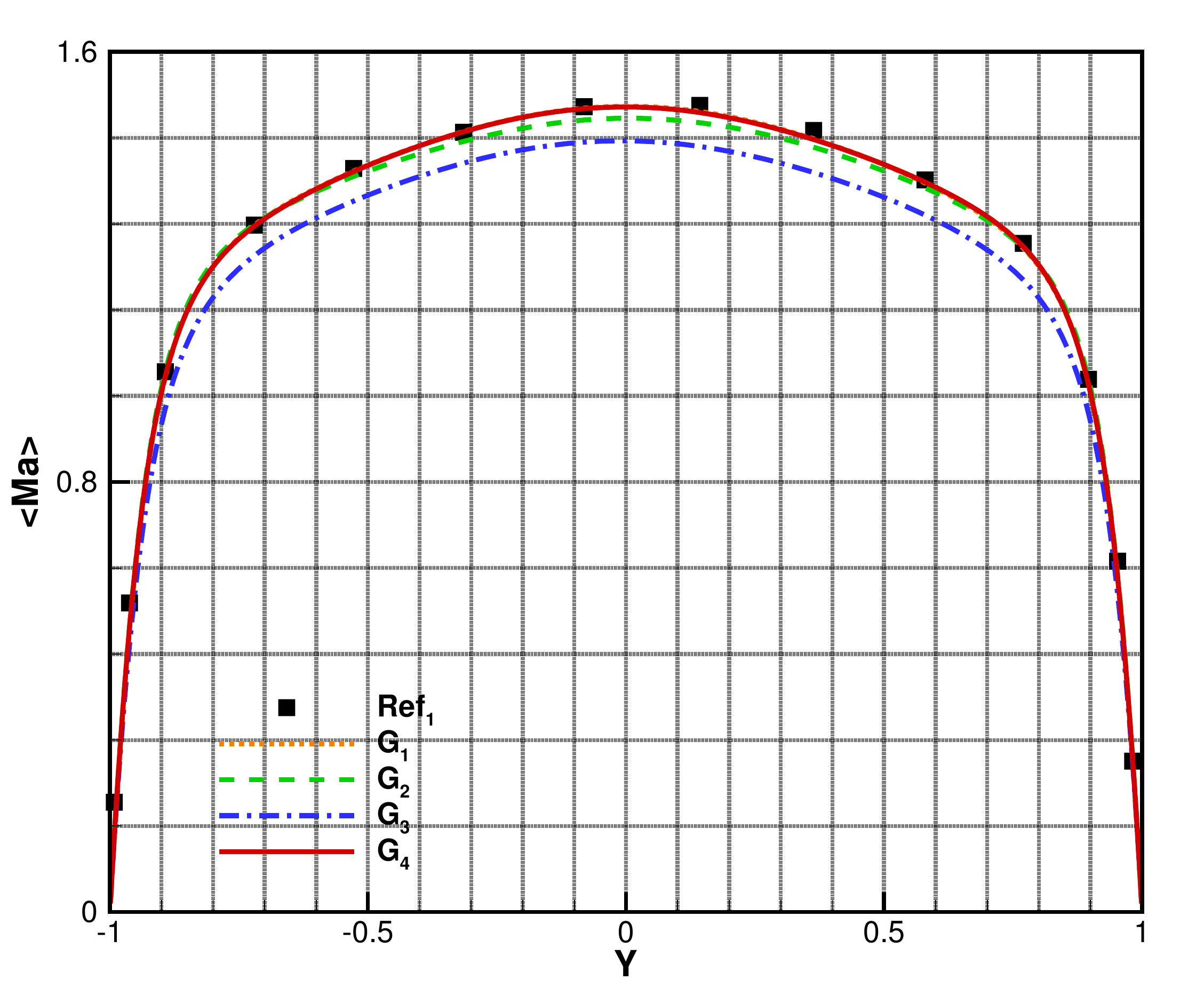}
	\includegraphics[width=0.455\textwidth]{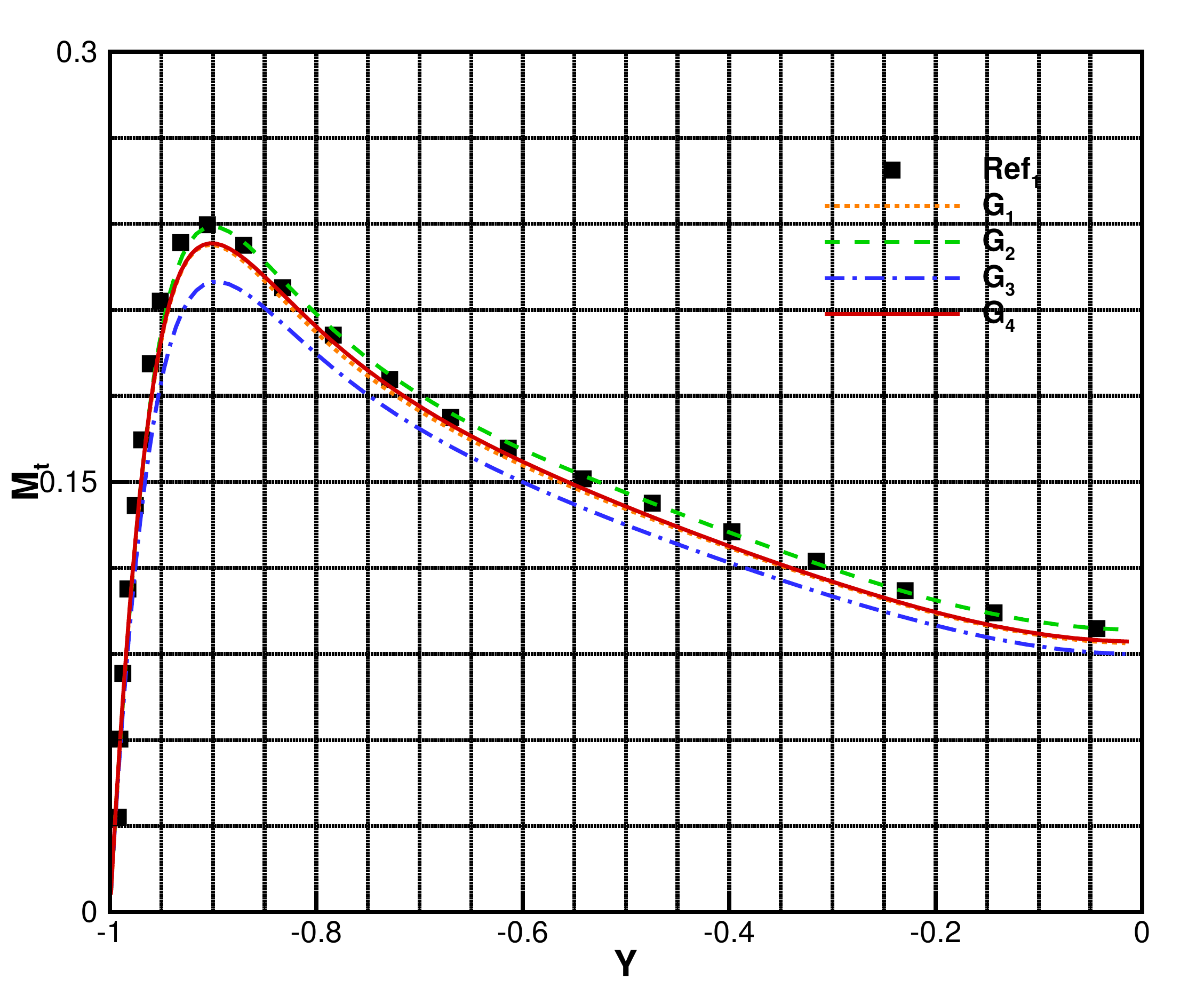}
	\caption{\label{channel_rhotma} Compressible turbulent channel flow: the mean density $\langle \rho \rangle$, temperature $\langle T \rangle$ and Mach number $\langle Ma \rangle$ (left column), and the normalized
		root-mean-square of density $\langle \rho_{rms}^{'}\rangle / \langle \rho \rangle$, temperature $\langle T_{rms}^{'}\rangle / \langle T \rangle$, turbulent Mach number $M_t$ (right column).}
\end{figure}

To further quantify the performance of HGKS-cur, the mean density $\langle \rho \rangle$, temperature $\langle T \rangle$ and Mach number $\langle Ma \rangle$, the normalized root-mean-square of density $\langle \rho_{rms}^{'}\rangle / \langle \rho \rangle$, temperature $\langle T_{rms}^{'}\rangle / \langle T \rangle$, and the turbulent Mach number $M_t$ are presented in Fig.\ref{channel_rhotma}. 
The turbulent Mach number is defined as $M_t = q/\left\langle c \right
\rangle$, where $q^2 = \left\langle U_i^{'} U_i^{'} \right \rangle$,
$U_i^{'} = U_i - \left\langle U_i \right \rangle$, and $c$ is the
local sound speed. The root mean square is defined as
$\phi_{rms}^{'} = \sqrt{(\phi - \langle \phi \rangle)^2}$, where
$\phi$ represents the density, temperature and velocity.
For current iLES with HGKS-cur, Fig.\ref{channel_rhotma} shows that case $G_1$ converges to case $G_4$, and both of them agree well with the refereed DNS solutions. 
The smaller streamwise computational domain as case $G_2$ slightly changes
the first-order statistical quantities but deviates the
root-mean-square of density and temperature obviously. The numerical
behavior of case $G_2$ indicates the streamwise computational size
should be adopted as previous study \cite{coleman1995numerical},
where the one-dimensional Fourier spectral has been used to validate
the physical domain is large enough to resolve the streamwise
turbulent structures. In terms of the effect of Prandtl number, the
large Prandtl number $Pr = 1$ changes the mean density,
temperature and Ma number profiles greatly. The large Prandtl number
also enlarges the peak of root-mean square of density and
temperate, while reduces the peak value of turbulent Mach number. For
compressible turbulence simulation using HGKS-cur, it is necessary to
modify the Prandtl number to the targeted one \cite{GKS-Xu2}.
Otherwise, the statistical thermodynamic and kinematic quantities will
deviate from the expected values greatly. For current supersonic
turbulent channel flow, the Mach number is $Ma = 1.5$, while the
peak values of the turbulent Mach number $M_t$ is less than $0.25$.
This means no strong shock-lets in such case, so spectral method \cite{coleman1995numerical} works well.
The performance of case $G_1$ and $G_4$ confirms the high-accuracy flow-fields has been obtained by the HGKS-cur with non-uniform grids.

\begin{figure}[!h]
	\centering
	\includegraphics[width=0.675\textwidth]{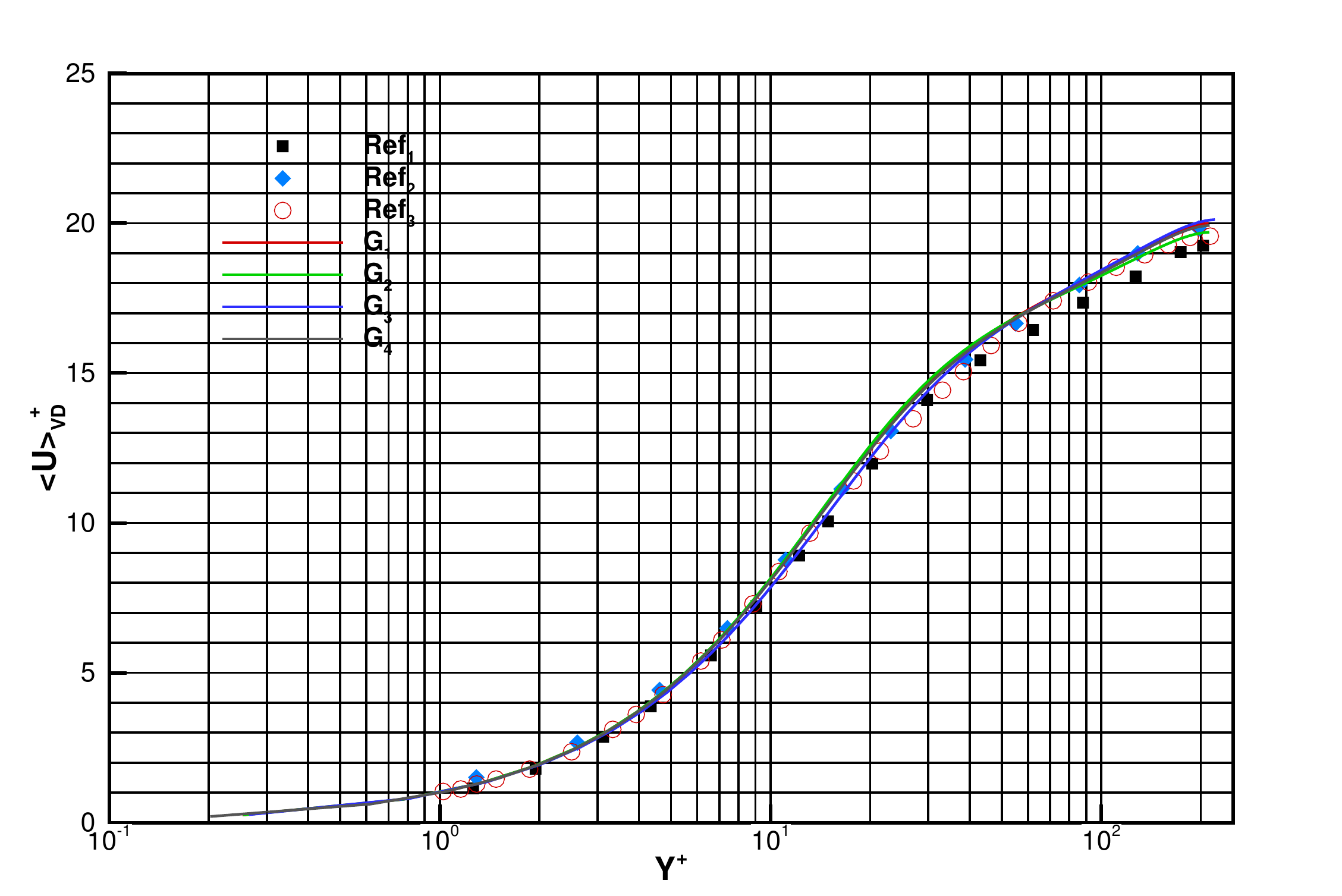}
	\caption{\label{channel_velocity}  Compressible turbulent channel flow: VD transformation of streamwise velocity profiles ${\left\langle U \right\rangle}_{VD}^+$.}
\end{figure}

\begin{figure}[!h]
    \centering
    \includegraphics[width=0.455\textwidth]{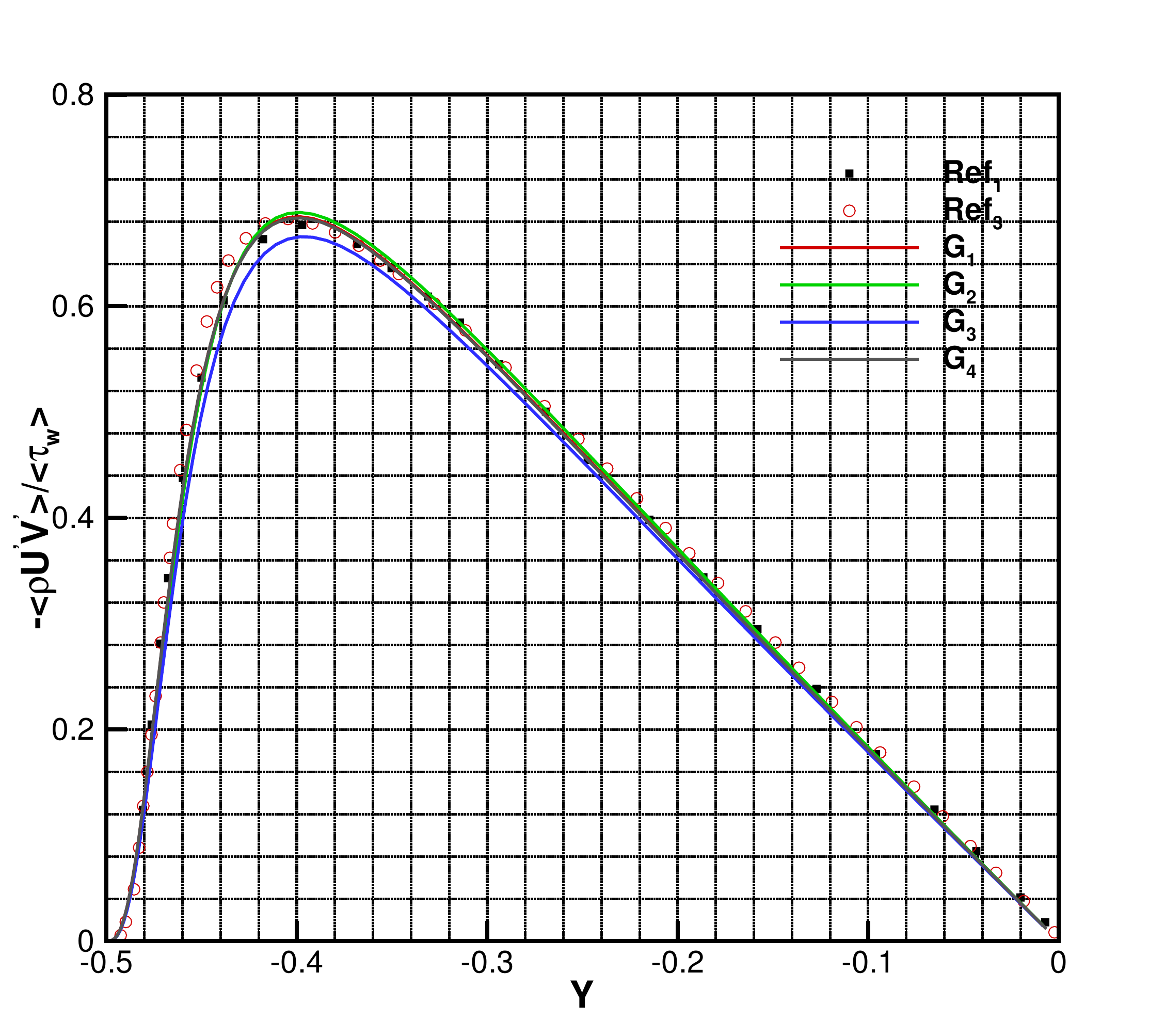}
    \includegraphics[width=0.455\textwidth]{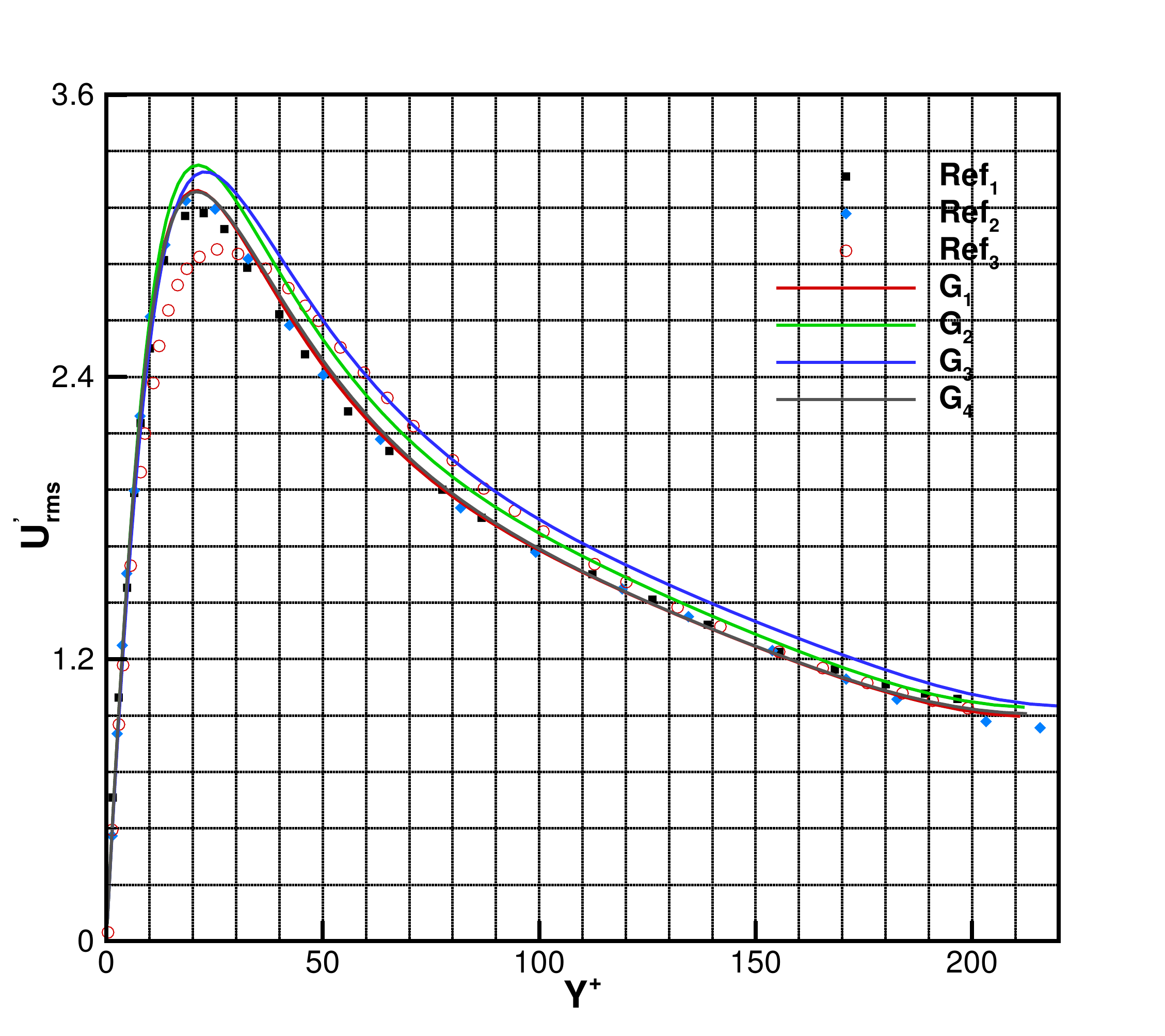}
    \includegraphics[width=0.455\textwidth]{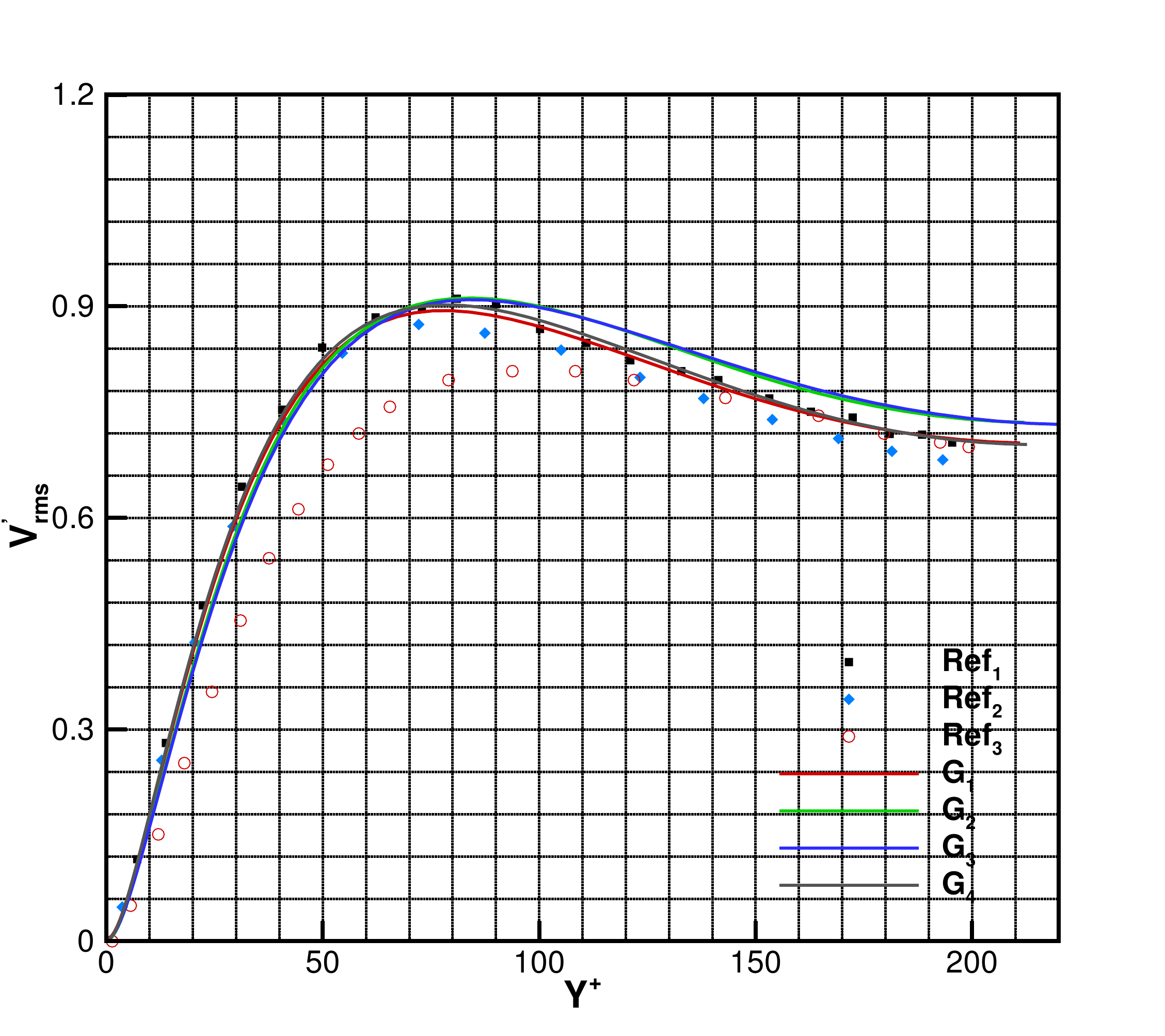}
    \includegraphics[width=0.455\textwidth]{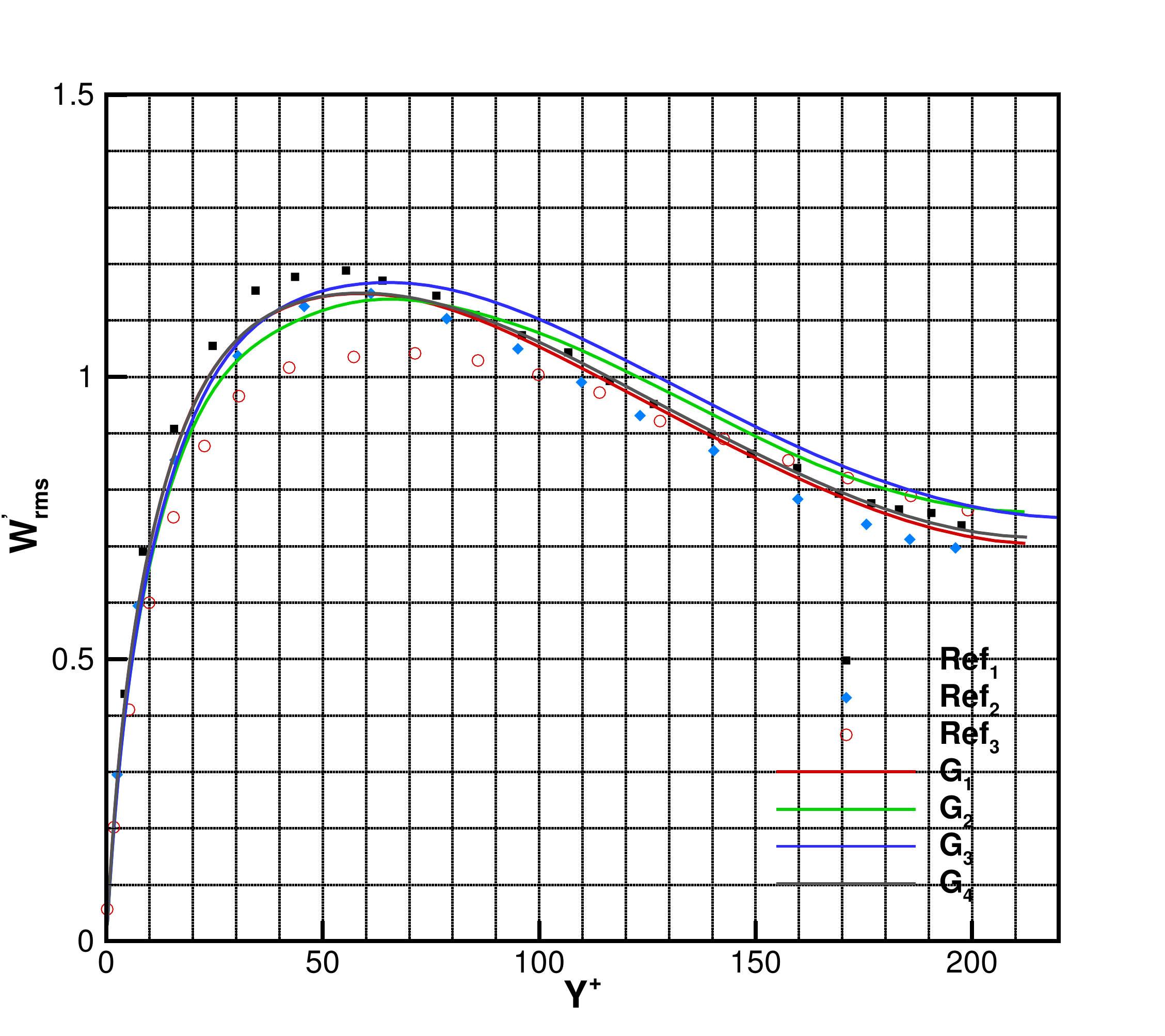}
    \caption{\label{channel_fluctuation} Compressible turbulent channel flow: profiles of normalized Reynolds stress $-\langle \rho U^{'} V^{'} \rangle/\langle \tau_{w} \rangle$ and turbulence intensities $U_{rms}^{'}$, $V_{rms}^{'}$, $W_{rms}^{'}$.}
\end{figure}
In order to account for the mean property of variations caused by
compressibility, the Van Driest (VD) transformation \cite{vdtransformation} for the mean velocity, i.e., density-weighted velocity, is considered 
\begin{align*}
    {\left\langle U \right\rangle}_{VD}^+ = \int_{0}^{{\left\langle U \right\rangle}^+} \large \bigg (\frac{\left\langle \rho \right\rangle}{\left\langle \rho_w \right\rangle} \bigg )^{1/2} \text{d} {\left\langle U \right\rangle}^+,
\end{align*}
where the transformed velocity is expected to satisfy the incompressible log law
\cite{coleman1995numerical}. 
The streamwise velocity profiles $\langle U \rangle_{VD}^+$ with VD transformation are given in Fig.\ref{channel_velocity}. 
Overall, the iLES with HGKS-cur is in reasonable agreement with the reference DNS solutions, and CLES also performs very well on coarse grids. 
The profiles of normalized Reynolds stress $-\langle \rho U^{'} V^{'} \rangle/\langle \tau_{w} \rangle$ and the turbulence intensities (the root-mean-square
velocities as $U_{rms}^{'}$, $V_{rms}^{'}$, $W_{rms}^{'}$) are presented in Fig.\ref{channel_fluctuation}.
Case $G_1$ converges to the case $G_4$, and both of them agree well
with the refereed solutions. The smaller streamwise computational domain
as case $G_2$ and the large Prandtl number case $G_3$ deviate
obviously from the refereed solutions. This confirms again that the enough
streamwise computational size and targeted Prandtl number are essential in 
compressible wall-bounded turbulence simulations. The computational
domain and the Prandtl number should be stressed for the iLES of
compressible turbulent channel flow. 
The total Reynolds stress from CLES (containing the mean modeled SGS stress) coincides  well with the DNS data.
Even though CLES underestimates the values of turbulence intensities in the inner layer of flow, it still performs slightly better than the explicit LES with Smagorinsky model \cite{jiang2013constrained}. 
Based on the reasonable performance of case $G_1$ and $G_4$, it can be concluded that iLES with current HGKS-cur on non-uniform grids offers the high-accuracy flow-fields for compressible turbulent channel flow.

\begin{figure}[!h]
	\centering
	\includegraphics[width=0.55\textwidth]{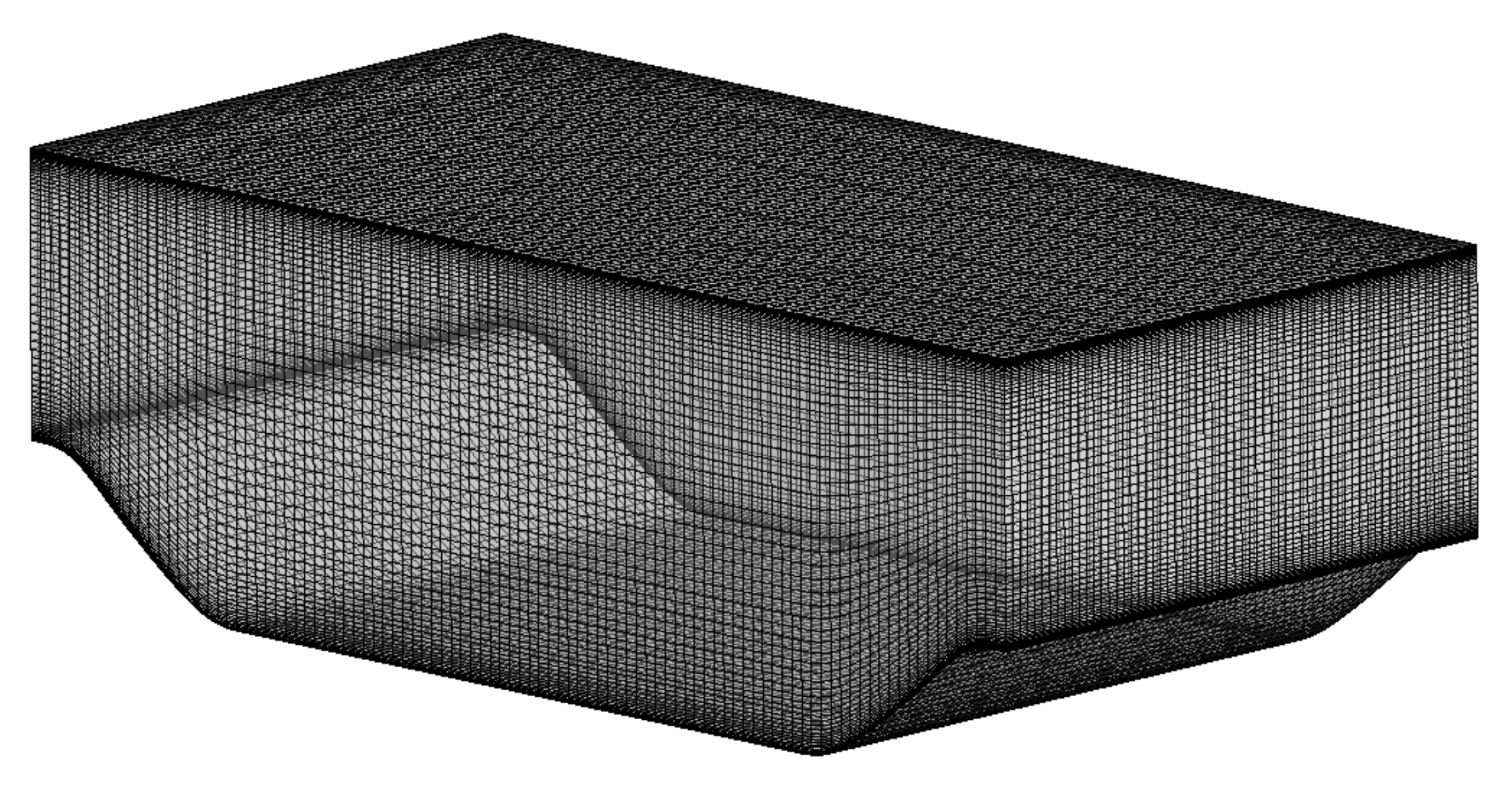}
	\includegraphics[width=0.65\textwidth]{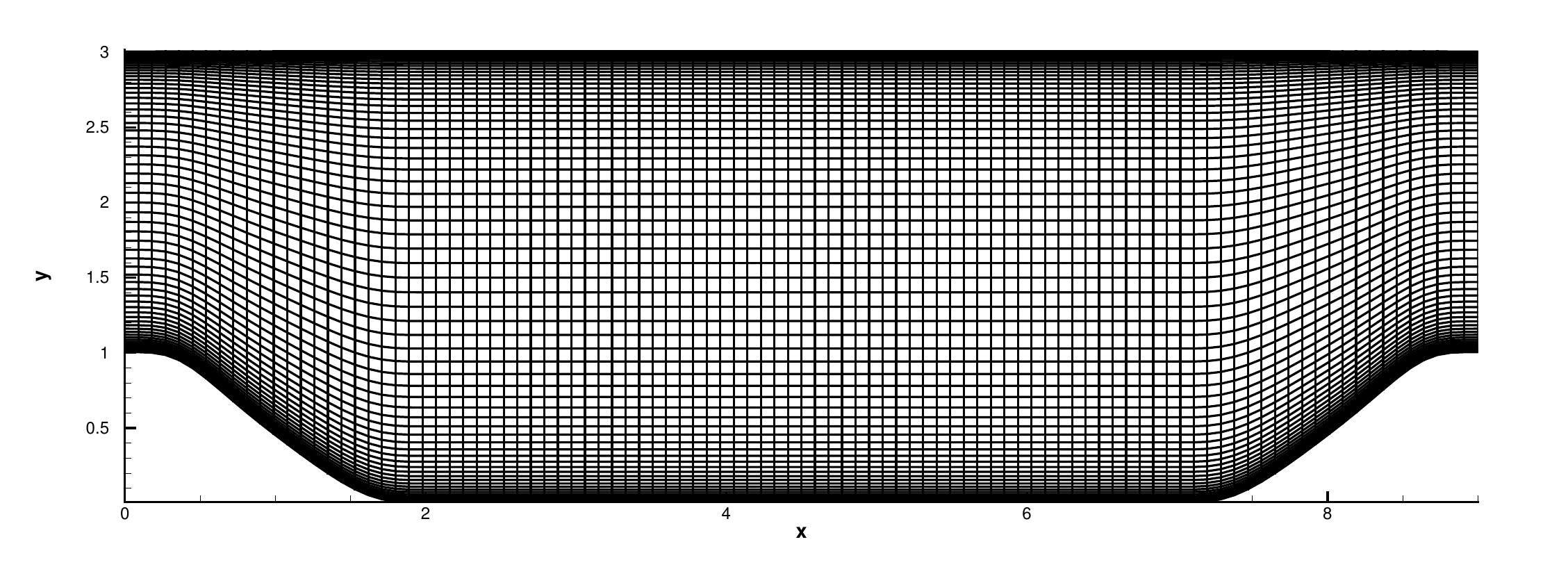}
	\caption{\label{periodic-hill-mesh} Compressible turbulent flow over periodic hills: the curvilinear 3D physical mesh with $100^3$ cells (upper) and a side view of 2D mesh with $100^2$ cells (bottom).}
\end{figure}

\subsection{Compressible turbulent flow over periodic hills}
The turbulent flow over periodically arranged hills in a channel
\cite{peller2004dns,ziefle2008large,breuer2009flow,balakumar2015dns} has been widely utilized to study the massive flow separation.  
In this section, the compressible turbulent flow over periodic hills with volumetric Mach number $Ma_v = 0.2$ and cross-sectional Reynolds number $Re_b = 2800$ is tested with curvilinear mesh.
The geometry and a side view of the periodic hill are shown in
Fig.\ref{periodic-hill-mesh}. 
The physical domain is irregular with curved bottom wall and flat upper wall. 
The physical box extends over $x \in [0,9H]$ in the streamwise direction, $y \in [0,3.036H]$ in the wall-normal direction, and $z \in [0,4.5H]$ in the spanwise
direction, respectively. The computational domain takes
$(\xi,\eta,\zeta) \in [0,9H] \times [-1.5\pi H,1.5\pi H] \times[0,4.5H]$ as a
regular cuboid.  
The hill height $H = 1$ is chosen as the reference length for normalization.
Equidistant grids are generated in spanwise and streamwise directions. The grid points on the bottom is taken from Ref \cite{gloerfelt2019large}, and the wall-normal grids is given by
\begin{align*}
y=\frac{3-y_0}{2}\Big(\tanh(b_g(\frac{\eta}{1.5\pi}-1))/\tanh(b_g)\Big)+\frac{3+y_0}{2},
\end{align*}
where $y_0$ is $y$-coordinate of  bottom grid points and $\eta$
distributes uniformly over $[-1.5\pi,1.5\pi]$. The curvilinear mesh is given by the discretized grid points without analytical transformation. This case addresses
the performance of HGKS-cur for the separated turbulence from the
curved surface. The periodic boundary conditions are used in both
streamwise $x$-direction and spanwise $z$-direction, and the
non-slip and isothermal boundary conditions are used in upper wall
and bottom wall. 

\begin{figure}[!h]
	\centering
	\includegraphics[width=0.475\textwidth]{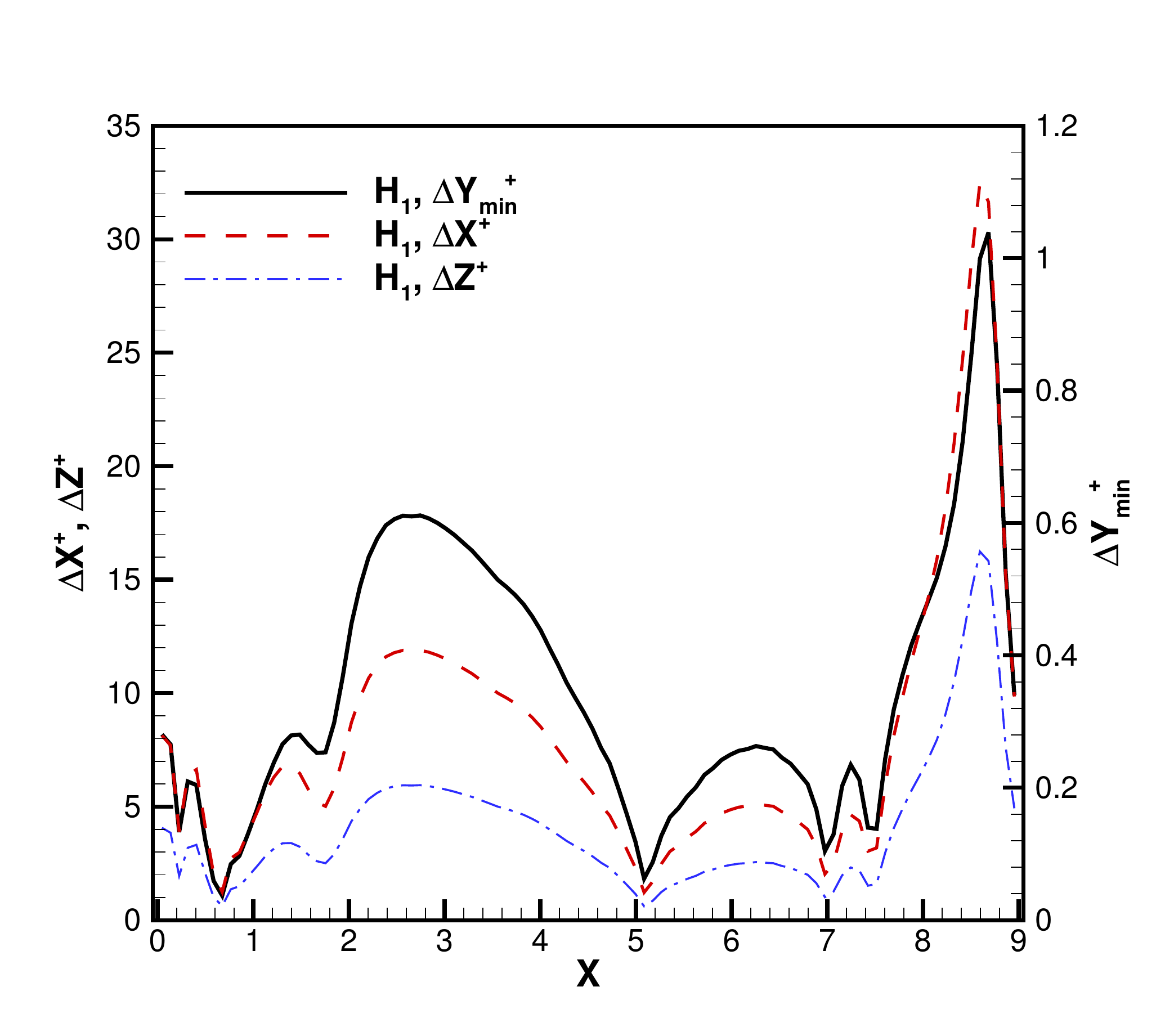}
	\includegraphics[width=0.475\textwidth]{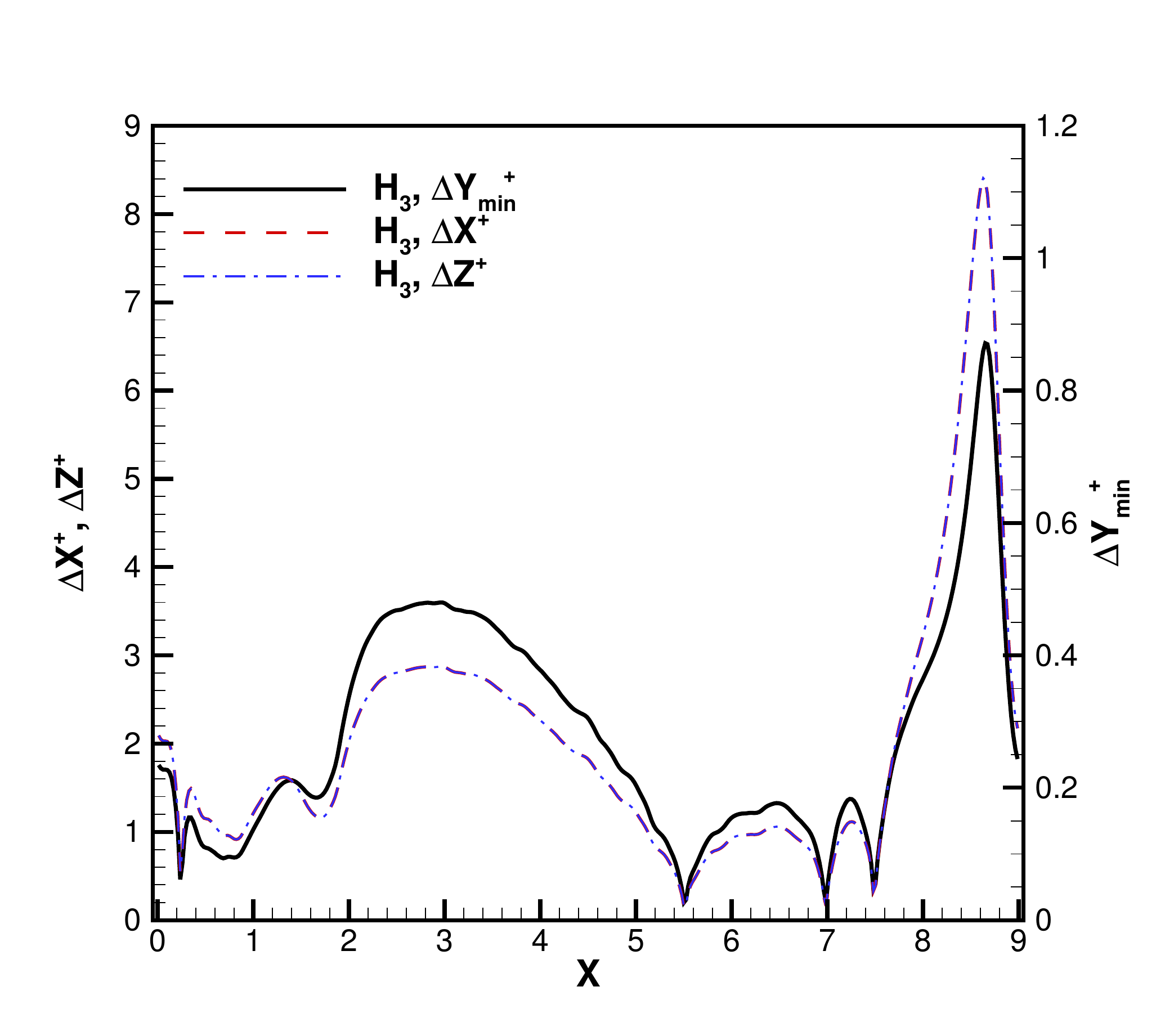}
	\caption{\label{periodic-hill-wall-unitplus} Compressible turbulent flow over periodic hills: grids spacing in the wall units for case $H_1$ and $H_3$. $\Delta Y_{min}^+$ is the first grid space off the bottom wall in the wall-normal direction.}
\end{figure}

In this study, the volumetric Mach number $Ma_v$ is defined as
\begin{align*}
	Ma_v = \frac{U_v}{c_w}, 
	~ U_v = \frac{1}{|\Omega|} \iiint_{\Omega} U\text{d}\Omega,
\end{align*}
where $|\Omega|$ is the volume of physical domain, $c_w = \sqrt{\gamma R T_w} $ is the wall sound speed and $T_w$
is the temperature at wall.
The volumetric Reynolds number $Re_{v}$ and the
cross-sectional Reynolds number $Re_{b}$ are defined as
\begin{align*}
Re_{v} = \frac{H}{\mu|\Omega|}\iiint_{\Omega}\displaystyle\rho U\text{d}\Omega, ~
Re_{b} = \frac{H}{\mu|S|}\iint_{S}\displaystyle\big(\rho U\big)\big|_{x=0}\text{d}S,
\end{align*}
where $|S|$ is the area
of inlet cross section at the crest of hill. 
The cross-sectional Reynolds number $Re_{b}$ can be determined by
\begin{align*}
Re_{b} =  \frac{Re_{v}}{\Gamma},
~\Gamma = \frac{L_x L_y |_{x = 0}}{\int_{0}^{L_x}L_y(x)\text{d}x} = 0.72,
\end{align*}
where $\Gamma$ is geometry factor, $L_x = 9$ and $L_y(x)$ the height of tunnel with respect to streamwise direction.
The constant dynamic viscosity is used, and Prandtl number takes $Pr = 0.72$.
In what follows, the mean average over the time and spanwise
$Z$-direction is denoted by $\langle \cdot \rangle$. 
The mean friction coefficient $C_f$ reads
\begin{align*}
	C_f = \frac{\tau_{w}}{\langle \rho_b\rangle \langle U_b \rangle^2},
	~\tau_{w} = \mu_w \frac{\partial \langle U \rangle}{\partial
		n}\big|_w,
\end{align*}
where cross-sectional density $\rho_b$ and cross-sectional velocity $U_b$ are given by
\begin{align*}
	\rho_b = \frac{1}{|S|} \iint_{S} \rho|_{x=0} \text{d}S,
	~ U_b = \frac{1}{|S|} \iint_{S} U|_{x=0} \text{d}S.
\end{align*}
The pressure coefficient $C_p$ is defined as
\begin{align*}
	C_p= \frac{\langle p \rangle - \langle p_x \rangle}{\langle
		\rho_b\rangle \langle U_b \rangle^2},  
	~ p_x = \frac{1}{L_x} \int_{0}^{L_x}p(x)\text{d}x,
\end{align*}
where $p_x$ is the average pressure along the bottom wall.
In this computation, cases $H_1-H_3$ are implemented by HGKS-cur as iLES. 
Details of volumetric Mach number $Ma_v$, numerical parameters and separation/reattachment locations $X_{sep}/X_{reatt}$ are presented in Table.\ref{hill_parameters}.
DNS with immersed boundary technique on a non-equidistant staggered Cartesian mesh in conjunction with an incompressible second-order finite-volume
solver \cite{peller2004dns} is referred as Ref$_1$.
Ref$_2$ is equipped with the fourth-order finite-volume scheme \cite{ziefle2008large} for compressible LES. 
The approximate deconvolution model (ADM) is
used for the compressible LES. 
$\Delta$t is the fixed time step used in simulations.
Grids spacing in the wall
units for case $H_1$ and $H_3$ are presented in
Fig.\ref{periodic-hill-wall-unitplus}, where $\Delta Y_{min}^+$ is the
first grid space off the bottom wall in the wall-normal direction.
The unit plus is computed based on the post-processed mean flow fields, where
each wall point has a local friction velocity. For current iLES
study, the grids spacing in the wall units of case $H_1$ is comparable with that
in previous iLES \cite{balakumar2015dns}. While, the girds of case $H_3$ is much finer, to implement the grid convergence study of iLES.
\begin{table}[!h]
	\centering
	\begin{tabular}{c|c|c|c|c|c|c|c}
		\hline \hline
		Case       &Run  &$Ma_v$     &$N_x \times N_y \times N_z$  &$1\times10^6$ cells &$\Delta$t/$10^{-3}$    &$X_{sep}$  &$X_{reatt}$\\
		\hline
		Ref$_1$    &DNS  &N/A     &$464 \times 304 \times 338$  &47.68 &2.0 &0.21 &5.41\\
		\hline
		Ref$_2$    &LES &0.2    &$128 \times 72 \times 69$     &0.64  &1.0 &0.21 &5.30\\
		\hline
		$H_1$      &iLES &0.2    &$100 \times 100 \times 100$   &1.0   &0.6 &0.23 &5.18\\
		\hline
		$H_2$      &iLES &0.2    &$200 \times 100 \times 100$   &2.0   &0.6 &0.20 &5.15 \\
		\hline
		$H_3$      &iLES &0.2    &$400 \times 200 \times 200$   &16.0  &0.55 &0.24 &5.52 \\
		\hline \hline
	\end{tabular}
	\caption{\label{hill_parameters} Compressible turbulent flow over periodic hills: volumetric Mach number, numerical simulation parameters, and separation/reattachment locations of the present and reference simulations. "N/A" means no volumetric Mach number resulting from the incompressible simulation.}
\end{table}

\begin{figure}[!h]
	\centering
	\includegraphics[width=0.485\textwidth]{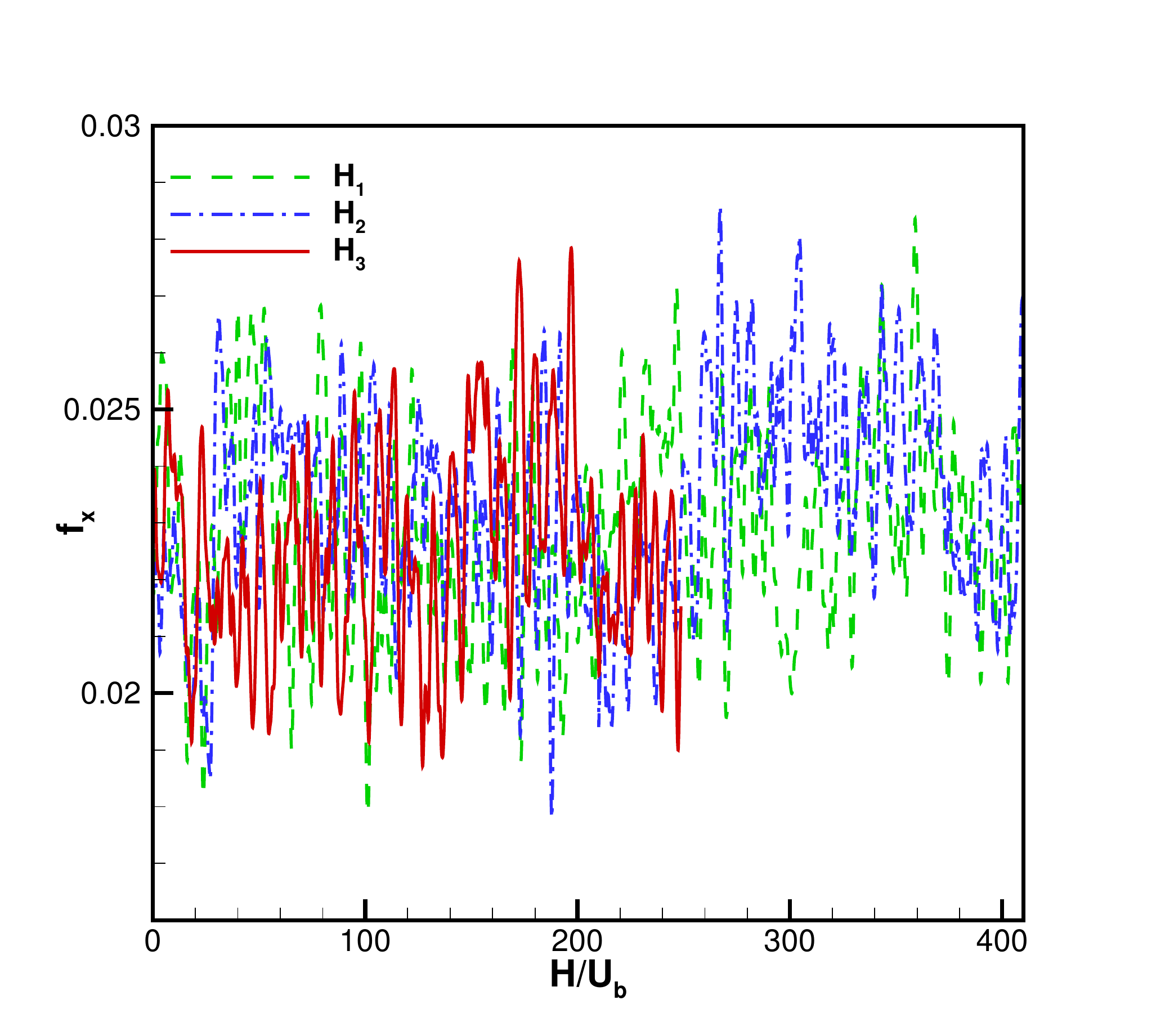}
	\includegraphics[width=0.485\textwidth]{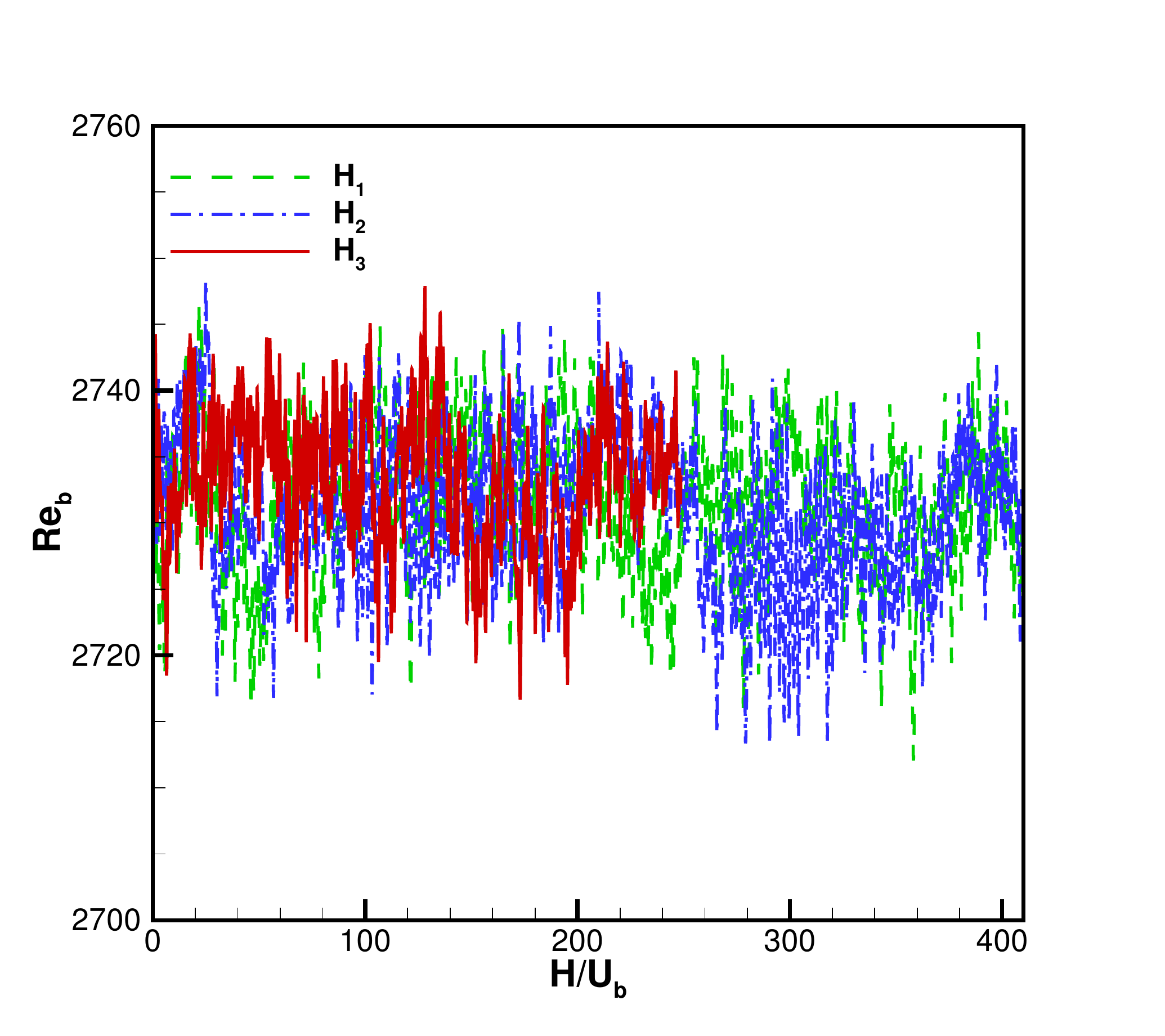}
	\caption{\label{hill_force} Compressible turbulent flow over periodic hills: external force $f_x$ and cross-sectional Reynolds number $Re_b$ for cases $H_1-H_3$.}
\end{figure}
\begin{figure}[!h]
	\centering
	\includegraphics[width=0.98\textwidth]{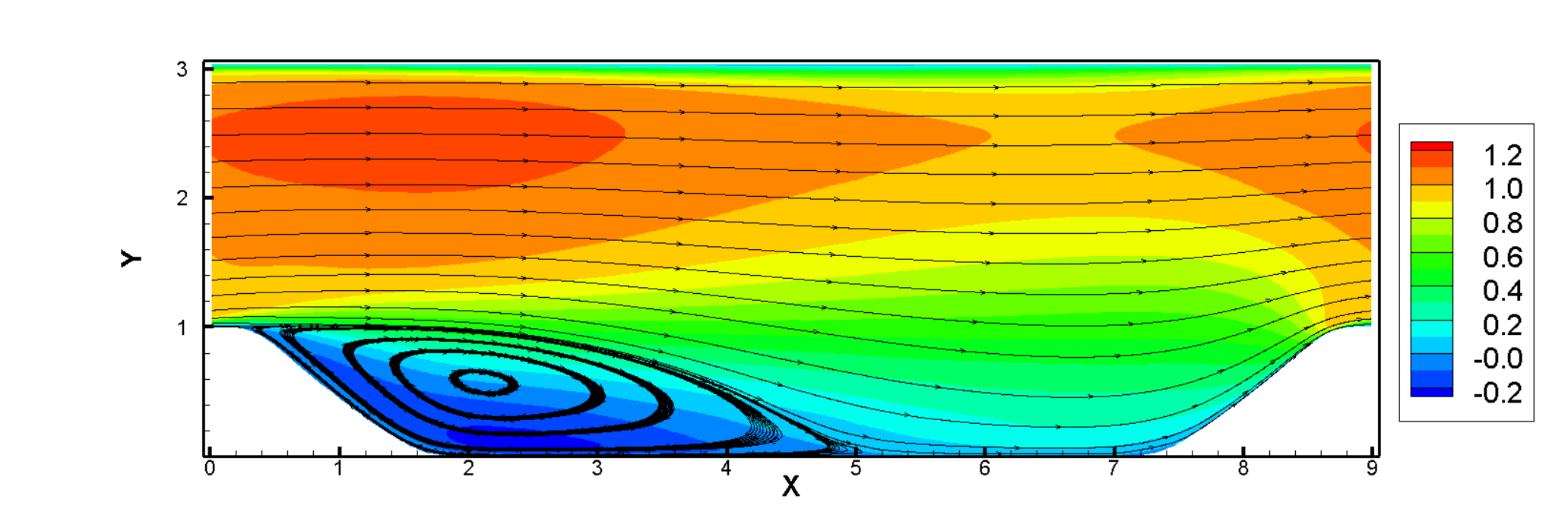}
	\caption{\label{hill_3d2d} Compressible turbulent flow over periodic hills: contour of mean streamwise velocity and streamlines.}
\end{figure}
\begin{figure}[!h]
	\centering
	\includegraphics[width=0.75\textwidth]{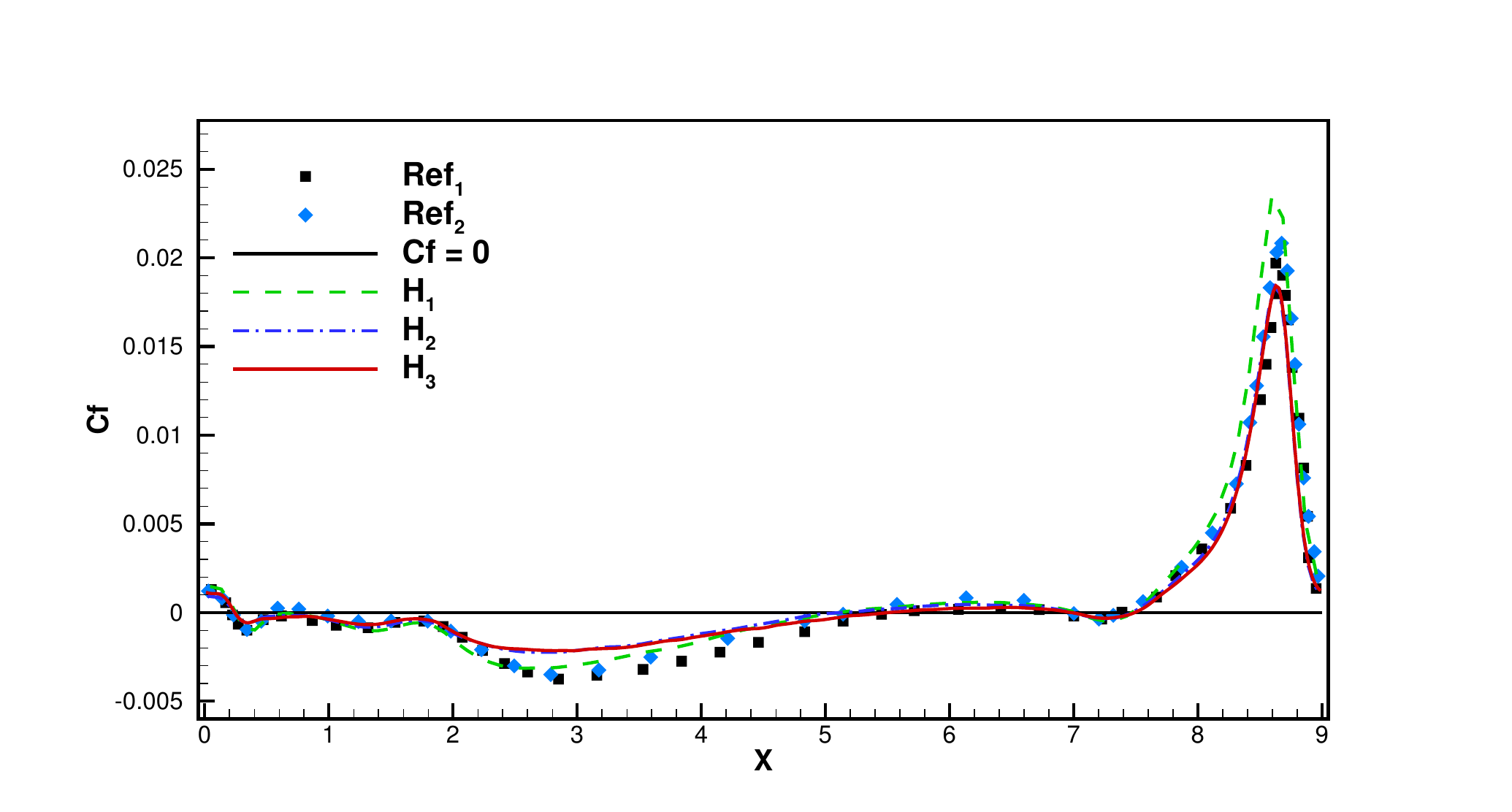}
	\includegraphics[width=0.75\textwidth]{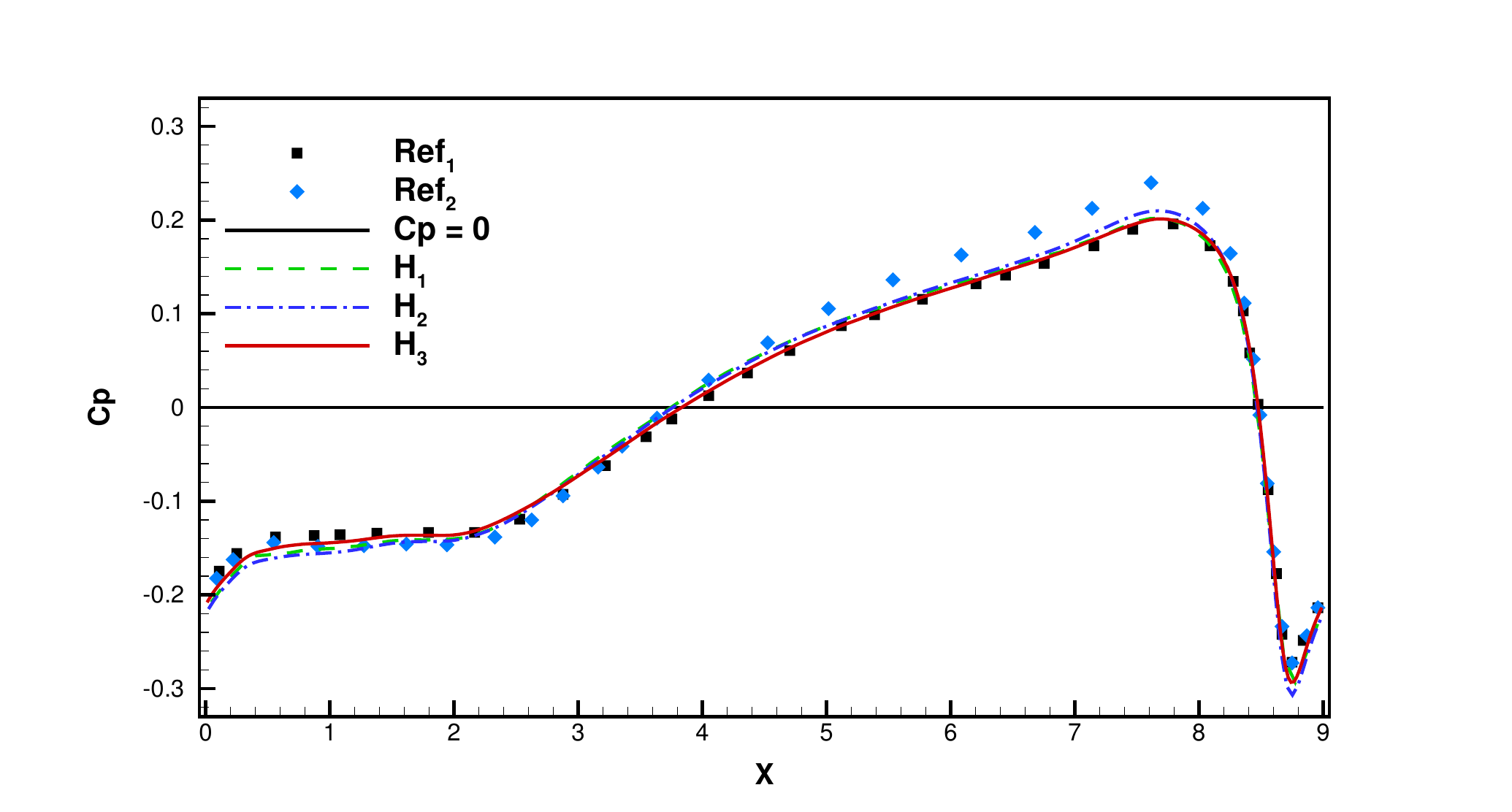}
	\caption{\label{hill_cfcp}
		Compressible turbulent flow over periodic hills: frictional coefficients $C_f$
		and pressure coefficients $C_p$ along the bottom wall.}
\end{figure}
To keep the constant streamwise moment flux, the force is
implemented as a spatially constant but temporally varying volume
force in the streamwise direction \cite{GKS-high-cao-2}. 
The external force $f_x$ and the cross-sectional
Reynolds number $Re_b$ after transition for cases $H_1 - H_3$ are
presented in Figure.\ref{hill_force}. 
Due to the variations of mass flux over cross section, $Re_{b}$ is a function of time and fluctuates around $2730$,
which is slightly smaller than the targeted values $2800$.
These highly unsteady flow properties, lead to long sampling times to obtain sufficiently converged statistics. $400$ characteristic periodic time is used for
obtaining the statistically stationary turbulence for cases $H_1$
and $H_2$. For case $H_3$ with finest grids, more than $250$ characteristic
periodic time is adopted for a converged statistical study. The
averaging time is comparable to that in the refereed paper
\cite{ziefle2008large}.  
The contour of mean streamwise velocity and streamlines for case $H_3$
is presented as Fig.\ref{hill_3d2d}. The instantaneous flow shows
a periodic shedding of smaller vortices that are convected
downstream, and the resulting separation bubble can be recognized
clearly in the mean flow field. 
The friction coefficients and pressure coefficients along the bottom wall are
presented in Fig.\ref{hill_cfcp}. The friction coefficients $C_f$ of the
current iLES and the explicit LES with ADM deviates slightly from the refereed DNS solution. The separation and reattachment locations shown in
Table.\ref{hill_parameters} are obtained based on these profiles of friction
coefficient. The locations from current iLES are close to the
refereed explicit LES and DNS solutions. For pressure coefficient $C_p$, current iLES agrees well with the refereed DNS solution, better than
the explicit LES. Considering the less grids are used than DNS, the
iLES of current HGKS-cur performs reasonably and provides efficient tool
for compressible separated flow simulations.

 \begin{figure}[!h]
    \centering
    \includegraphics[width=0.24\textwidth]{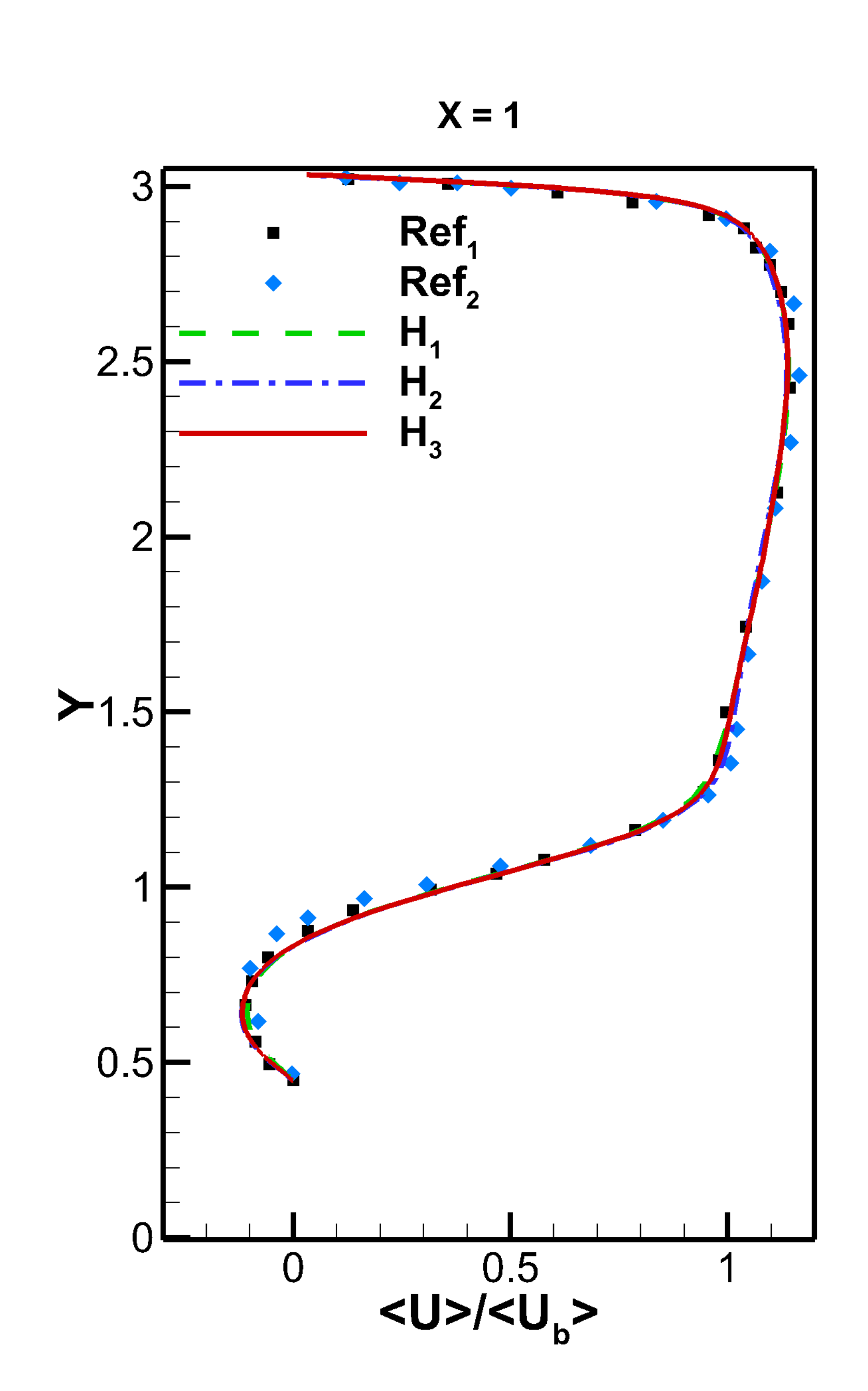}
    \includegraphics[width=0.24\textwidth]{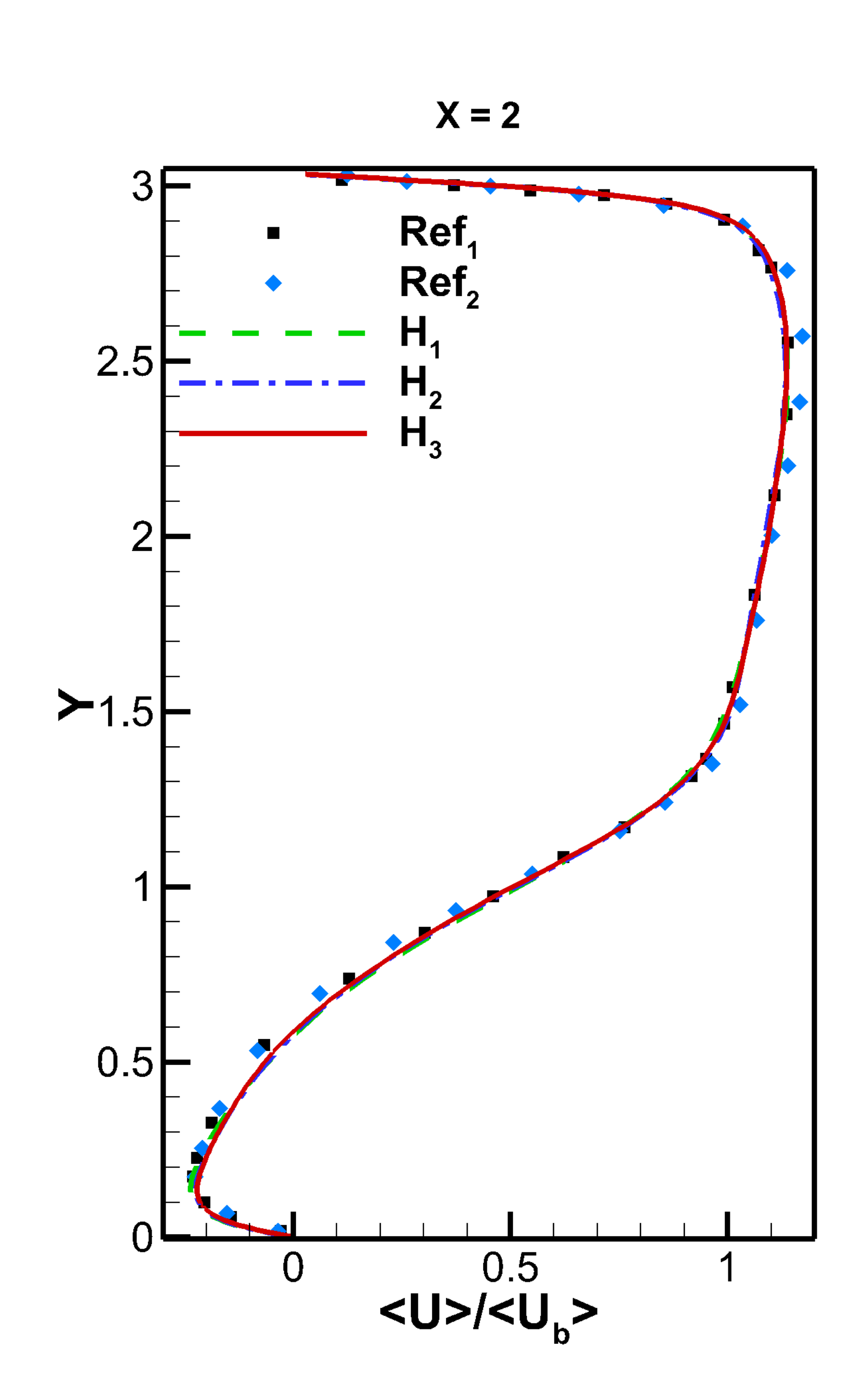}
    \includegraphics[width=0.24\textwidth]{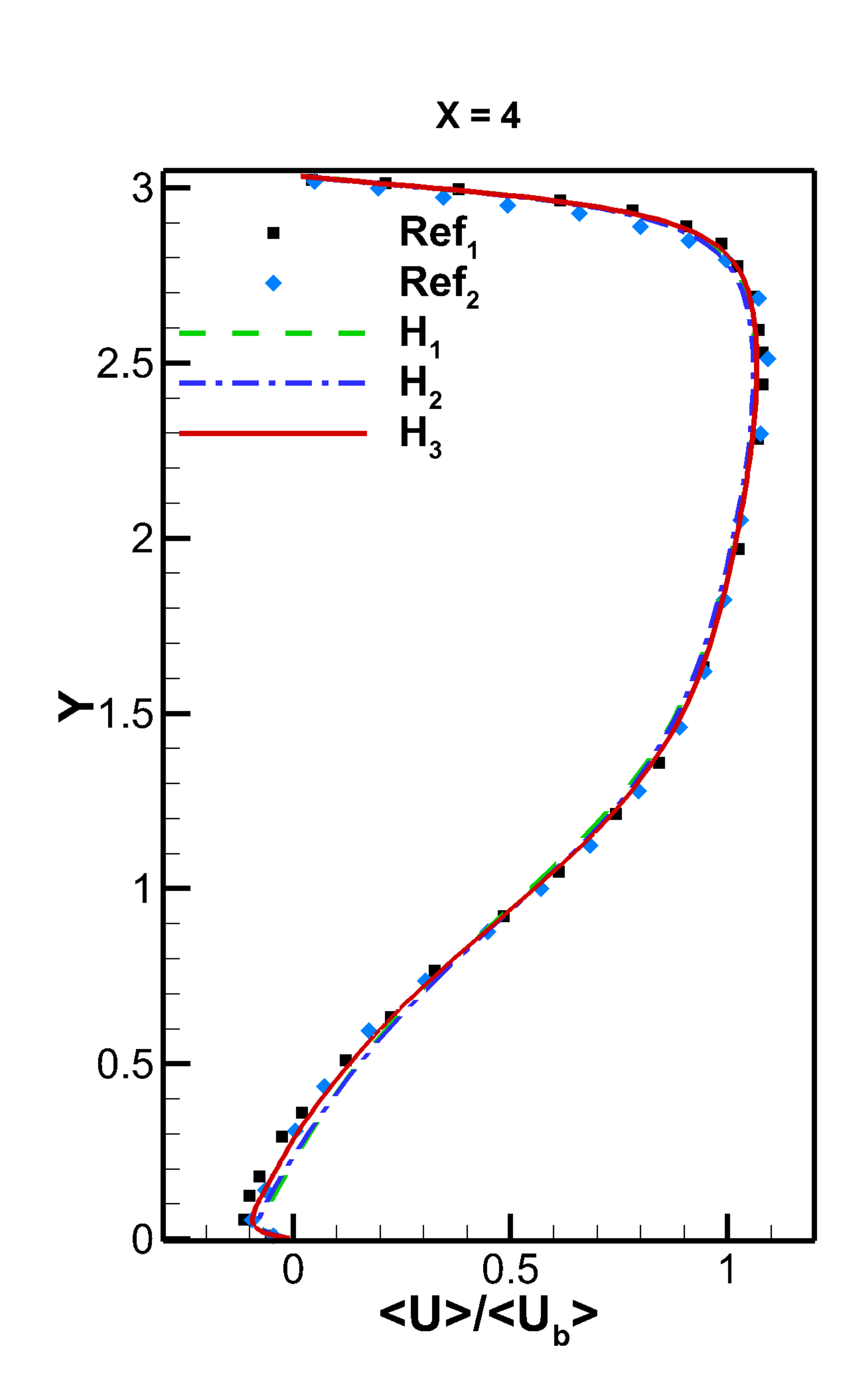}
    \includegraphics[width=0.24\textwidth]{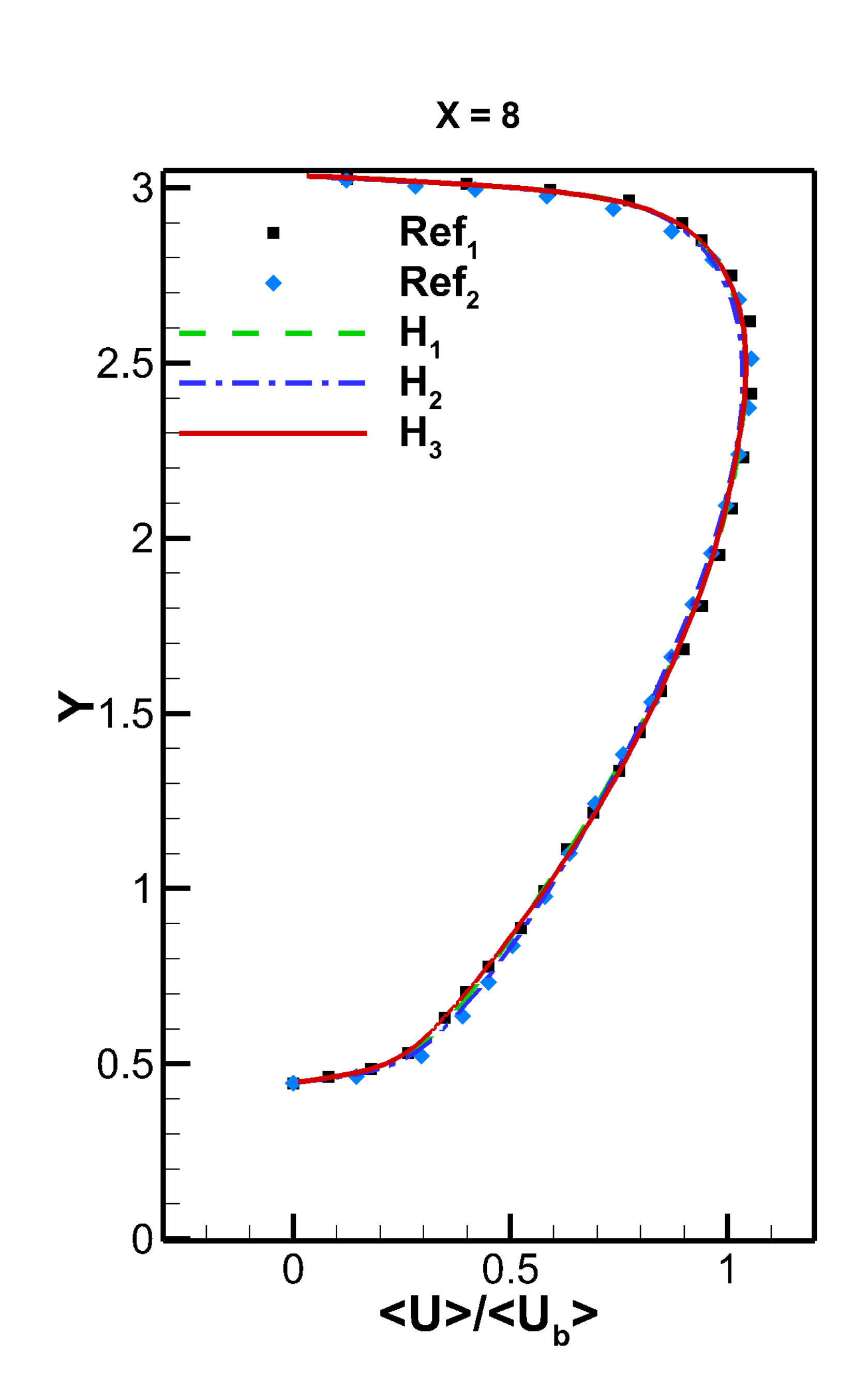}
    \includegraphics[width=0.24\textwidth]{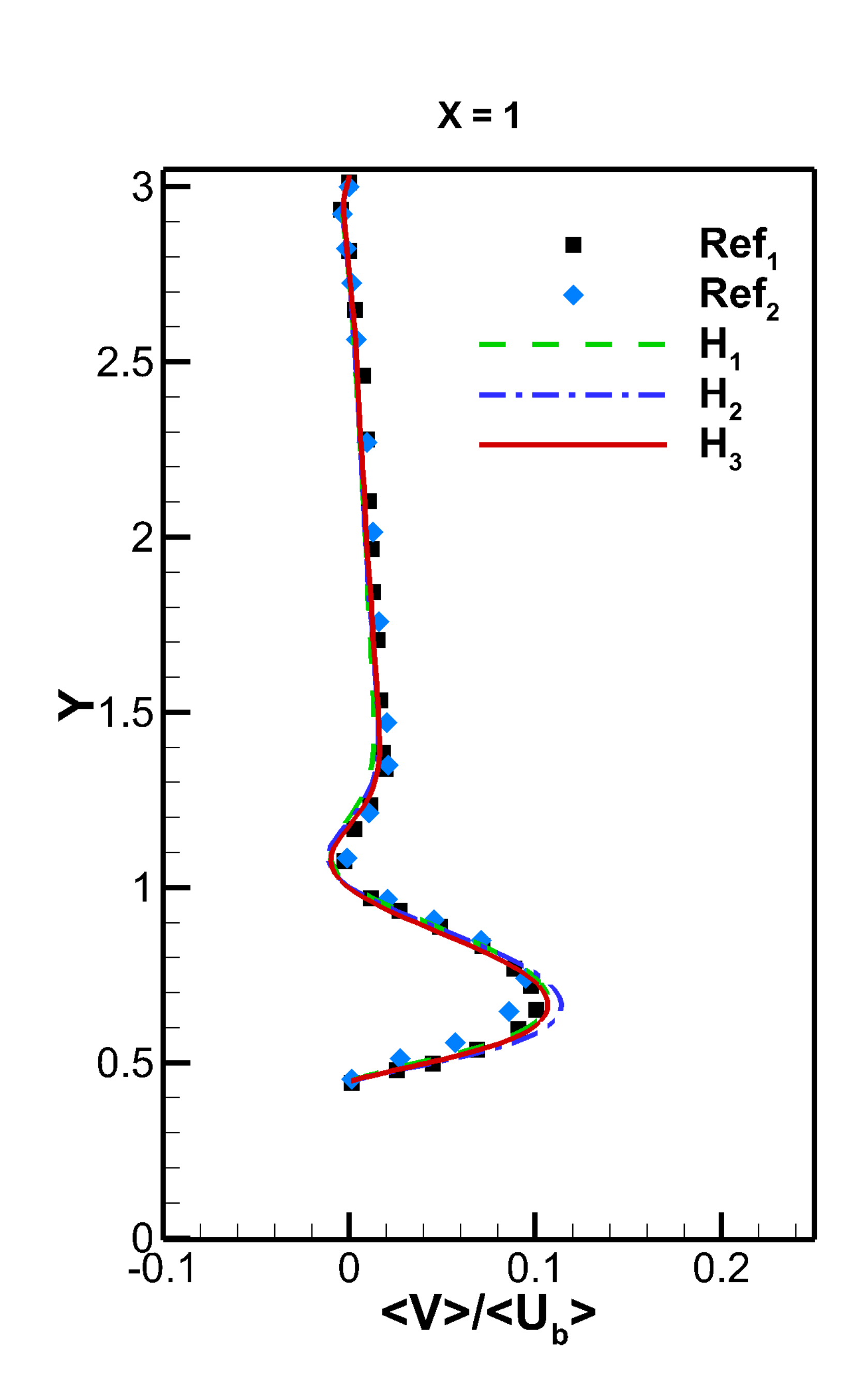}
    \includegraphics[width=0.24\textwidth]{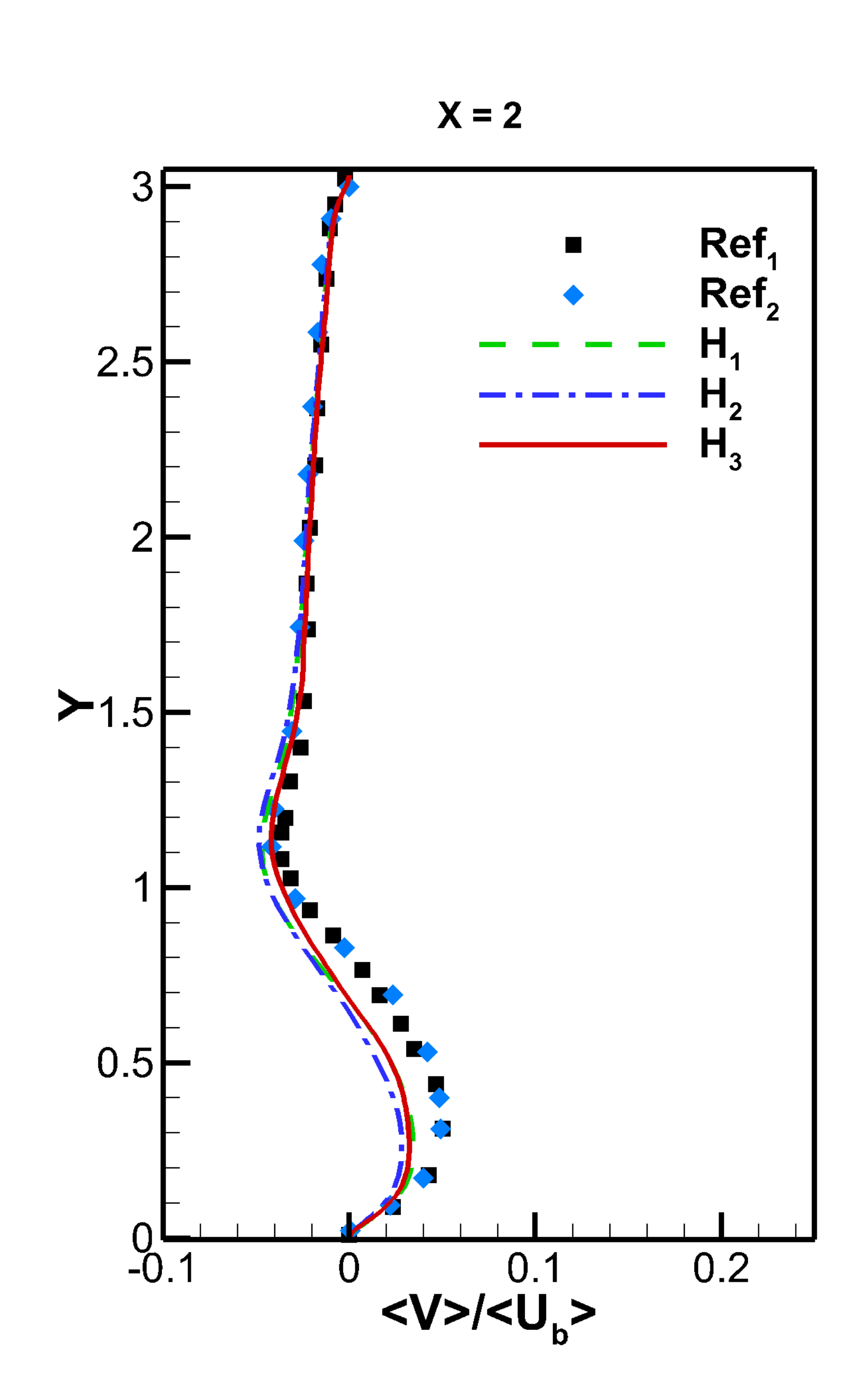}
    \includegraphics[width=0.24\textwidth]{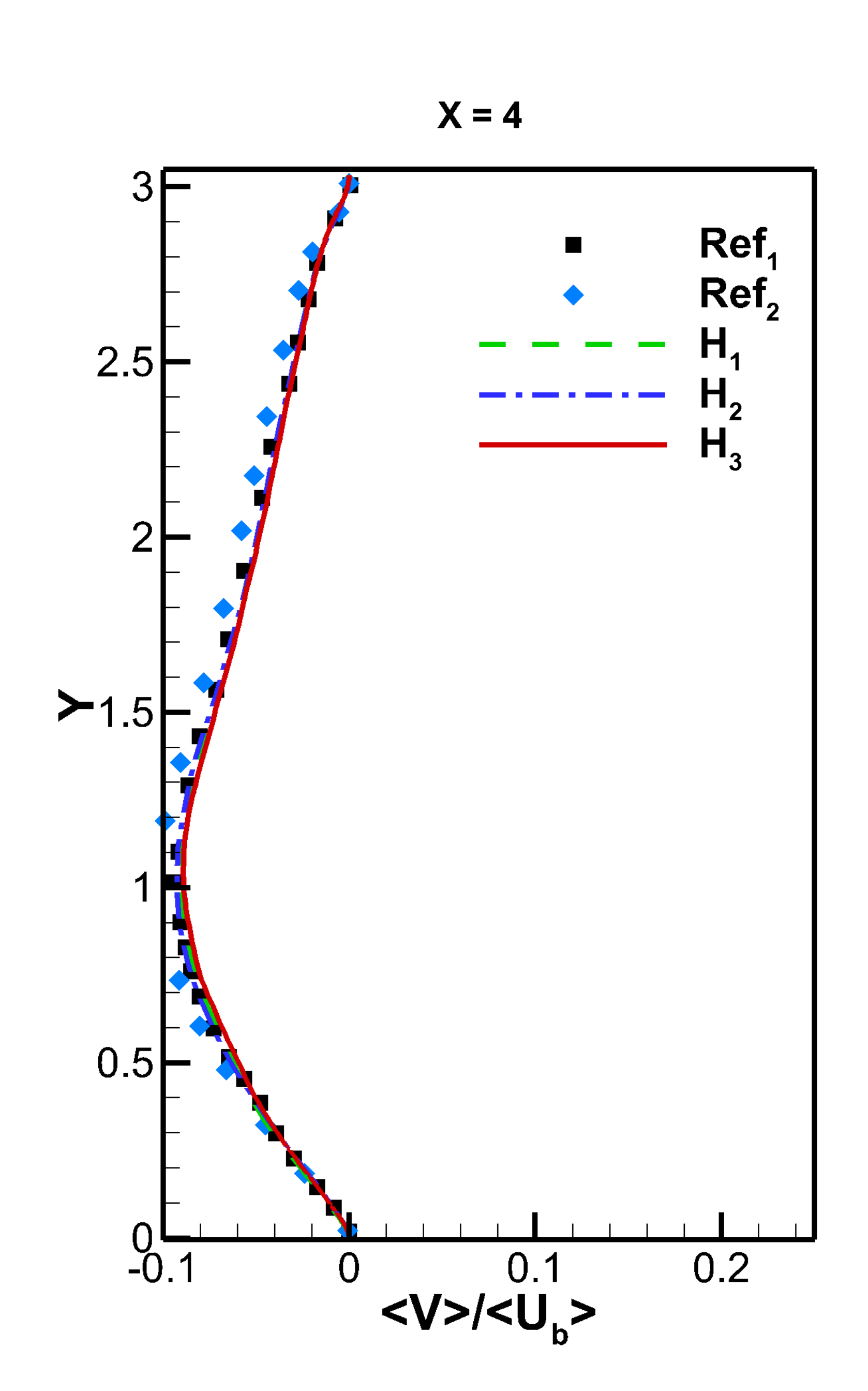}
    \includegraphics[width=0.24\textwidth]{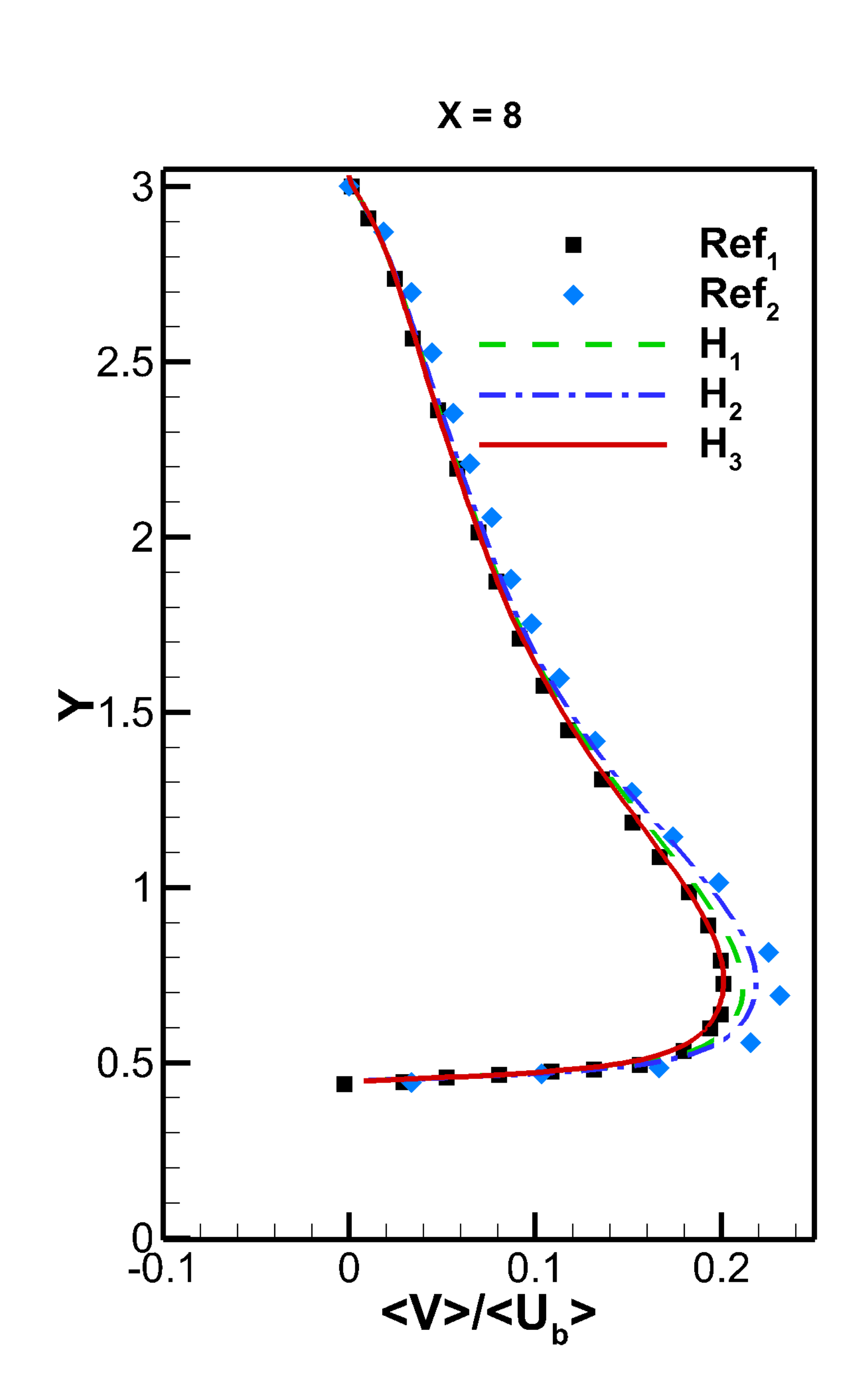}
    \caption{\label{hill_u_v_averge}
Compressible turbulent flow over periodic hills: profiles of normalized mean
streamwise velocity $\langle U \rangle /\langle U_b \rangle$ and normalized mean wall-normal velocity $\langle V \rangle /\langle U_b \rangle$.}
\end{figure}
\begin{figure}[!h]
    \centering
    \includegraphics[width=0.24\textwidth]{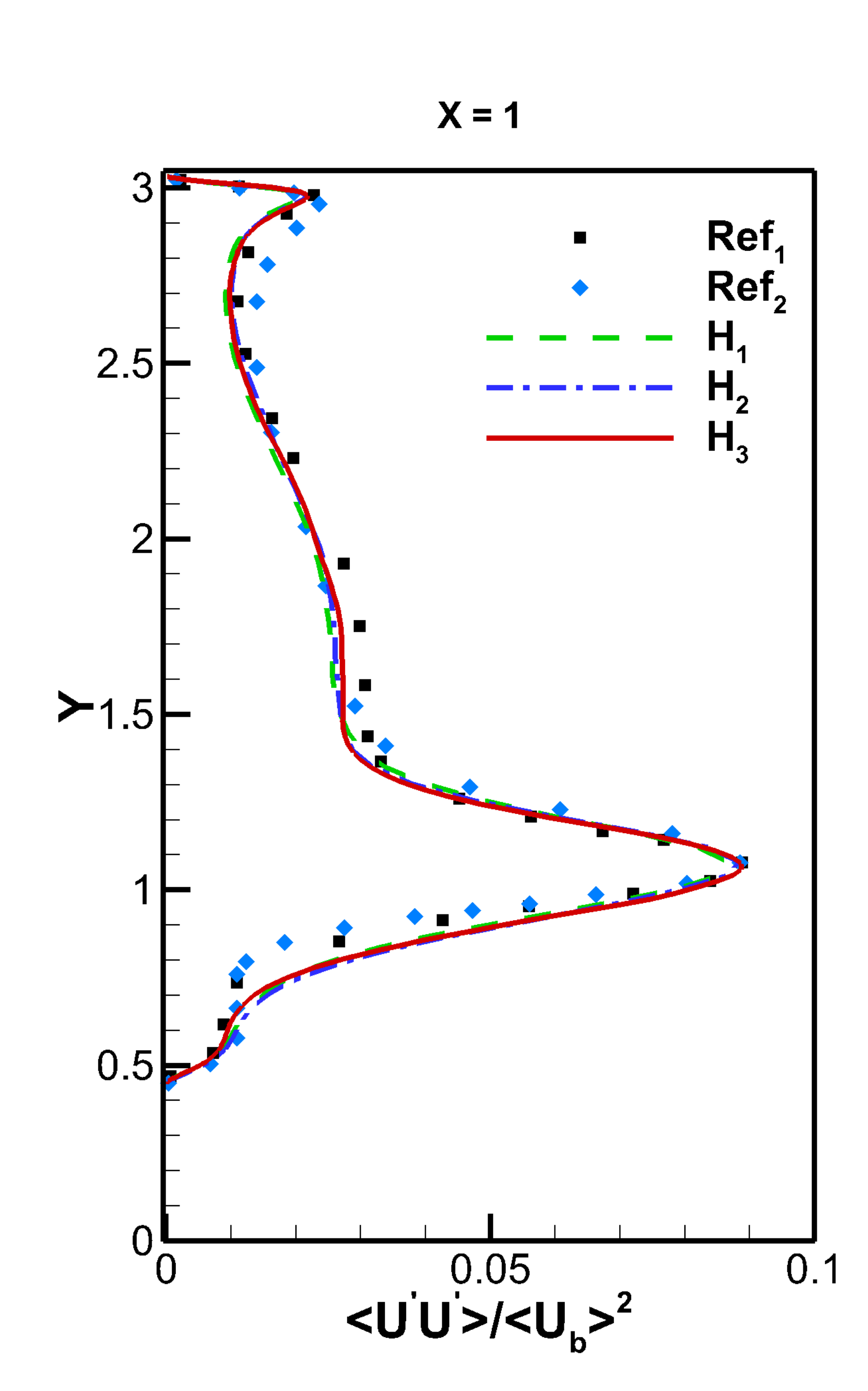}
    \includegraphics[width=0.24\textwidth]{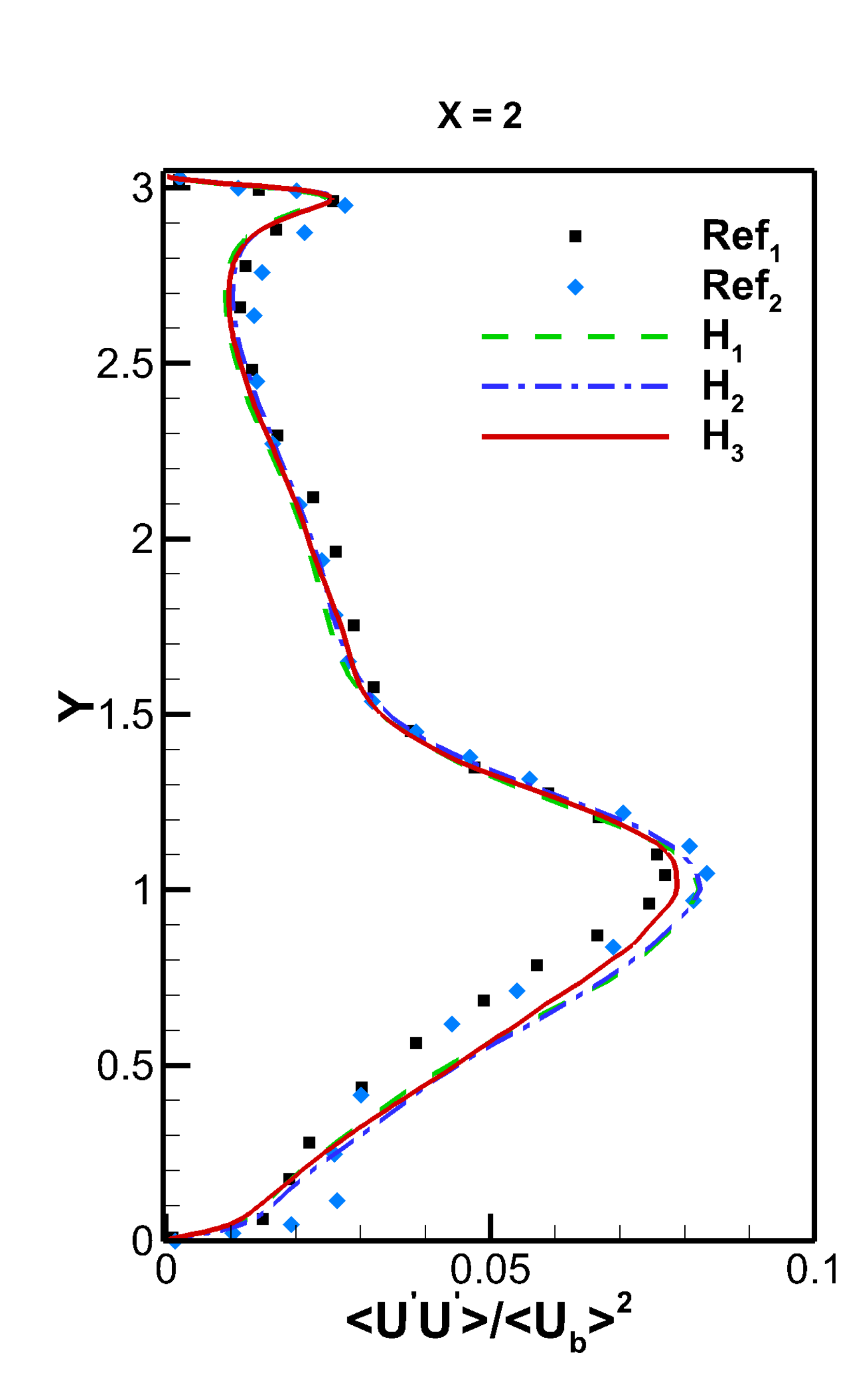}
    \includegraphics[width=0.24\textwidth]{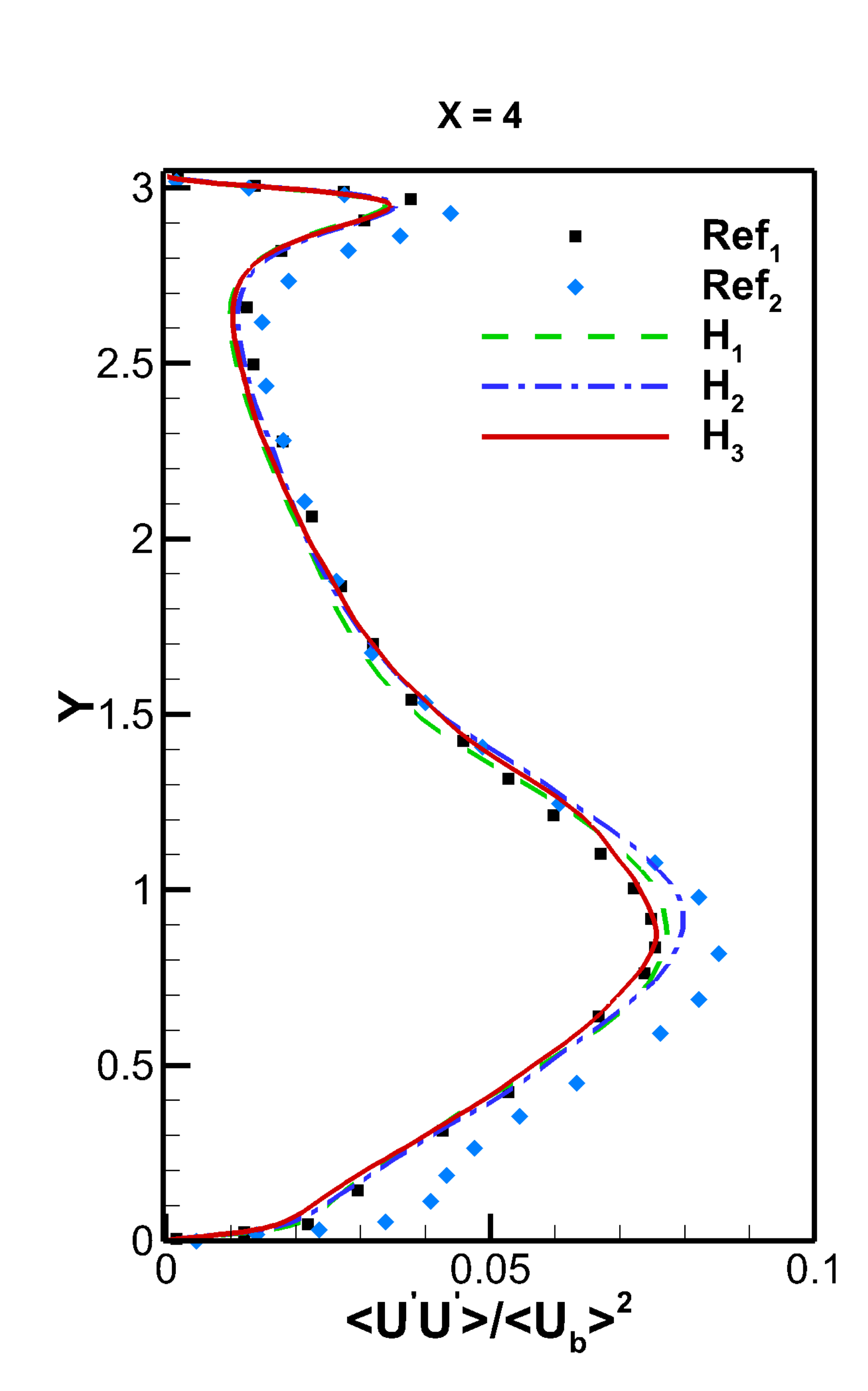}
    \includegraphics[width=0.24\textwidth]{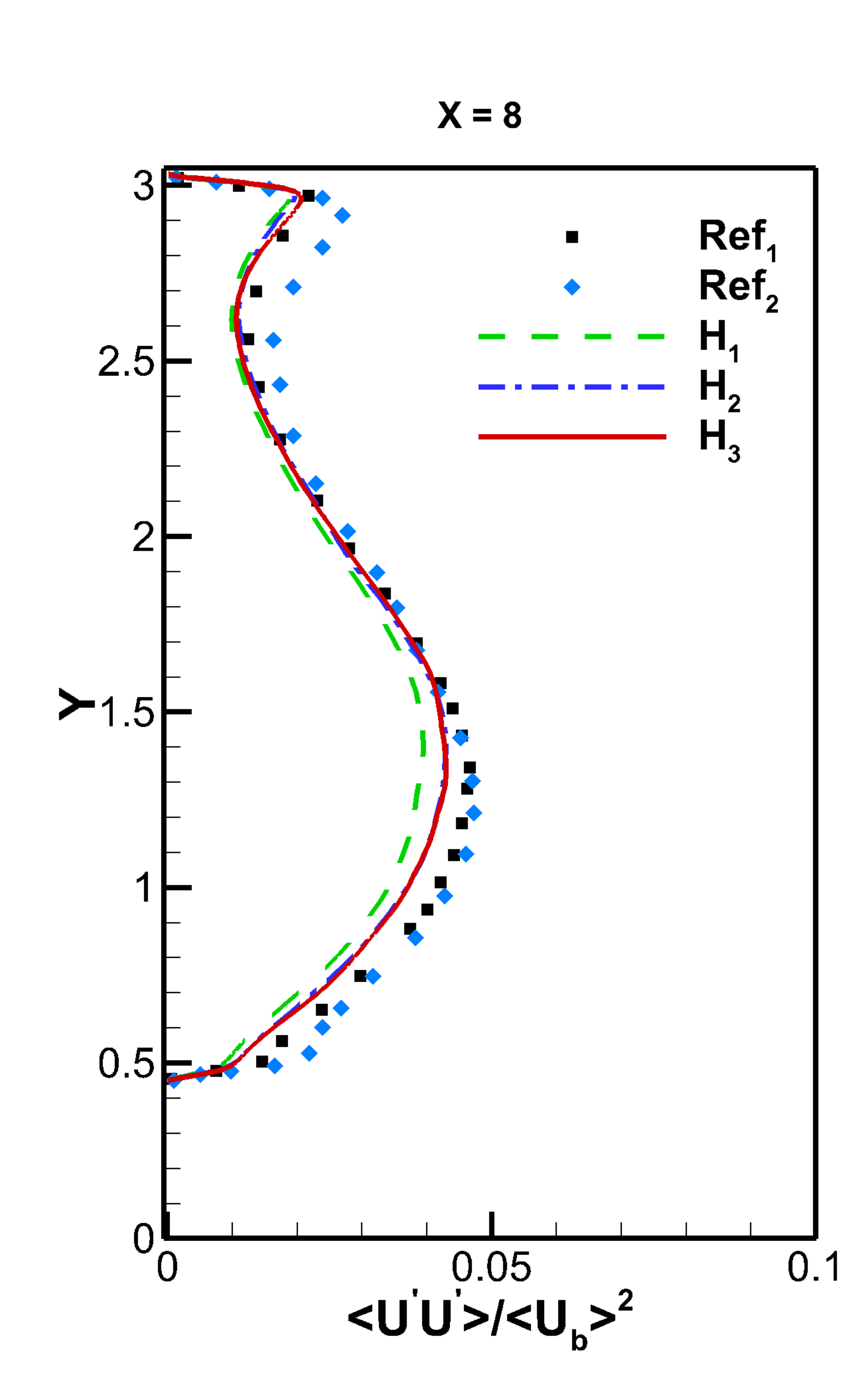}
    \includegraphics[width=0.24\textwidth]{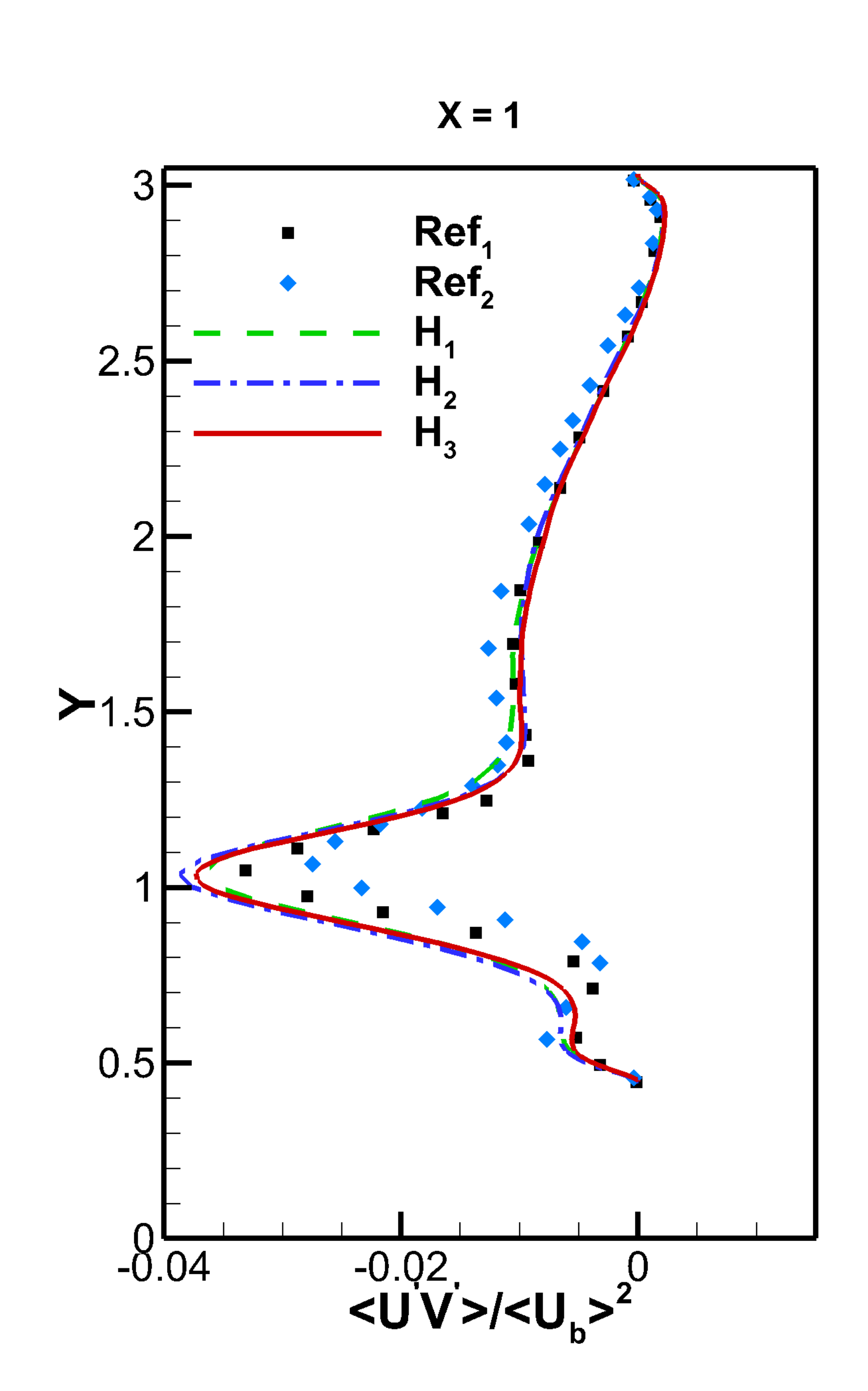}
    \includegraphics[width=0.24\textwidth]{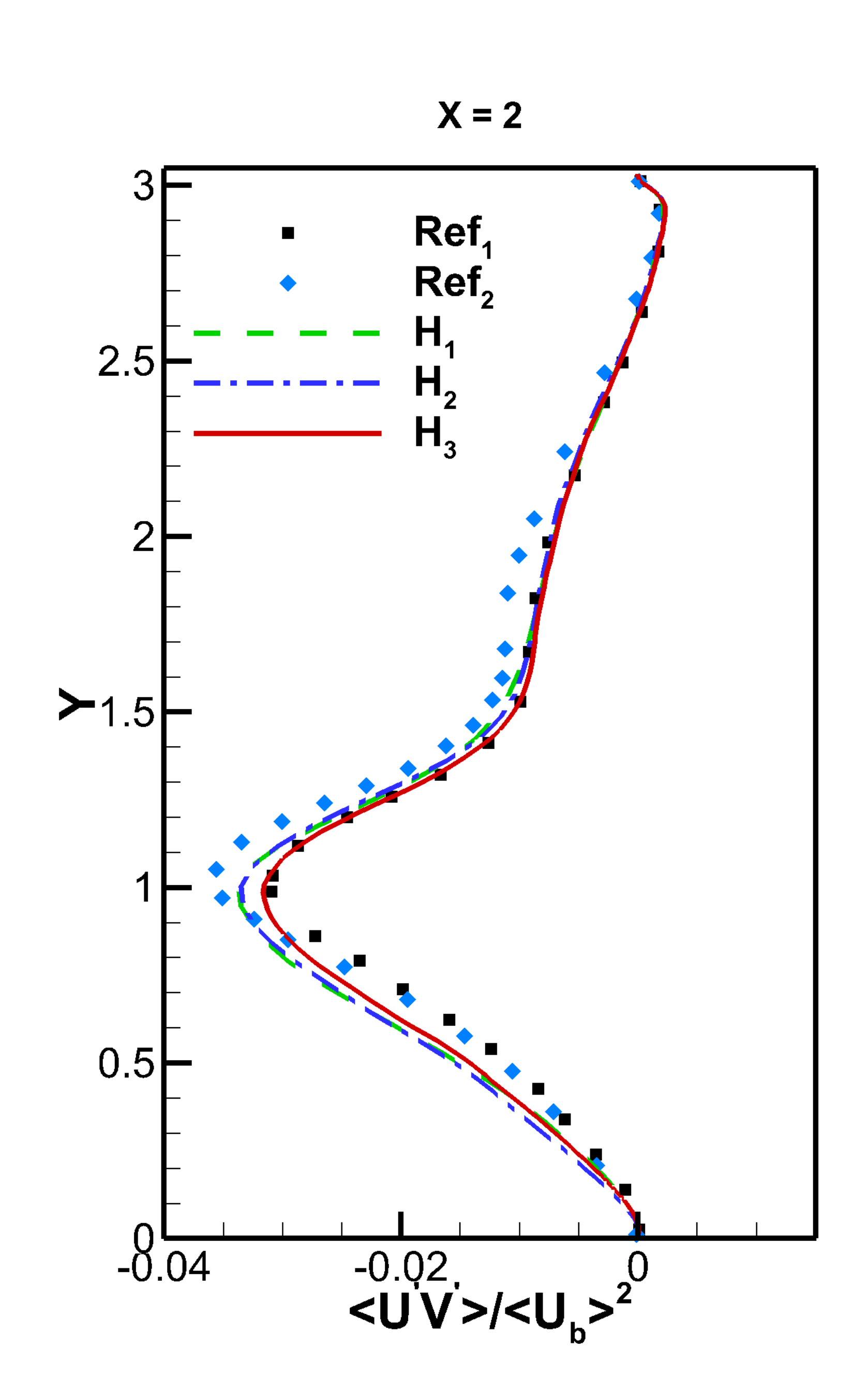}
    \includegraphics[width=0.24\textwidth]{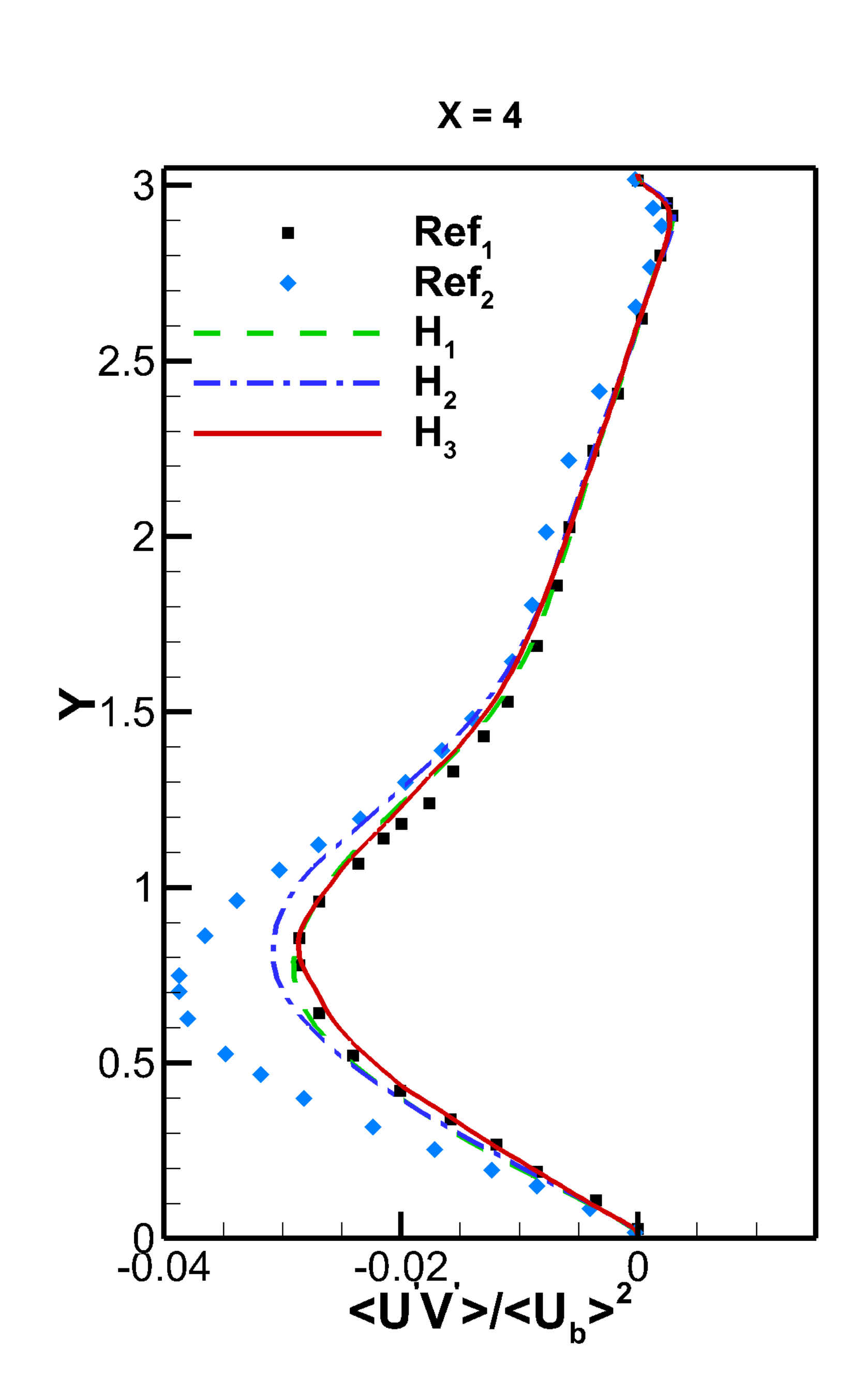}
    \includegraphics[width=0.24\textwidth]{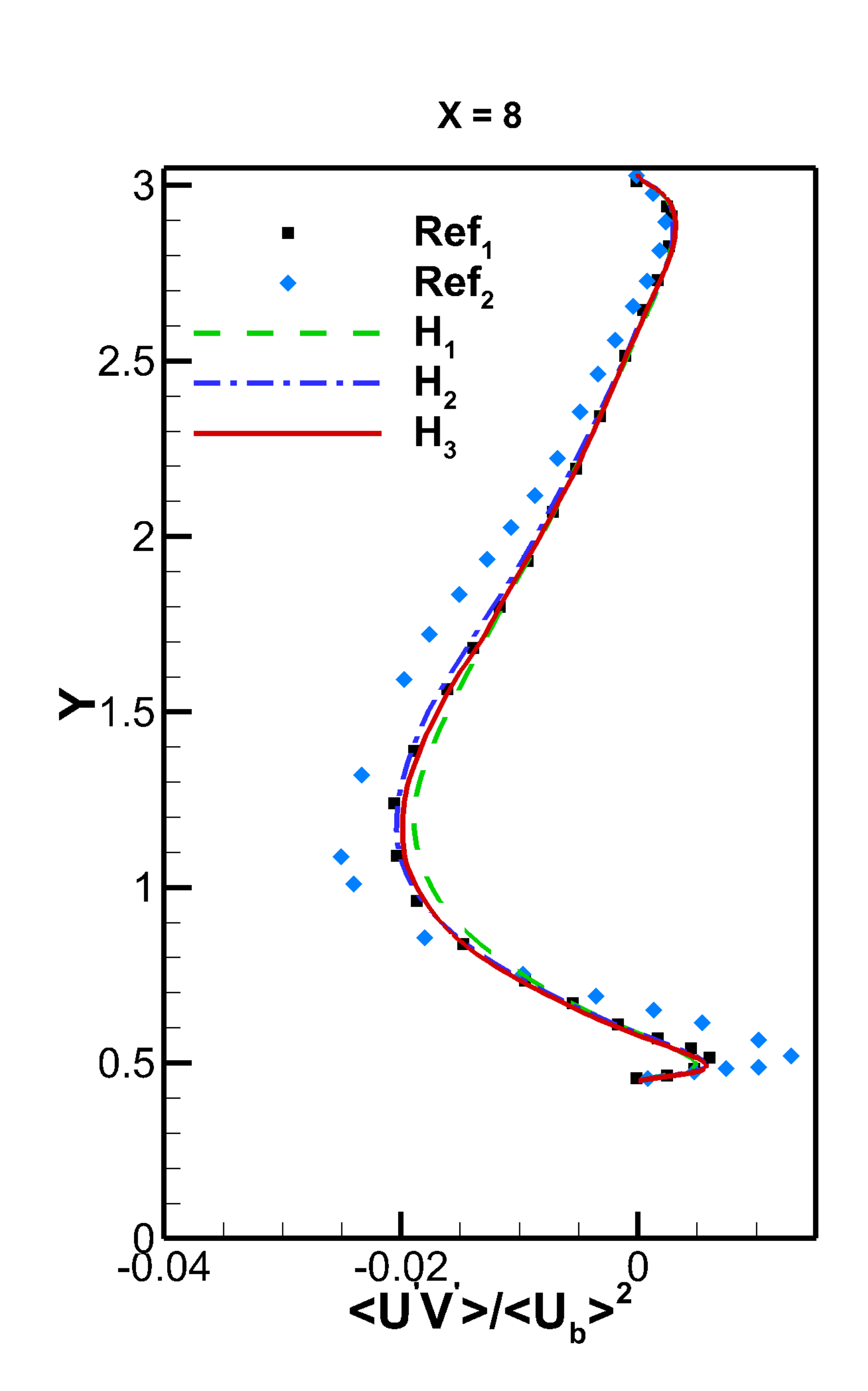}
    \caption{\label{hill_uv_averge} Compressible turbulent flow over periodic hills: profiles of normalized Reynolds stresses $\langle U^{'} U^{'} \rangle /\langle U_b \rangle^2$ and $\langle U^{'} V^{'} \rangle /\langle U_b \rangle^2$.}
\end{figure}
Profiles at $X=1$, $X=2$, $X=4$ and $X=8$ of normalized mean velocities $\langle U \rangle /\langle U_b \rangle$, $\langle V \rangle /\langle U_b \rangle$ and normalized Reynolds stresses $\langle U^{'} U^{'} \rangle /\langle U_b \rangle^2$, $\langle U^{'} V^{'} \rangle /\langle U_b \rangle^2$ are presented in Fig.\ref{hill_u_v_averge} and Fig.\ref{hill_uv_averge}, respectively.
For explicit LES with ADM, the density-weighted velocity and Reynolds
stress are presented, while density-weighted procedure is not adopted in DNS and current iLES. 
The normalized mean streamwise velocity profiles $\langle U \rangle / \langle U_b \rangle$ of HGKS-cur are in good agreement with the DNS solutions. 
The normalized mean wall-normal velocity profiles $\langle
V \rangle / \langle U_b \rangle$ of HGKS-cur are comparable
with the results from the explicit LES. For second-order statistical
Reynolds stresses, the iLES of HGKS-cur is comparable with the
explicit LES. 
However, the explicit LES overpredicts the normalized Reynolds
stresses, especially for $\langle U^{'} V^{'} \rangle/\langle U_b\rangle^2$. 
For this separated turbulent flow, it can be concluded that the explicit LES provides much stronger turbulent fluctuation information than the DNS. Thus, the
explicit LES model may pollute current low-Reynolds number separated turbulent flow. 
While, the solutions from current iLES agree well with the DNS results, and
the over-predicted behaviour seldom appears. 
For Reynolds stresses, case $H_3$ with the finest grids indeed performs better than cases $H_1$ and $H_2$.
However, considering the computational costs of case $H_3$, the improvement is not so worthwhile.
It is implied that coarse grids is enough for iLES when simulating low-Reynolds number separated turbulent flows. 
Overall, current iLES with HGKS-cur is comparable with the explicit LES with ADM using fourth-order finite-volume method \cite{ziefle2008large}. 
HGKS-cur provides a confident numerical tool for compressible
separated flow simulations.

\section{Conclusion}
Within the two-stage fourth-order framework, HGKS  in the general curvilinear coordinate (HGKS-cur) is developed to simulate the compressible wall-bounded turbulent flows. 
Based on the coordinate transformation, the BGK equation is transformed from physical space to computational space. 
To deal with the meshes given by discretized points,  the geometrical metrics
need to be reconstructed at quadrature points of control volumes and cell interfaces by the dimension-by-dimension Lagrangian interpolation. 
To achieve high-order accuracy, WENO reconstruction is implemented to reconstruct the cell averaged Jacobian and the Jacobian-weighted conservative variables.  
The two-stage fourth-order method, which was developed for spatial-temporal
coupled flow solvers, is used for temporal discretization. 
The numerical tests for inviscid and laminar flows validate the accuracy and
geometrical conservation law of HGKS-cur. 
As a direct application, current scheme is implemented for iLES in compressible wall-bounded turbulence, 
including the compressible turbulent channel flow and compressible turbulent flow over periodic hills. 
The simulation results are in good agreement with the refereed spectral method and the high-order finite-volume method.
Current work demonstrates the capability of HGKS-cur as a powerful tool for the numerical simulation in compressible 
wall-bounded turbulent flows and massively separated flows. 
More challenging examples using HGKS-cur at higher Mach numbers and different flow configurations will be investigated in the future.

\section*{Ackonwledgement}
This research is supported by National Natural Science Foundation of China (11701038, 91852114, 11772281),
the Fundamental Research Funds for the Central Universities, and the National Numerical Windtunnel project.
The authors would like to thank TaiYi supercomputers in the SUSTech for providing high performance computational resources.

\end{document}